\documentclass[twocolumn]{aastex6}
\usepackage{multirow}
\usepackage{color}
\usepackage{amsmath}

\newcommand{\vzero}{\mathbf{0}}

\newcommand{\vb}{\mathbf{b}}

\newcommand{\vW}{\mathbf{W}}
\newcommand{\vX}{\mathbf{X}}

\newcommand{\vx}{\mathbf{x}}
\newcommand{\vy}{\mathbf{y}}

\newcommand{\vbeta}{\boldsymbol{\beta}}
\newcommand{\tbeta}{\tilde{\beta}}
\newcommand{\vtheta}{\boldsymbol{\theta}}

\newcommand{\vdelta}{\boldsymbol{\delta}}
\newcommand{\vgamma}{\boldsymbol{\gamma}}

\newcommand{\pec}{\mathrm{pec}}

\newcommand{\sitom}{Si~{\small II}~\lambda5972}

\newcommand{\siyem}{Si~{\small II}~\lambda6355}

\newcommand{\sito}{Si~{\small II}~$\lambda$5972}
\newcommand{\sirp}{Si~{\small II}~$\lambda$4000}
\newcommand{\siye}{Si~{\small II}~$\lambda$6355}
\newcommand{\cahk}{Ca~{\small II}~H\&K}
\newcommand{\mgii}{Mg~{\small II}}
\newcommand{\feii}{Fe~{\small II}}
\newcommand{\siW}{Si~{\small II}~`W'}
\newcommand{\oit}{O~{\small I}~triplet}
\newcommand{\caii}{{Ca~{\small II}}}

\newcommand{\tmax}{\mathrm{max}}

\fullcollaborationName{The Friends of AASTeX Collaboration}

\begin{document}


\title{Characterization of Type Ia  Supernova Light Curves
Using Principal Component Analysis of Sparse Functional Data} 

\author{Shiyuan He\altaffilmark{1}, Lifan Wang\altaffilmark{2,3,4} \&
  Jianhua Z. Huang\altaffilmark{1}} 

\altaffiltext{1}{Institute of Statistics and Big Data, Renmin University of China, Beijing, China}
\altaffiltext{2}{George P.~and Cynthia W.~Mitchell Institute for
  Fundamental Physics \& Astronomy, Department of Physics \&
  Astronomy, Texas A\&M University, College Station, TX, USA} 
\altaffiltext{3}{Purple Mountain Observatory, Nanjing, China} 
\altaffiltext{4}{Corresponding author: wang@physics.tamu.edu}
\altaffiltext{5}{Department of Statistics, Texas A\&M University, College Station, TX, USA}

\begin{abstract}
 With growing data from ongoing and future supernova surveys it is
 possible to empirically quantify the shapes of SNIa light curves in
 more detail, and to quantitatively relate the shape parameters with
 the intrinsic properties of SNIa. Building such relationship is
 critical in controlling systematic errors associated with supernova
 cosmology. Based on a collection of well-observed SNIa samples
 accumulated in the past years, we construct an empirical SNIa light
 curve model using a statistical method called the functional
 principal component analysis (FPCA) for sparse and irregularly
 sampled functional data. Using this method, the entire light curve of
 an SNIa is represented by a linear combination of principal component
 functions, and the SNIa is represented by a few numbers called
 principal component scores. These scores are used to establish
 relations between light curve shapes and physical quantities such as
 intrinsic color, interstellar dust reddening, spectral line strength,
 and spectral classes. These relations allow for descriptions of some
 critical physical quantities based purely on light curve shape
 parameters. Our study shows that some important spectral feature
 information is being encoded in the broad band light curves, for
 instance, we find that the light curve shapes are correlated with the
 velocity and velocity gradient of the Si II $\lambda$6355 line. This is
 important for supernova surveys, e.g., LSST and WFIRST. Moreover, the
 FPCA light curve model is used to construct the entire light curve
 shape, which in turn is used in a functional linear form to adjust
 intrinsic luminosity when fitting distance models. 
\end{abstract}






\section{Introduction}

Type Ia supernovae (SNIa) are ``standardizable'' candles for cosmology
study. They have relatively
uniform intrinsic peak luminosity after explosion, and thereby provide
us a very important tool for cosmological distance 
measurement. Observations of SNIa provided the first direct evidence of 
the accelerating expansion of the universe
\citep{riess1998observational, perlmutter1999measurements}.  

SNIa have small inhomogeneity in their peak 
magnitude, which can be further reduced by correlating their intrinsic
luminosity with  their
color at peak brightness and optical light curve width \citep{Pskovskii:1977,Phillips:1993ng}.  
Generally, brighter supernovae
have broader light curves and bluer colors, and dimmer supernovae have
narrower light curves and redder colors. Adjusting these effects
usually reduces the dispersion of distance modulus prediction to
$\sim$ 0.15 mag \citep{Tripp:1999}.   
The remaining residual scatter is due to a combination of observational error and
intrinsic supernova magnitude dispersion, and is difficult to disentangle.
Several methods have been published to further reduce this
magnitude dispersion. \cite{blondin2011spectra} included an additional
spectral feature in the classical distance prediction
model. \cite{WangX:2009} sub-classified SNIa 
into two groups   using the expansion velocity inferred from \siye\ 
line, and found significant reduction 
of peak magnitude dispersion.  
These improvements of distance
prediction models still depend largely on the 
$\Delta M_{15}$ \citep{Phillips:1993ng} or an equivalent shape parameter that
measures the width of the light curves. 
From the color magnitude 
evolution, \cite{wang2003multicolor} developed the CMAGIC magnitude to
substitute the peak magnitude.
Compared with other published distance models, the CMAGIC method is
strikingly  successful \citep{2006ApJ...644....1C,WangStrovink}.
\replaced{Recent}{Some} studies reveal also that the peak-to-tail ratio is correlated
with the intrinsic luminosity of SNIa and that the success of CMAGIC is expected from theoretical models of SNIa explosion and radiative transfer \citep{2010ApJ...710..444H,2017ApJ...846...58H}.  

\cite{2017ApJ...846...58H} discussed the recent improved
understanding of the physical processes in SNIa explosion. 
Theoretical models  suggested that different chemical layers and 
radiative processes are active at different light curve phases
\citep{1996ApJ...472L..81H,kasen2016magnetar}. 
In particular, the CMAGIC linear region 
\citep{wang2003multicolor,2006ApJ...641...50W} corresponds to $\sim$ 5-30
days past optical maximum for a normal SNIa. During these epochs,
SNIa spectroscopic data are \deleted{remarkably} more uniform than that of
pre-maximum or around maximum. 
For light curves in the $R$ and redder bands,
their secondary bump is caused by
a very different physical process than the first peak
\citep{kasen2016magnetar}. It is desirable that the corresponding
strengths of these features can be extracted robustly from empirical
modeling of the light curves as well.

An essential step in cosmology studies with SNIa is the construction of an
empirical multicolor light curve model. 
This light curve model can be used to fit observed SNIa light curves, calculate
 their peak brightnesses, and characterize their light curve shapes.  
Existing methods include light curve stretch method
\citep{Goldhaber:2001}, MLCS \citep{riess1996precise, 
  jha2007improved}, SiFTO \citep{conley2008sifto}, 
SALT \citep{guy2005salt}, and SALT II \citep{guy2007salt2}. 
Among these methods, the MLCS directly
models light curve data with 
vectorized templates on a hypothesized grid. The SiFTO, SALT and SALT~II
are based on spectral modeling.  

In this paper we develop a purely data-driven light curve model using
observations of only light curves but not of spectra, and without
using any pre-specified structure. \added{Since the light curves are not
observed at a common grid, we can not directly apply the standard
principal component analysis to build an empirical light curve
model. On the other hand, interpolation of the original light curve
observations to a grid will create errors, which can be very large
when the light curve is very sparsely observed.} To overcome these
challenges, we treat SNIa light curves as sparse and 
irregularly observed functional data and apply the functional
principal component analysis 
\citep[FPCA,][]{james2000principal,  zhou2008joint} 
 developed in the statistics literature. Using FPCA, an SNIa light
 curve is represented  as a linear combination of a mean function and
 a few principal  component functions. The coefficients appeared in this linear
combination are called principal component scores or scores for
short. We train the mean function and the principal component
functions using a training dataset from a collection of observed SNIa
light curves. After the mean and principal component functions are
learned using the training data, we can characterize a new light curve
using its corresponding scores. 

The principal component scores thus obtained parametrize the shape of the
light curve and provide abundant information of SNIa. The primary goal
of this paper is to explore the potential of the extracted scores in
explaining important physical quantities, such as intrinsic color,
interstellar dust reddening, spectral line strength, and spectral
classes. 
We show that our FPCA light curve model provides a flexible and
effective light curve shape characterization. One direct application
is \deleted{a simple parametrization of the intrinsic light curves of SNIa, which
leads to well constrained intrinsic properties of SNIa, and} a method for 
robust color excess determination. Moreover, 
by exploring the relations between the principal component scores and 
spectral features, we open the 
possibility of inferring SNIa spectral information from purely light
curve data and therefore provide the opportunity of more precise
K-correction  and distance prediction within subgroups of SNIa. Finally, our light curve
model is used to produce a new light curve shape parameterization
as a better constrained alternative to the classical  $\Delta M_{15}$
parameter in distance prediction.

This paper consists of eight sections. 
\S\ref{sec:data} describes the dataset used in this study. 
The mathematical framework of FPCA is given
in \S\ref{sec:model}. Then, \S\ref{sec:lcfits} reports the 
result of applying FPCA to the dataset in \S\ref{sec:data},
where data from each of $BVRI$ bands are used to train a FPCA model
separately.  After that, \S\ref{sec:scorecolor} presents
\replaced{the possibility of nonlinear dimension reduction and}{the
  result of estimating color excess}. 
In \S\ref{sec:specfits}, we study the relations
between FPCA light curve scores and spectral properties of SNIa. 
At last, in \S\ref{sec:distance}, we examine the precision of distance
determination based on our FPCA light curve model, using
both magnitudes at $B$ maximum and magnitudes deduced with the
CMAGIC method as distance indicators. \S\ref{sec:joint} presents the
results of training the FPCA model using data from $BVRI$ bands
together. Major results and conclusions of this paper are
summarized in \S\ref{sec:discussions}.

\section{The Dataset}\label{sec:data}

Our data sample derives from SNIa light curves published by 
the Lick Observatory Supernova  Search (LOSS)
\citep{ganeshalingam2010results}, the Carnegie Supernova Project (CSP)
\citep{contreras2010carnegie, stritzinger2011carnegie}, and the
Harvard-Smithsonian Center for Astrophysics (CfA)
\citep{hicken2009cfa3, hicken2012cfa4}. We restrict our studies to
light curves from these publications, and in $BVRI$ bands only. 

The selection criterion is based on the light curve time and color coverage. 
Each SN in our sample must have at least one observation within 5 days before the light 
curve maximum, and at least one observation within 5 days after the
maximum. This is required for all four filter bands.
These selected supernovae are nearby supernovae, with CMB redshift
$z<0.1$.  
\replaced{K-correction is}{Both K-correction and S-correction are} applied to the data so that all the light curve
magnitudes are transformed to the standard rest-frame Kron-Cousins
$BVRI$ bands from their closest natural system filters. These
corrections are performed using the SNooPy package of
\cite{burns2010carnegie}, 
\added{and the employed SED model is from the work of \cite{hsiao2007k}}
All data are corrected for the Galactic extinction using the
dust map of \cite{schlafly2011measuring}, \added{while the extinction of
the hosting galaxy is not corrected.}

When constructing the proposed FPCA model, we removed 
SN~2008bf from the LOSS observation, due to an inconsistency of $V$
band magnitude with the CfA observation \citep{ganeshalingam2010results}. 
SN~1999ej is removed as it is too faint for the observed redshift for unknown reasons
\citep{reindl2005reddening, wang2006determination, jha2007improved}.
Two peculiar supernovae are also removed:
SN2005hk \citep{stanishev2006peculiar} and 
SN2008ae \citep{foley2009sn}. 
Nonetheless, these ``peculiar'' SNIa can still be analyzed by FPCA model
although they do not contribute to the construction of the model.  
After the preprocessing, a total number of 111 supernovae remain for analysis.

\section{The FPCA Model} \label{sec:model}

This section presents the construction, model training, and new light
curve fitting  of the functional
principal component model (FPCA) for SNIa. 

\subsection{Model construction} 
\label{sec:modelconstruction}
Fix a supernova indexed by $s$, let $l_{s\lambda}(t)$ be its light curve as a function
of Julian date $t$ for filter (or band) $\lambda\in \{B,V,R,I\}$. 
Let $b_{s\lambda}$  be the epoch of its peak magnitude, i.e., 
\begin{equation} \label{equ2:max}
m_{s\lambda} = \min_t l_{s\lambda}(t)=l_{s\lambda}(b_{s\lambda}),
\end{equation}
with the peak magnitude $m_{s\lambda}$. 
Since the light curve around the peak epoch is of central interest for
cosmology, we transform $l_{s\lambda}(t)$ to a function of phase $q$.
Specifically, let $z_s$ be the redshift of the corresponding
supernova, and   define the transformation of time into phase 
by  $q=(t-b_{s\lambda})/(1+z_s)$. 
Then the light curve as a function of phase $q$ is
\begin{equation}\label{eq:standardization}
g_{s\lambda}(q) = l_{s\lambda}(t) = l_{s\lambda}\left(q(1+z_s)+b_{s\lambda}\right).
\end{equation}
We usually concentrate on the phase range from $q_{\min}(<0)$ 
to $q_{\max} (>0)$ around the peak epoch ($q = 0$),
and so $q\in [q_{\min}, q_{\max}]$ in the above expression.

Using a truncated version of the functional principal component
expansion \citep{james2010functional}, we represent each light
curve as a basis expansion
\begin{equation}\label{equ2:basics}
g_{s\lambda}(q)=m_{s\lambda} +\phi_{0\lambda}(q)+
\sum_{k=1}^K \beta_{s\lambda}^{(k)}\phi_{k\lambda}(q),
\end{equation}
where $\phi_{0\lambda}(q)$ is the mean function, 
$\phi_{1\lambda}(q), \cdots, \phi_{K\lambda}(q)$ are 
$K$ fixed principal component functions for filter $\lambda$. 
\added{Since these functions are are unique to each filter and trained
  separately, we refer (\ref{equ2:basics}) as a filter-specific FPCA model, and fs-FPCA for short.} 
The $\beta_{s\lambda}^{(1)}, \cdots,
\beta_{s\lambda}^{(K)}$ are the coefficients in the basis expansion called
principal component scores or scores for short.  
To ensure that $m_{s\lambda}$ is the peak magnitude defined in equation (\ref{equ2:max}),
we require that $\phi_{k\lambda}(0)=0$ for $k=0,\cdots,K$.
For identifiability, we also require that $\phi_{k\lambda}(s)$'s  are
orthonormal to each other for $k \ge 1$. This means 
$\int \phi_{k\lambda}(q) \phi_{k'\lambda}(q) \,dq=\delta_{kk'}$, with
$\delta_{kk'} =1$ if $k=k'$ and $\delta_{kk'}=0$ otherwise. 
It is a common practice in applying FPCA that
 the $\phi_{1\lambda}(q), \phi_{2\lambda}(q), \cdots, \phi_{K\lambda}(q)$ 
are ordered by decreasing importance, where the ``importance'' is
measured by the ability to explain total variability of the data.

Since we have aligned all light curves at their peaks when applying
Equation~\eqref{equ2:basics}, the mean curve $\phi_{0\lambda}(q)$
 specifies the average light curve shape for filter $\lambda$, 
while the principal component functions $\phi_{k\lambda}(q)\, (k =
1,\cdots, K)$
 provide additional adjustments for filter $\lambda$, with the
amount of adjustment controlled by the parameters
$\beta^{(k)}_{s\lambda}$. 

With the proposed model~(\ref{equ2:basics}), when the mean and
principal component functions are given, each supernova light curve is characterized
by a group   of parameters:  (1) the peak magnitude $m_{s\lambda}$;
(2) the date  $b_{s\lambda}$ at peak magnitude 
(implicitly coded in phase $q$); and (3) the $K$-vector of shape parameters
(or principal component scores)
$\boldsymbol{\beta}_{s\lambda} 
=(\beta_{s\lambda}^{(1)},\beta_{s\lambda}^{(2)},\cdots,\beta_{s\lambda}^{(K)})^T$.
The parameters $m_{s\lambda}$, $b_{s\lambda}$ and
$\vbeta_{s\lambda}$'s are unique to each light curve. 

In the FPCA model~(\ref{equ2:basics}), the mean and principal component functions are
specific to each filter. We can force these functions to be the same for
all filters, leading to the filter-vague FPCA model (abbreviated as fv-FPCA),  
\begin{equation} \label{eqn:jointbandmodel}
g_{s\lambda}(q)=m_{s\lambda} +\phi_{0}(q)+
\sum_{k=1}^K \beta_{s\lambda}^{(k)}\phi_{k}(q),
\end{equation}
and the methodology developed below still applies with a
straightforward modification. 
Notice the functions $\phi_0(q),\cdots, \phi_K(q)$ are common to all
filters.  We will focus on the fs-FPCA model from now on, 
but see Section~\ref{sec:joint} for some results and
discussion of the merit of the fv-FPCA model.

Next we show how to estimate the mean and principal component
functions in Equation~(\ref{equ2:basics}) using the observed data from
a collection of SNIa.   This procedure is referred to as model training. 

\subsection{Model training}\label{sec:training}
The observed light curves are usually recorded at 
sparsely sampled time points and affected by noise. This can be
described as a signal-plus-noise model as follows. Suppose, 
for the light 
curve indexed by $s$ and in the filter $\lambda$, there are totally $n_{s\lambda}$ 
observations: at time points $t_{s\lambda j}$ with magnitude
$y_{s\lambda j}$ 
and magnitude uncertainty $\sigma_{s\lambda j}$ for $j = 1,2, \cdots,
n_{s\lambda }$. 
This time series of magnitude observations can be decomposed 
as the summation of the underlying light curve function $l_{s\lambda }(t)$ 
plus noise, i.e., 
\begin{equation}\label{eq:s+n}
y_{s\lambda j}
= l_{s\lambda}(t_{s\lambda j})+\sigma_{s\lambda j}\, \epsilon_{s\lambda j},
\end{equation}
where $l_{s\lambda }(t) = g_{s\lambda} ((t-b_{s\lambda})/(1+z_s)) $ with
$g_{s\lambda}(q)$ following 
the basis expansion \eqref{equ2:basics}, 
$\sigma_{s\lambda}$ is the standard deviation of measurement uncertainty, 
and $\epsilon_{s\lambda j}$ is a random variable with zero mean and unit variance. 

Let $ \mathbf{b}(q)= (b_1(q),b_2(q),\cdots, b_P(q) )^T$
be a known orthonormal basis system for functions defined
on $[q_{\min}, q_{\max}]$,  with
the dimension $P$ much larger than $K$. We represent
the $K+1$ unknown functions in \eqref{equ2:basics} by this
rich basis so that
$\phi_{0\lambda}(q)=\mathbf{b}(q)^T\boldsymbol{\theta}_{0\lambda}$,
and $ \boldsymbol{\phi}_{\lambda}(q)^T=(\phi_{1\lambda}(q),\cdots, \phi_{K\lambda}(q))=
\mathbf{b}(q)^T\boldsymbol{\Theta}_{\phi\lambda}$, where
$\boldsymbol{\theta}_{0\lambda}$ is a $P$ dimensional vector,  
and $\boldsymbol{\Theta}_{\phi\lambda}$ is a $P\times K$ matrix.
As a consequence, model (\ref{equ2:basics}) becomes
\begin{equation} \label{equ2:splineform}
\begin{split}
g_{s\lambda}(q)&=m_{s\lambda}+\phi_{0\lambda}(q)  +\sum_{k=1}^K
\beta_{s\lambda}^{(k)}\phi_{k\lambda}(q) \\
&=m_{s\lambda}+\mathbf{b}(q)^T\boldsymbol{\theta}_{0\lambda}+
\mathbf{b}(q)^T\boldsymbol{\Theta}_{\phi\lambda}\boldsymbol{\beta}_{s\lambda}
\end{split}
\end{equation}
We require that 
$\boldsymbol{\Theta}_{\phi\lambda}^T 
\boldsymbol{\Theta}_{\phi\lambda}=\mathbf{I}_K$
to guarantee the orthonormality of the principal component functions.
Using the representation in equation~\eqref{equ2:splineform}, we reduce
the problem of estimating $K+1$ unknown functions to the estimation of 
the vector $\boldsymbol{\theta}_{0\lambda}$ and the matrix
$\boldsymbol{\Theta}_{\phi\lambda}$.  

Considering equations~\eqref{eq:s+n} and
\eqref{equ2:splineform} together, the fixed  
and unknown parameters $\boldsymbol{\theta}_{0\lambda}$ and 
$\boldsymbol{\Theta}_{\phi\lambda}$ are estimated from a training dataset
by minimizing the least squares criterion (or $\chi^2$ distance). 

In our implementation of the methodology, the rich basis $\vb(q)$  is
chosen such that it spans the space of cubic spline functions with
$P-4$ equally spaced interior knots on $[q_{min}, q_{max}]$. 
To ensure good statistical properties of the estimated functions,
we add two regularization penalty functions to the least squares criterion,
i.e., a roughness penalty \citep{james2010functional} 
to encourage the  smoothness of the estimated $\phi_{0\lambda}(q)$ and
$\phi_{k\lambda}(q)$ and
a nuclear-norm penalty \citep{cai2010singular} to encourage a small value of $K$.
More details are presented in the Appendix~\ref{appendix:algorithm}.

\subsection{Fitting a new light curve}

After the unknown functions
$\phi_{0\lambda}(q), \cdots, \phi_{K\lambda}(q)$
are estimated for each filter using the training data, 
they are considered as fixed  functions. 
The determination of the shape of a new light curve reduces to
the determination of a few parameters as pointed out at the end of
Section~\ref{sec:modelconstruction}.  
For a new sparsely observed light curve,
we can estimate the parameters $b_{s\lambda}$, $m_{s\lambda}$ and give
\replaced{a predicted}{an estimated} value of the score vector 
$\vbeta _{s\lambda} = (\beta^{(1)}_{s\lambda}, \cdots,
\beta^{(K)}_{s\lambda} )^T$. 
The following procedure can be applied to individual light
curve across different filters and different supernovae. 

We make the assumption that the noise term $\epsilon_{s\lambda j}$
in Equation~(\ref{eq:s+n})
follows the  standard normal distribution. We also assume that
$\boldsymbol{\beta}_{s\lambda}$  
has a zero-mean multivariate normal
distribution whose covariance matrix is estimated from the training data. This is
approximately correct for data with high signal to noise ratios. 
Under these assumptions, we can compute the joint multivariate normal
distribution of $\vy_{s\lambda} = (y_{s\lambda 1}, \cdots,
y_{s\lambda n_{s\lambda}} )^T$ and $\boldsymbol{\beta}_{s\lambda}$.
The new light curve fitting is done in two steps. Firstly, $b_{s\lambda}, m_{s\lambda}$ 
are determined by the method of generalized least squares,
considering equations~\eqref{eq:s+n} and \eqref{equ2:splineform} together
and treating the term involving $\boldsymbol{\beta}_{s\lambda}$ as part of
the error term. Secondly, given the estimated  $b_{s\lambda},
m_{s\lambda}$ and the vector of observations $\mathbf{y}_{s\lambda}$,  we can
\replaced{predict}{estimate}
$\boldsymbol{\beta}_{s\lambda}$ using the conditional expectation
$E(\boldsymbol{\beta}_{s\lambda}| \vy_{s\lambda})$. However, this calculation 
may produce a light curve with an incorrect shape at a region with sparse data.
We fix the problem by imposing several shape constraints 
on the reconstructed light curves---$g_{s\lambda}(q)$ is
monotonically decreasing (in terms of numerical magnitude values, 
i.e.,  monotonically brightening) 
and concave before the peak, and monotonically increasing 
in the phase interval $[35, q_{\rm max}]$. 
We do not impose a shape restriction between the peak epoch and phase
35,  because  the second peak of the $I$
band light curve may exist in this range.
We then maximize the conditional density of $\boldsymbol{\beta}_{s\lambda}$
given $\vy_{s\lambda}$ subject to these constraints.

From the above procedure,  we get the estimated parameters 
$\widehat{m}_{s\lambda}$, $\widehat{b}_{s\lambda}$ 
and $\widehat{\vbeta}_{s\lambda}$ from the actual observations.
The parameter uncertainty is determined by the parametric bootstrap
method \citep{efron1994introduction}. 
The parametric bootstrap procedure with $G$ bootstrap samples is as
follows.   For $g = 1,2,\cdots, G$, using  equations~\eqref{eq:s+n}
and \eqref{equ2:splineform} where the parameters are fixed at the
estimated values,
we generate a sequence of magnitude $y^{(g)}_{s\lambda 1}, \cdots,
y^{(g)}_{s\lambda n}$ at the original observation time, 
according to the model
$$
y^{(b)}_{s\lambda j} 
= \widehat{m}_{s\lambda} 
+ \phi_{0\lambda}(q_j) + \sum_{k=1}^K 
\widehat{\beta}_{s\lambda}^{(k)} \phi_{k\lambda}(q_{s\lambda j}) 
+ \epsilon^{(g)}_{s\lambda j}\, ,
$$
where $\epsilon^{(g)}_{s\lambda j}$'s are sampled from $\mathcal{N}(0,
\sigma _{s\lambda j}^2)$ 
for $j=1,\cdots, n_{s\lambda}$, and the $\sigma _{s\lambda j}$'s are the observation
uncertainties. For the $g$-th bootstrap sample, the
estimation \deleted{and prediction} procedure as described above is
applied to the  generated light curve mangitude $y _{s\lambda 1}^{(g)}, 
\cdots, y _{s\lambda n}^{(g)}$  with observation time $t _{s\lambda
  1},  \cdots, t _{s\lambda}$ 
and uncertainty $\sigma_{s\lambda 1},\cdots, \sigma_{s\lambda n}$.  
Denote the resulting parameter estimates as
$\widehat{m}^{(g)}_{s\lambda}, \widehat{b}^{(g)}_{s\lambda},
\widehat{\vbeta}^{(g)}_{s\lambda}$. The standard deviation of the $G$
bootstrapped values
$\widehat{m}^{(1)}_{s\lambda}, \widehat{m}^{(2)}_{s\lambda}, 
\cdots, \widehat{m}^{(G)}_{s\lambda}$
is an estimate  of the uncertainty of our actual estimate
$\widehat{m}_{s\lambda}$. 
The uncertainty of $\widehat{b}_{s\lambda}$ 
and $\widehat{\vbeta}_{s\lambda}$ is evaluated
in a similar way.

\subsection{Algorithm Iteration}
\added{After one round of  model training and fitting, we obtain better
estimated maximal epoch by finding the maximum of the fitted light
curve.  Given this better estimation, the model
training and fitting is repeated one more time, for more accurate FPCA
model and light curve fitting.}

\section{Model Training Results}\label{sec:lcfits}

\begin{figure}[t]
\centering
\includegraphics[width = 0.4\textwidth]{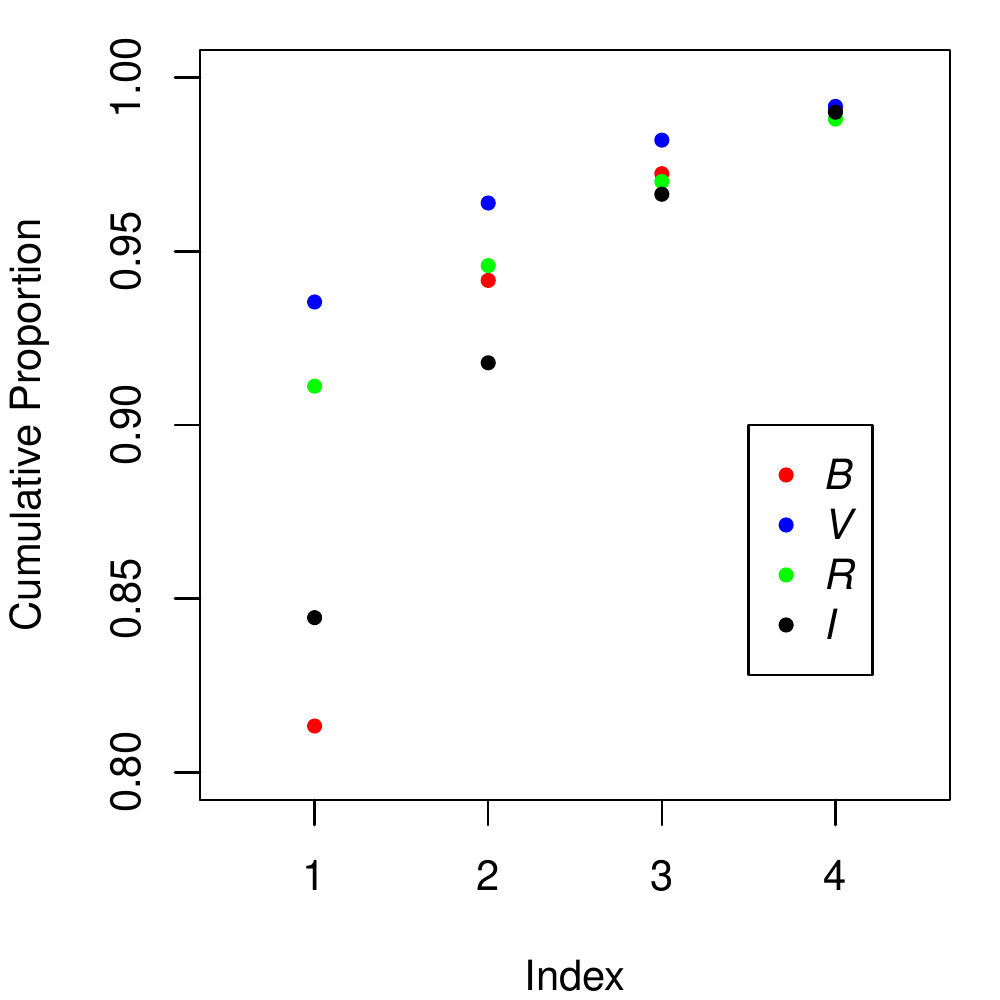}
\caption{The first seven eigenvalues and cumulatively explained
  proportion of variability. The first four principal components together accounts
  for 99.24\% variability in the dataset.
The vertical axis is on the logarithmic scale.} \label{fig:screeplot}
\end{figure}

This section presents the model training results using the selected
dataset as described in Section \ref{sec:data}. In particular, we will discuss
interpretation of the estimated principal component functions and
examine the correlation among scores, color and $\Delta M_{15}$. 
In the following, the $k$-th score for  $B,V,R,I$ band
 is denoted by $\beta^{(k)}_B, \beta^{(k)}_V, \beta^{(k)}_R,
\beta^{(k)}_I$, respectively, for $k = 1,2,\cdots, K$. 
The subscript $s$ is dropped to simplify notation.
Sometimes to further simplify the notation, we drop the band (or filter)
specification in the subscript and denote the $k$-th score as $\beta^{(k)}$.

The FPCA model is constructed in the phase range $(-10,50)$
around the peak epoch, trained for each filter separately.
\added{Figure~\ref{fig:screeplot} plots the cumulative proportion of
variability explained by principal component functions. Red, blue,
green and black points stand for  $B,V,R,I$ band, respectively. For
example, consider the red points for  $B$ band in
Figure~\ref{fig:screeplot}. The first principal component $\phi_{1B}(q)$
accounts for 81.33\% of the total variability of $B$ band data. The first
two principal components, $\phi_{1B}(q)$ and $ \phi_{2B}(q)$, together account
for 94.16\%; the first three explain 97.23\%;
and the first four together explain 99.07\%. We decided to use $K=4$
principal components.}

\deleted{Figure~\ref{fig:screeplot} plots their corresponding 
eigenvalues of the covariance $\mathbf{\Sigma}$ of the score vector $\vbeta$. 
The vertical axis is in the logarithmic scale.
The percentage number is the cumulative proportion of
variability explained in the dataset. For example, the first four
principal component functions
together explain 99.24\% of the total variability.}

\begin{figure*}[t]
\centering
\includegraphics[width=0.99\textwidth]{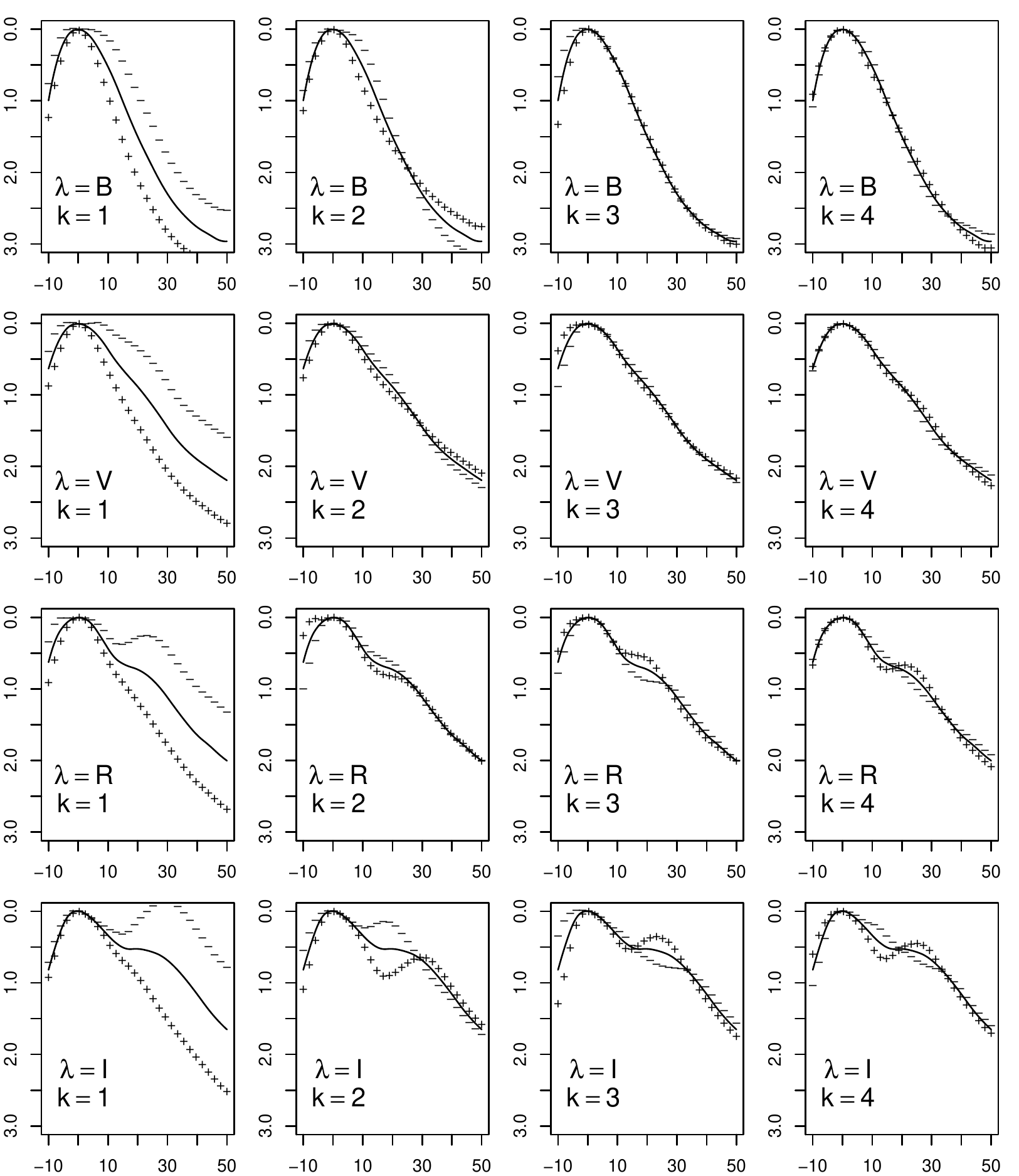}
\caption{Plots of the 
principal component functions for all bands $\lambda \in \{B,V,R,I\}$. The solid
  line in each panel is the mean function $\phi_{0\lambda}(q)$.
 The ``$+$''   points   represent $\phi_{0\lambda}(q) 
  + 2\sigma_{k\lambda}\phi_{k\lambda}(q)$, 
and the ``$-$'' points represent 
$\phi_{0\lambda}(q) -  2\sigma_{k\lambda}\phi_{k\lambda}(q)$. 
   \label{fig:bandpca} } 
\end{figure*}

\begin{figure*}
\centering
\includegraphics[width=0.32\textwidth]{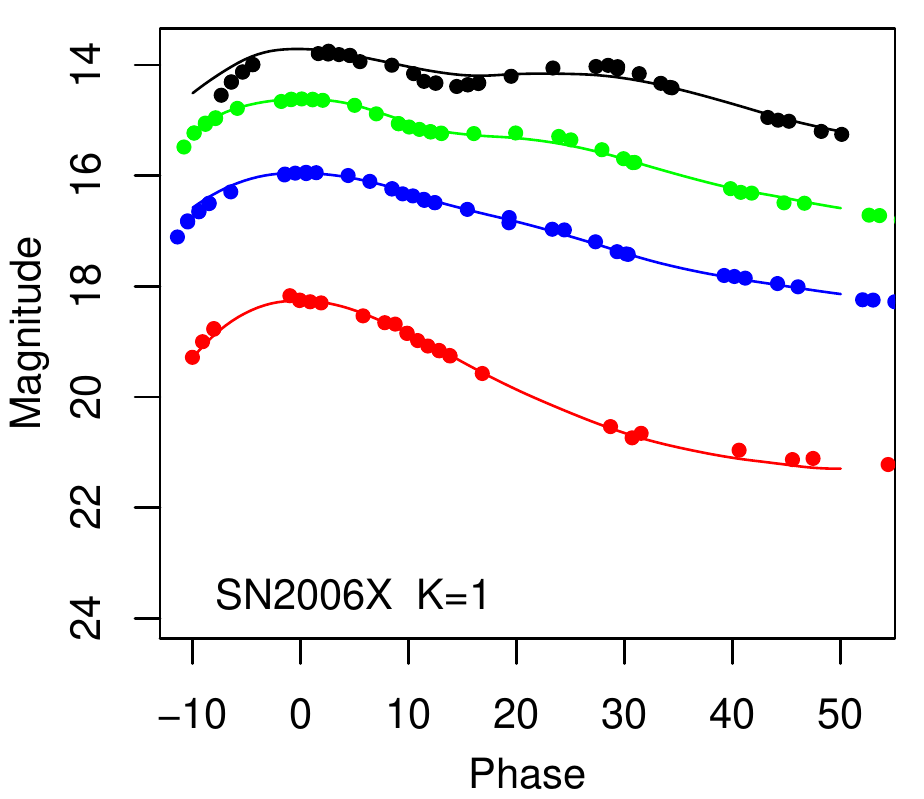} 
\includegraphics[width=0.32\textwidth]{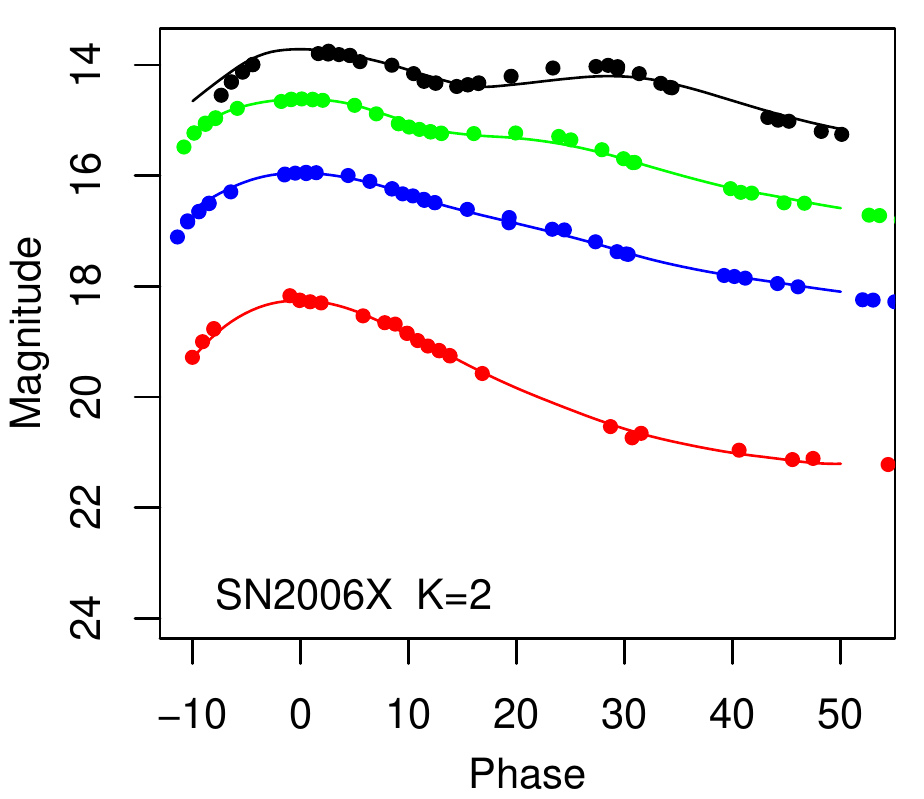} 
\includegraphics[width=0.32\textwidth]{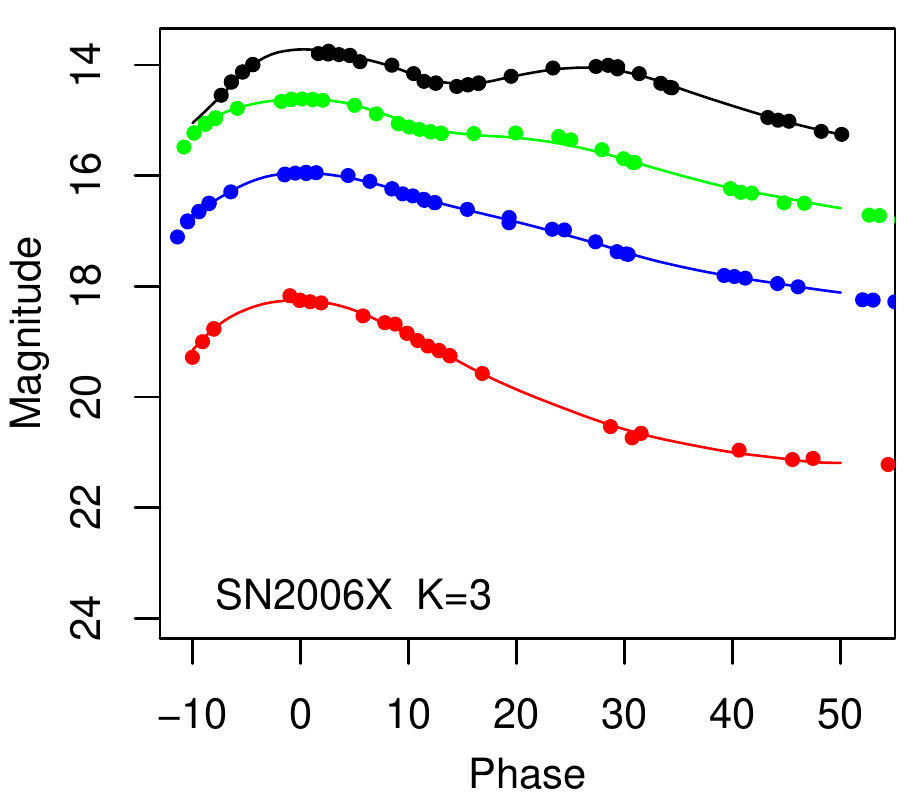} 
\includegraphics[width=0.32\textwidth]{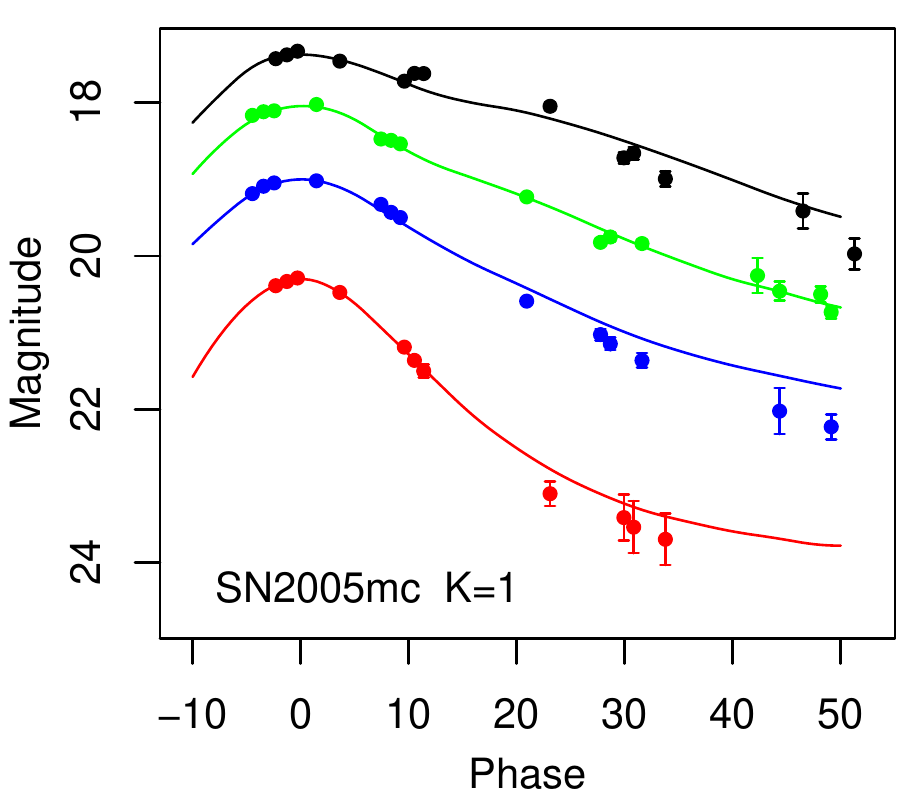} 
\includegraphics[width=0.32\textwidth]{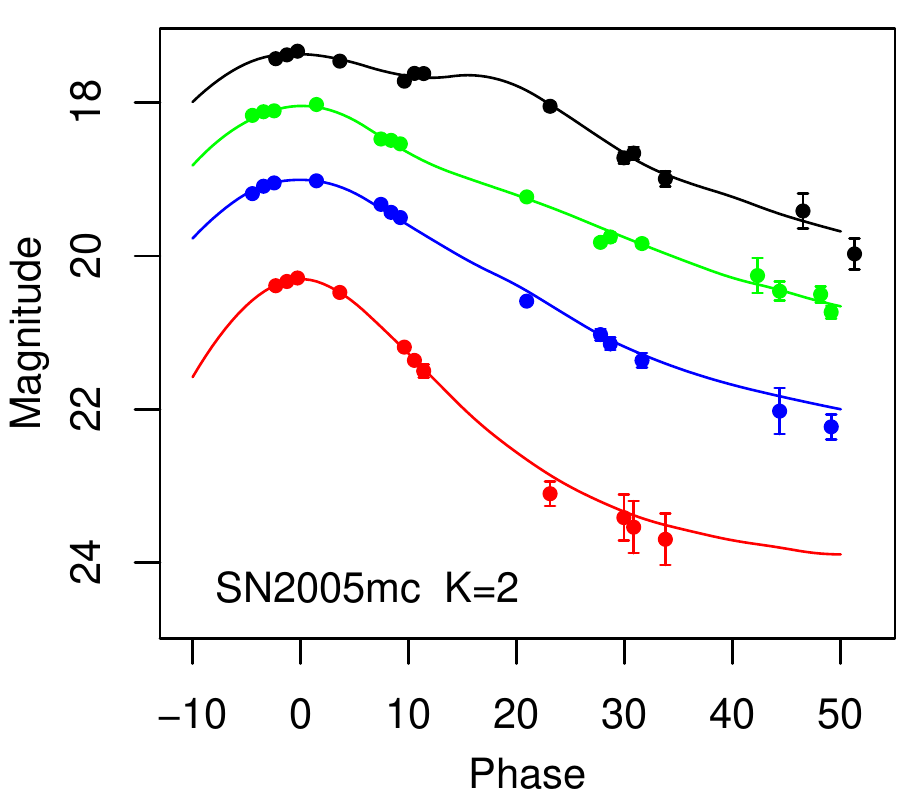} 
\includegraphics[width=0.32\textwidth]{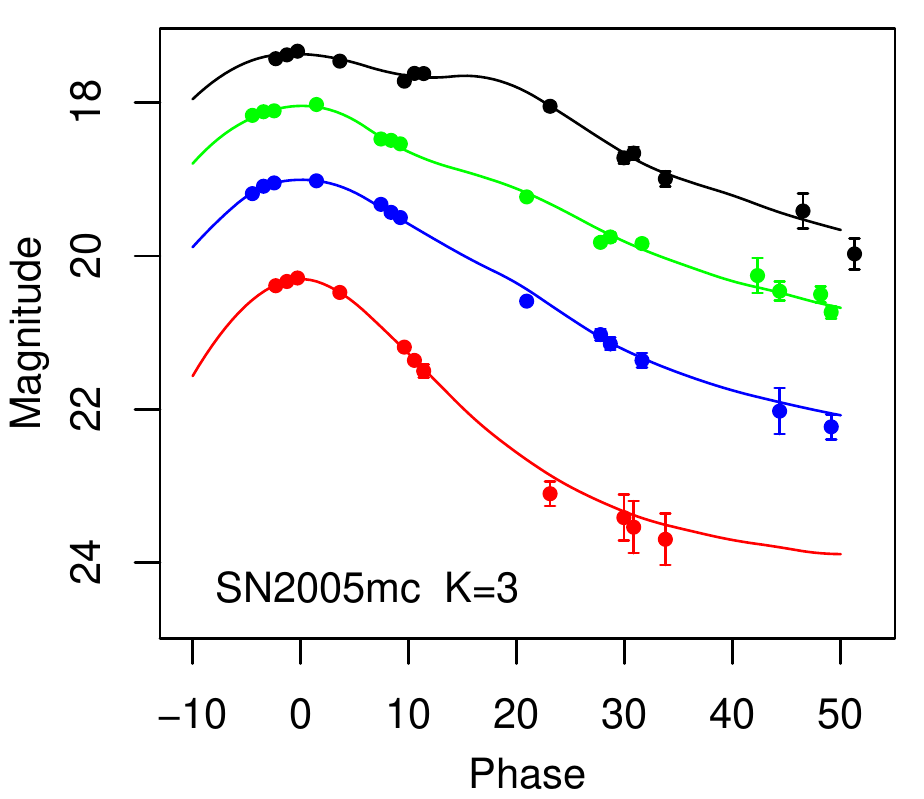} 
\includegraphics[width=0.32\textwidth]{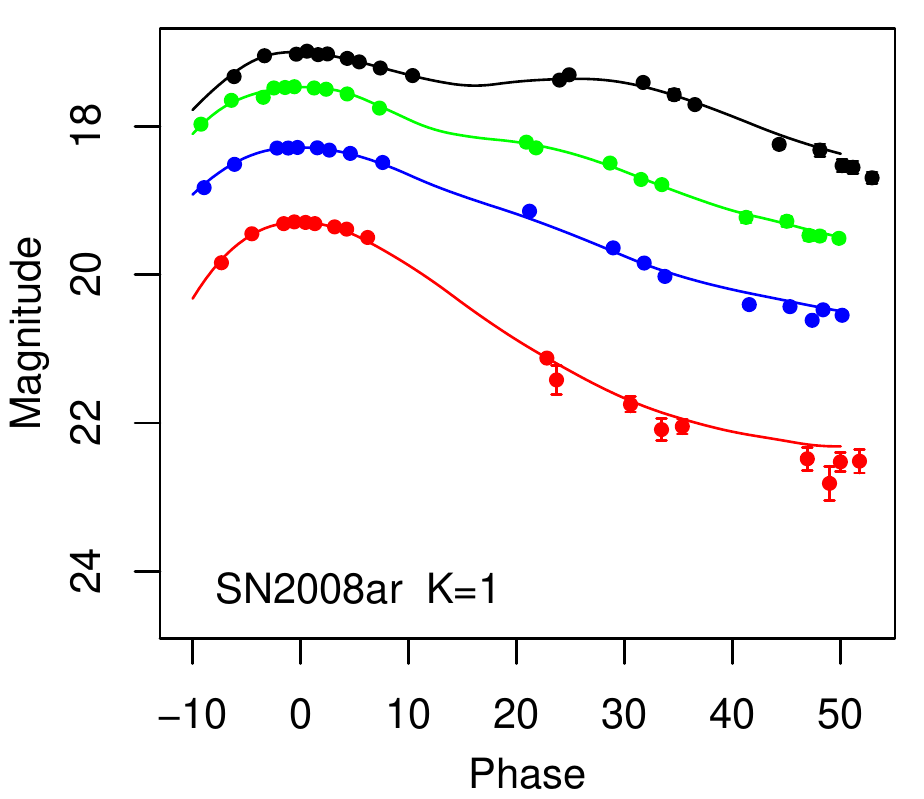} 
\includegraphics[width=0.32\textwidth]{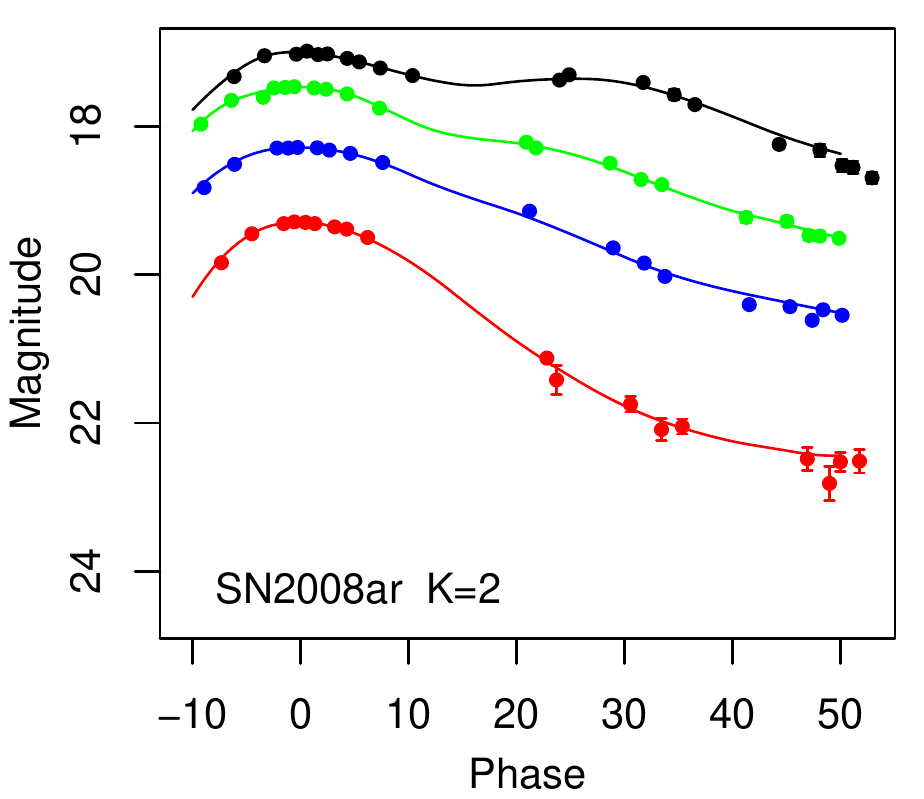} 
\includegraphics[width=0.32\textwidth]{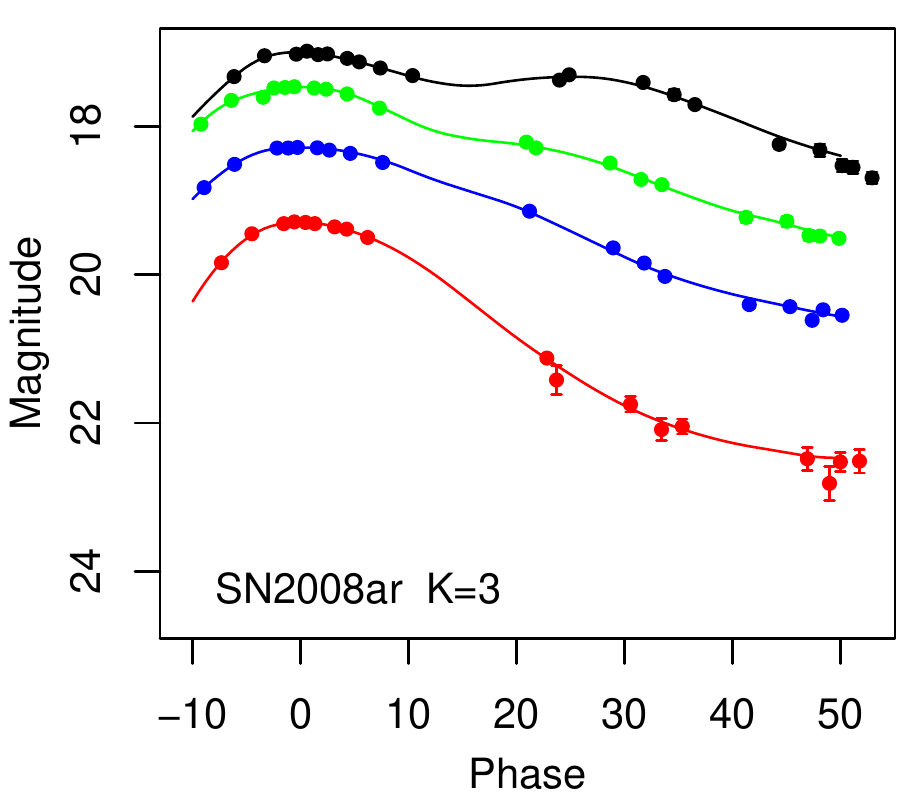} 
\includegraphics[width=0.32\textwidth]{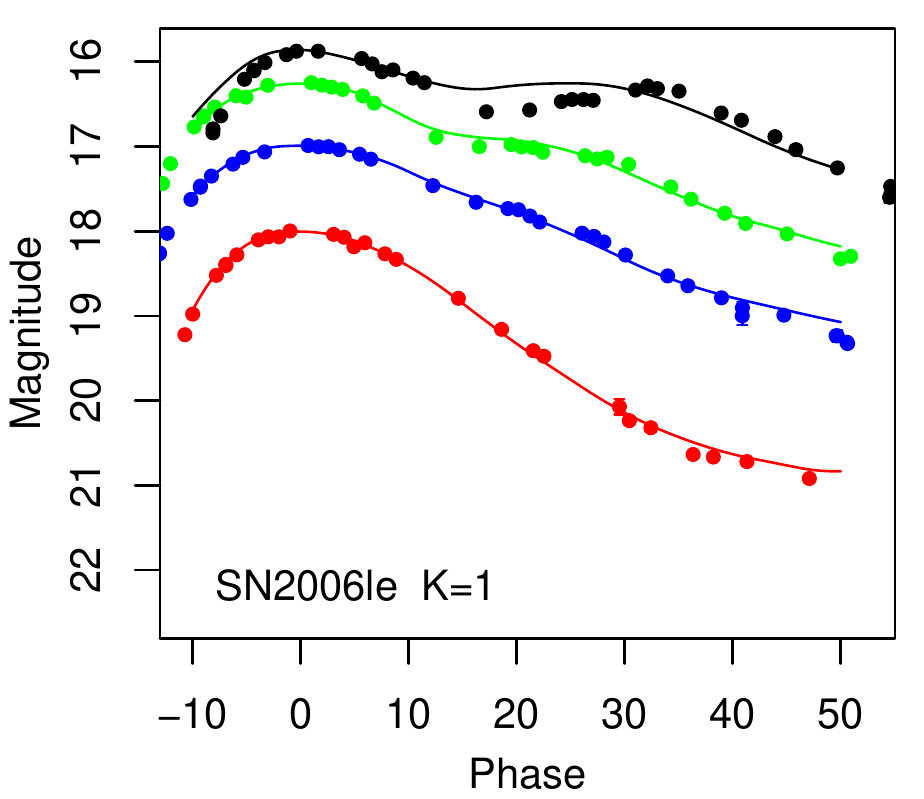} 
\includegraphics[width=0.32\textwidth]{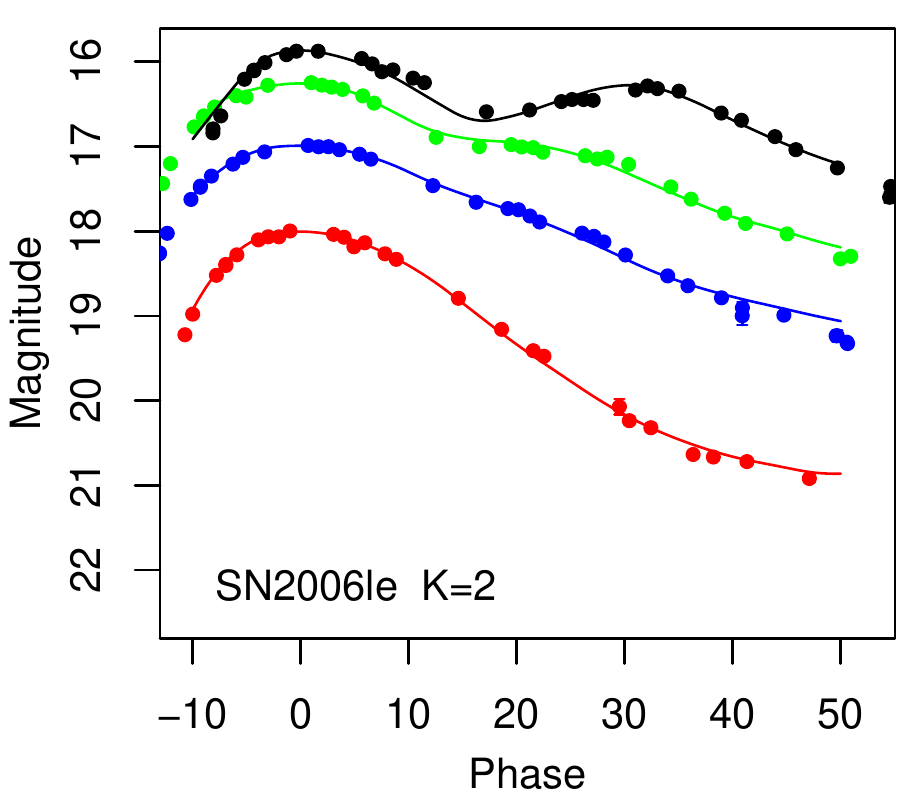} 
\includegraphics[width=0.32\textwidth]{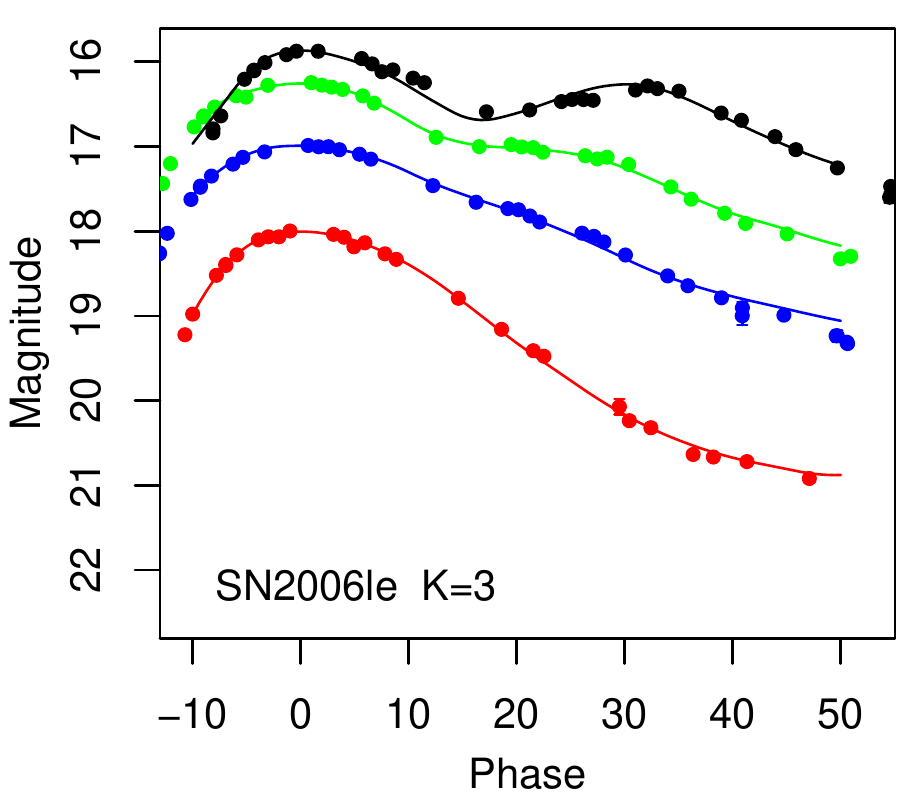} 
\caption{Light curve fitting examples with different component
  numbers $K$. Each row is for one SNIa with $K=1,2,3$. The red, blue, green and black points are $B+3$, $V+2$,
  $R+1$, $I$, respectively. Constants are added to the magnitude to separate different bands. \label{fig:lceg1}}  
\end{figure*}

\begin{figure*}
\centering
\includegraphics[width=0.32\textwidth]{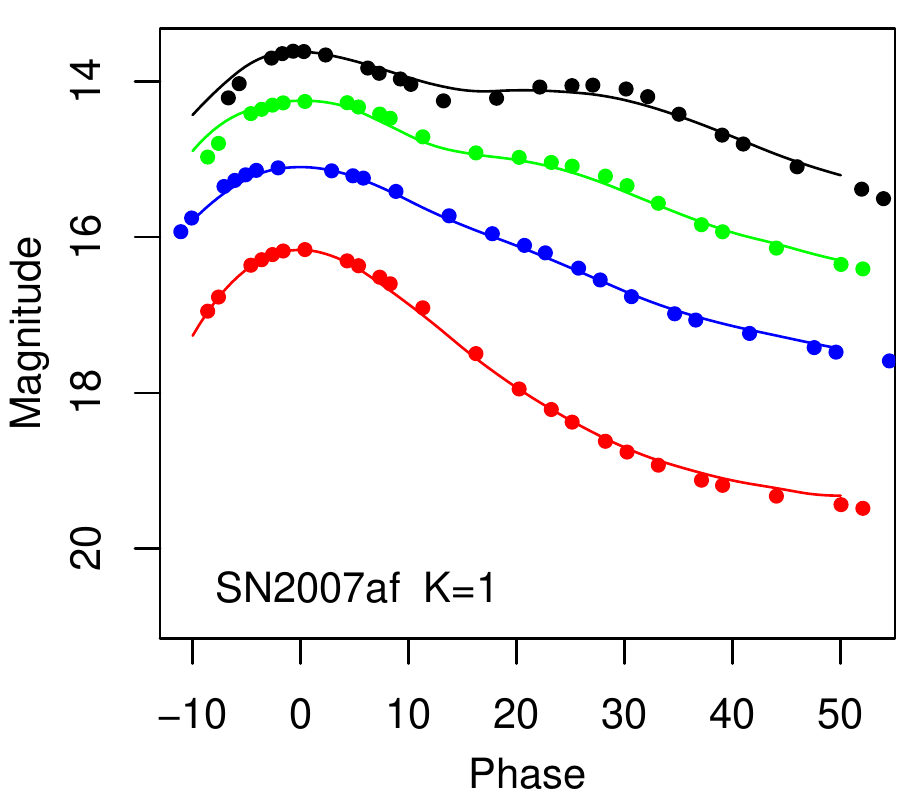} 
\includegraphics[width=0.32\textwidth]{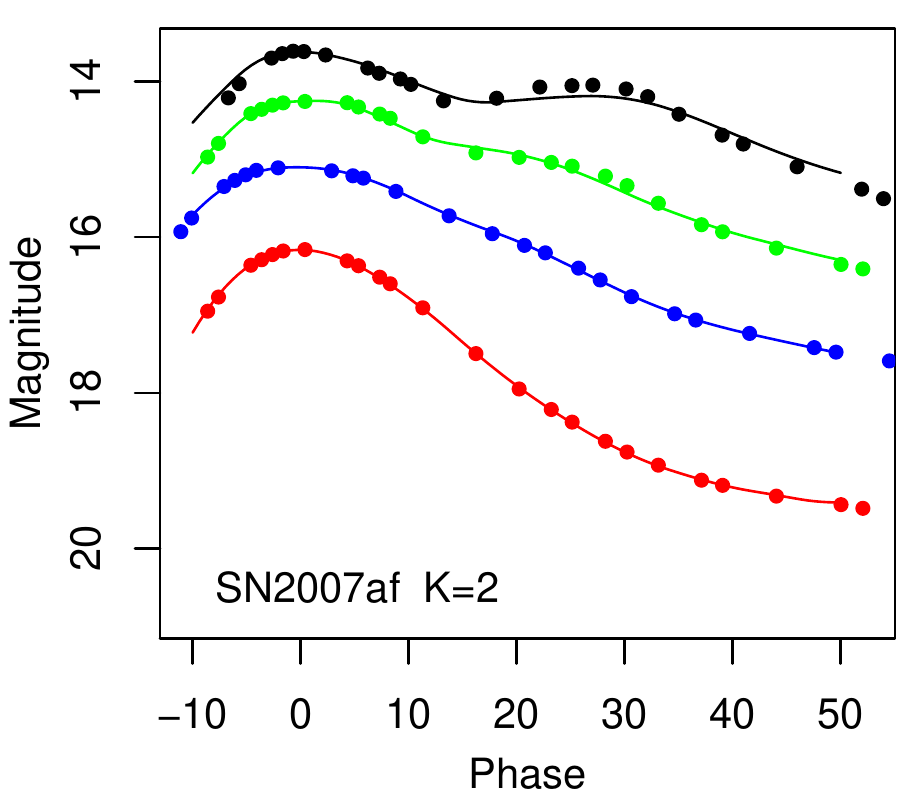} 
\includegraphics[width=0.32\textwidth]{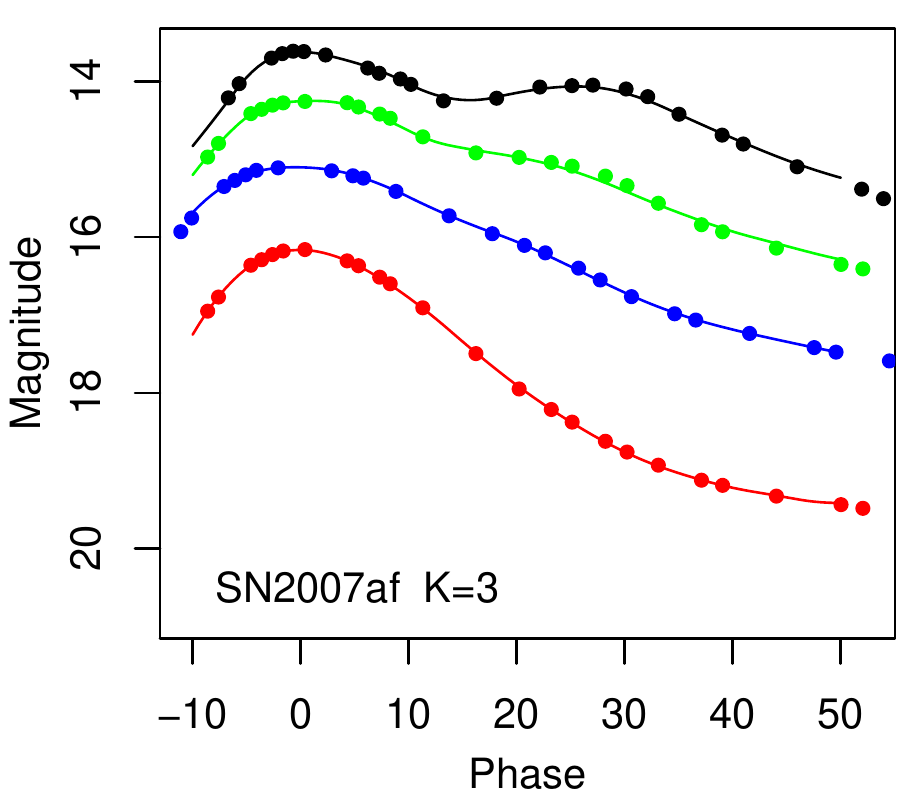} 
\includegraphics[width=0.32\textwidth]{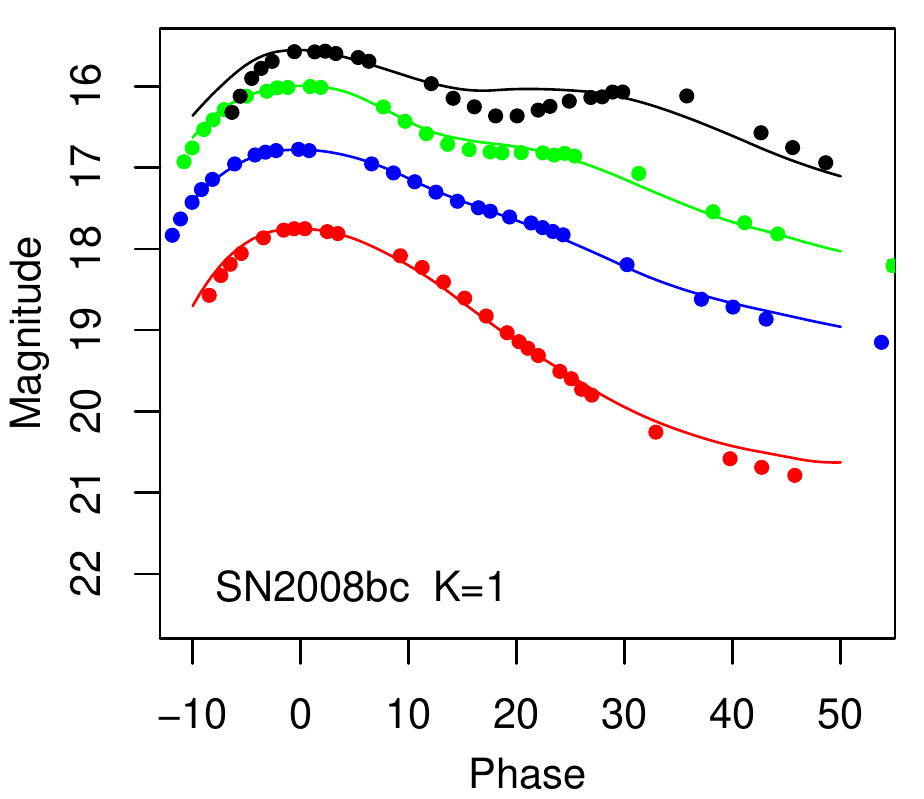} 
\includegraphics[width=0.32\textwidth]{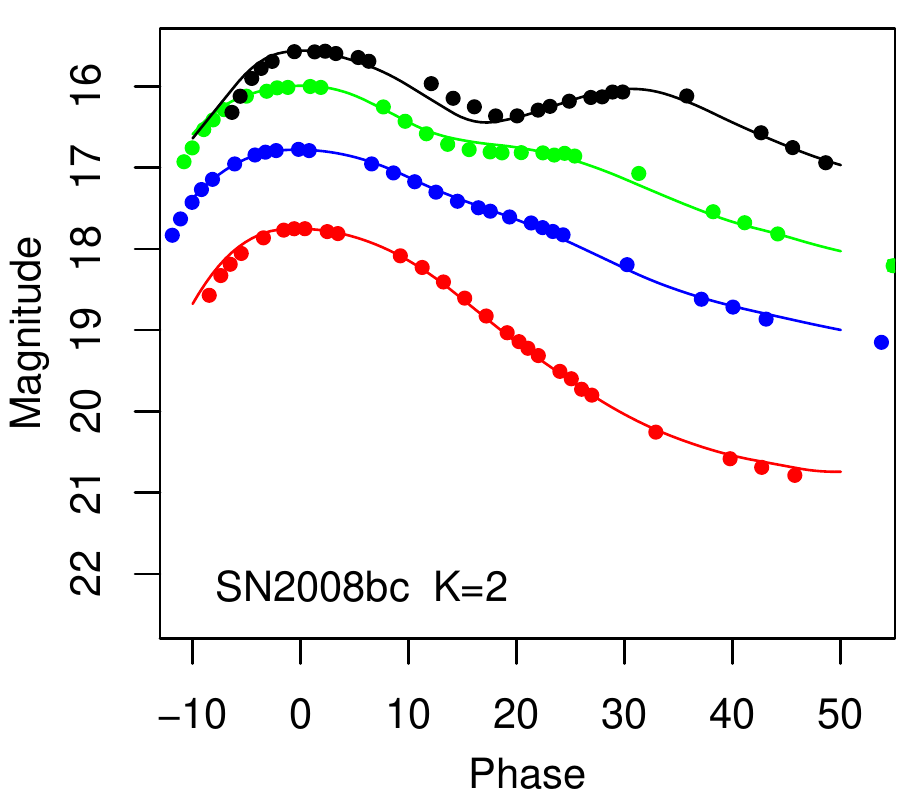} 
\includegraphics[width=0.32\textwidth]{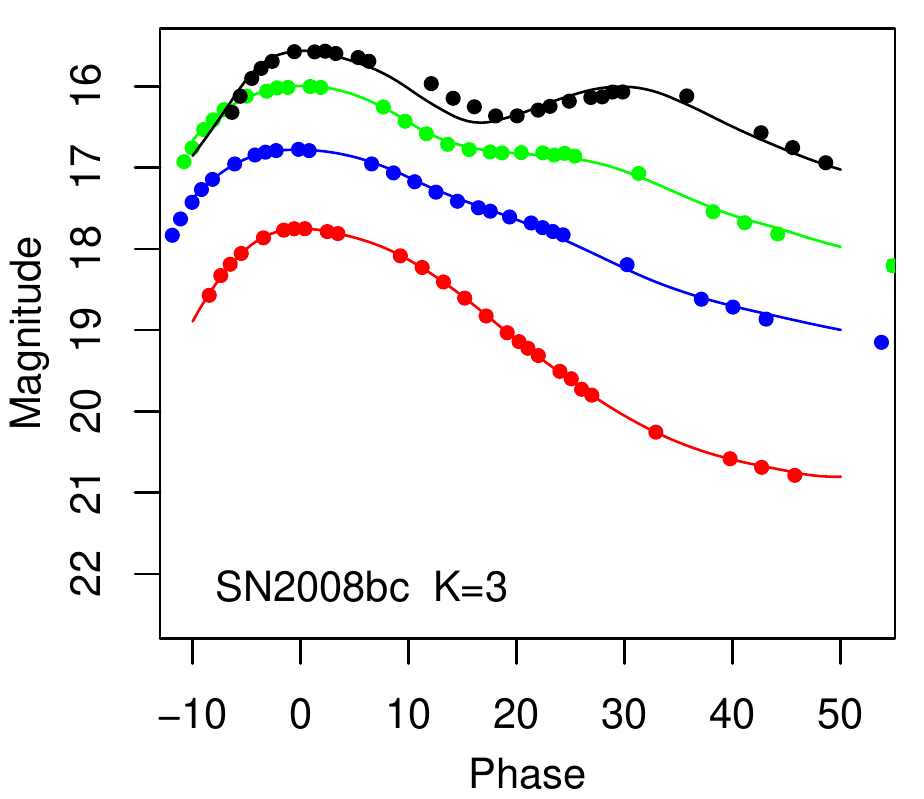} 
\includegraphics[width=0.32\textwidth]{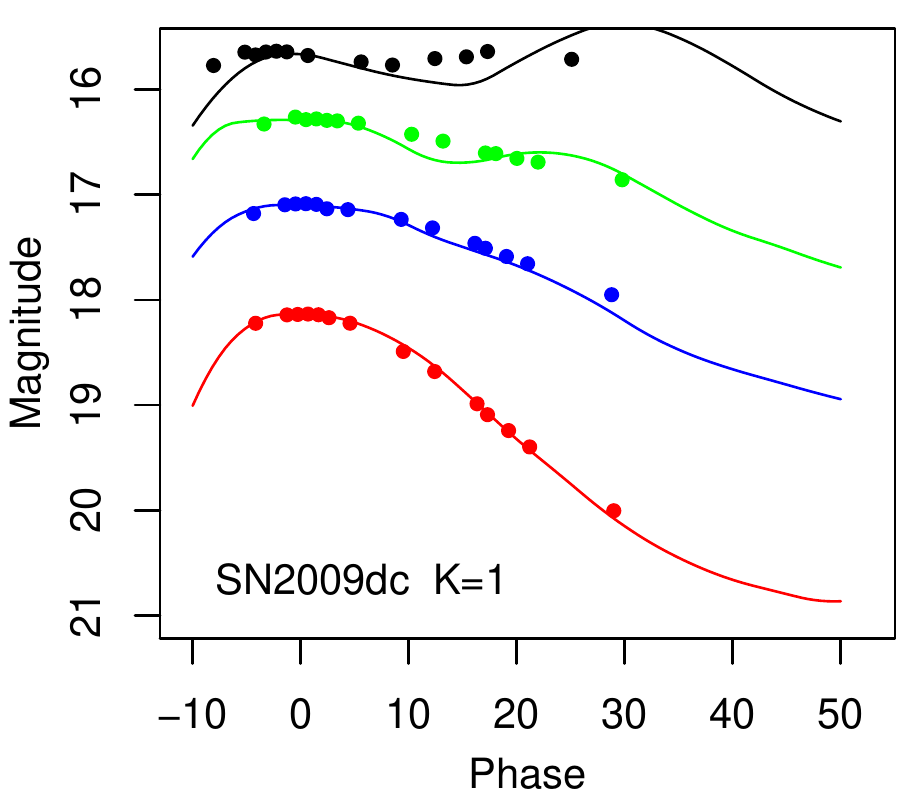} 
\includegraphics[width=0.32\textwidth]{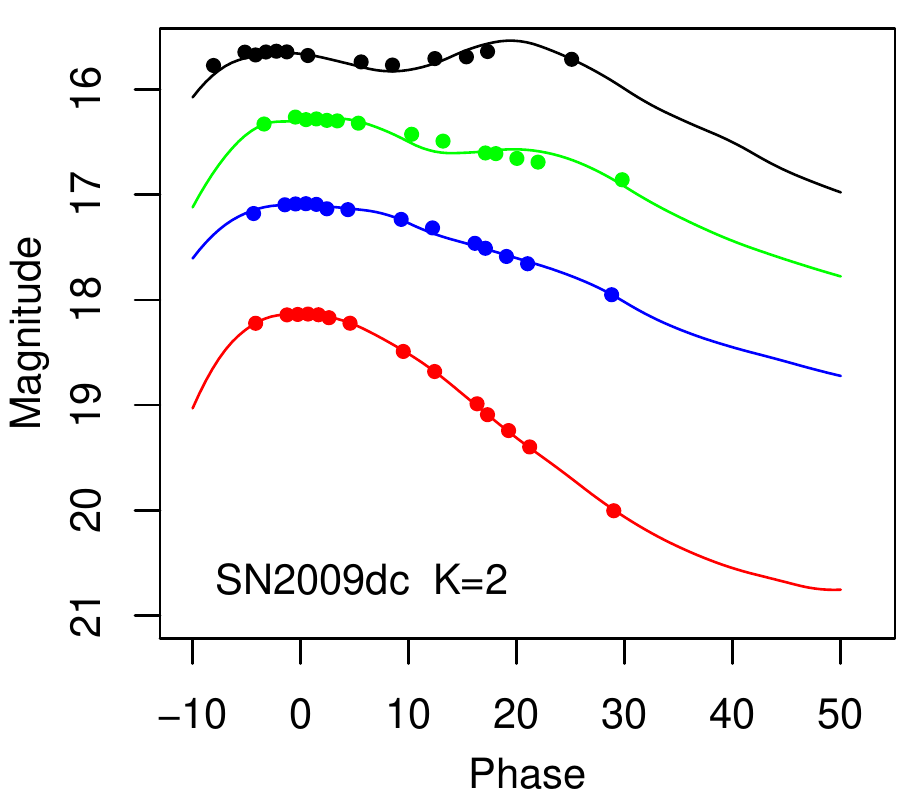} 
\includegraphics[width=0.32\textwidth]{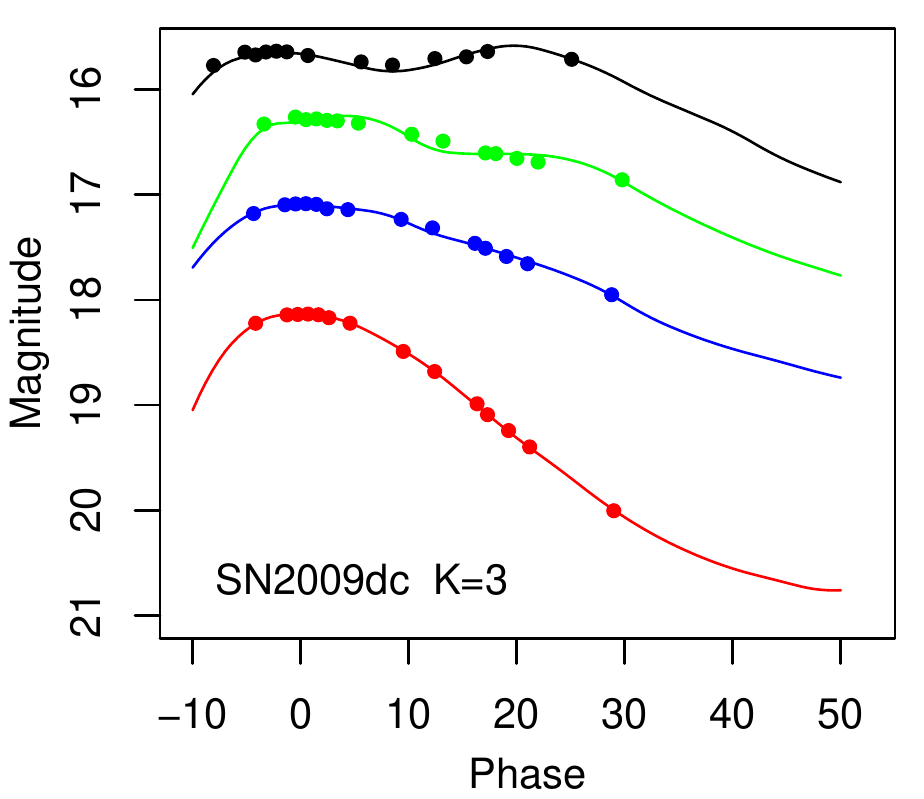} 
\includegraphics[width=0.32\textwidth]{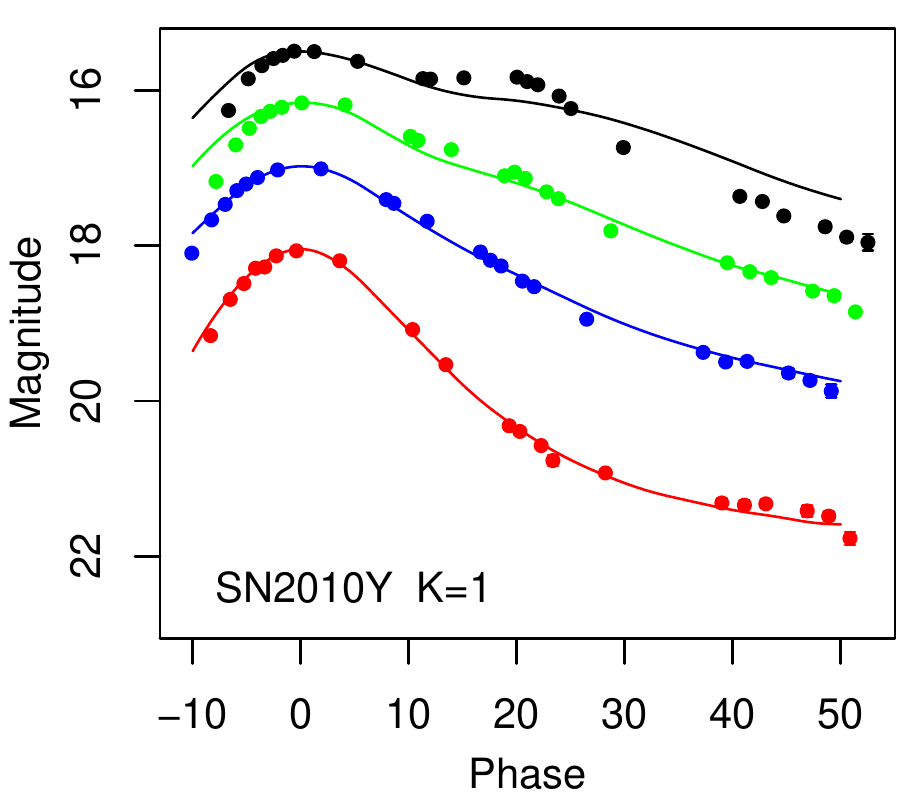} 
\includegraphics[width=0.32\textwidth]{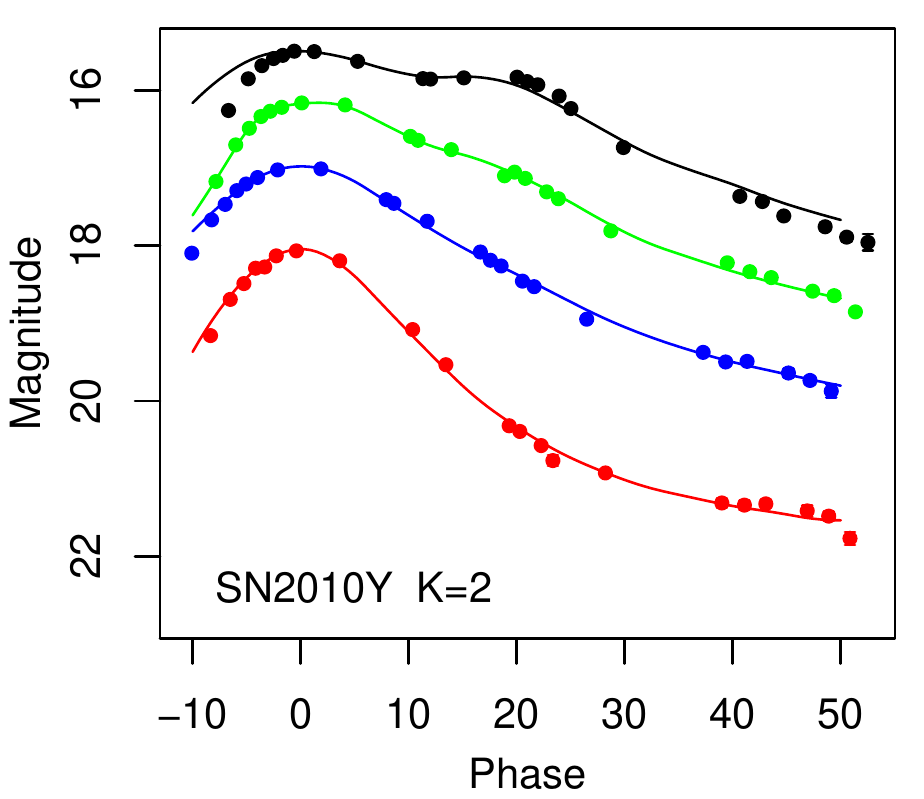} 
\includegraphics[width=0.32\textwidth]{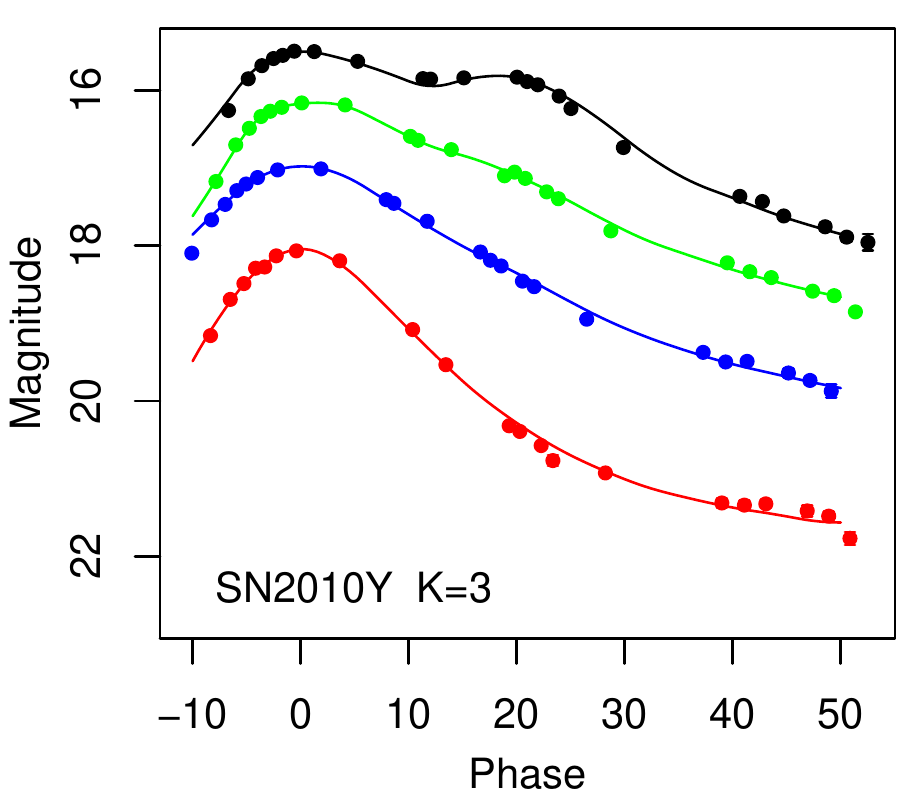} 
\caption{Light curve fitting examples with different component
  numbers $K$. Each row is for one SNIa with $K=1,2,3$. The red, blue, green and black points are $B+3$, $V+2$,
  $R+1$, $I$, respectively. Constants are added to the magnitude to separate different bands. \label{fig:lceg2}}  
\end{figure*}

 Figure~\ref{fig:bandpca}  shows the estimated
mean functions and the effects of the principal component
functions. Each row corresponds to one band $\lambda\in\{B,V,R,I\}$,
and each column corresponds to one component $k = 1,2,3,4$.
 In each panel, the solid line is the mean function $\phi_{0\lambda}(q)$. 
The ``$+$'' points  stand for $\phi_{0\lambda}(q) + 2\sigma_{k\lambda}
\phi_{k\lambda}(q)$,   and the ``$-$'' points stand for 
$\phi_{0\lambda}(q) - 2\sigma_{k\lambda} \phi_{k\lambda}(q)$  for $k =
1,2,3,4$.   
The $\sigma_{k\lambda}$ is the standard deviation of the score
$\beta_\lambda^{(k)}$ in the training data. Under the Gaussian assumption,
the  value $\pm  2\sigma_{k\lambda}$ reflect 95\% of the score
variability.  

Most of the principal component functions have a clear
interpretation. We take the $B$-band result (the first row in
Figure~\ref{fig:bandpca}) as an example,  
 and the other bands can be interpreted similarly.
 The first principal component function $\phi_{1B}(q)$ adjusts the
 width of the whole light curve, as the ``$+$'' points are all below
 the mean function while the ``$-$'' points are all above
 the mean function.
The second  principal component function
 $\phi_{2B}(q)$ adjusts the decline rate after the peak. Notice the
``$+$''  points are relatively flat around the peak, and then decline
fast after 10 days from the peak. 
The third principal component function  
 adjusts the brightening rate before the peak. 
Meanwhile, the fourth principal component function
contributes minor and  more complex adjustment.

\added{Several light curve fitting examples are presented in
  Figure~\ref{fig:lceg1} and Figure~\ref{fig:lceg2}. Each row is the
  fitting result of one SNIa with varying fitting component number, $K
  = 1,2,3$. Most of light curves have accurate fitting up to $2$
  components. In few cases, the $I$-band light curve may require up to
  $4$ components due to the varying phase of the secondary peak.} 
Note that our analysis does not
extend beyond 10 days before maximum.  
Important constraints from very early observations may become
available when more SNIa are discovered  
at very early phases. Moreover, since our light curve model construction
is fully data-driven, it can incorporate 
supernovae that do not agree with the
light curve stretch model \citep{goldhaber2001timescale}.

\begin{figure}[t]
\centering
\includegraphics[width=0.45\textwidth]{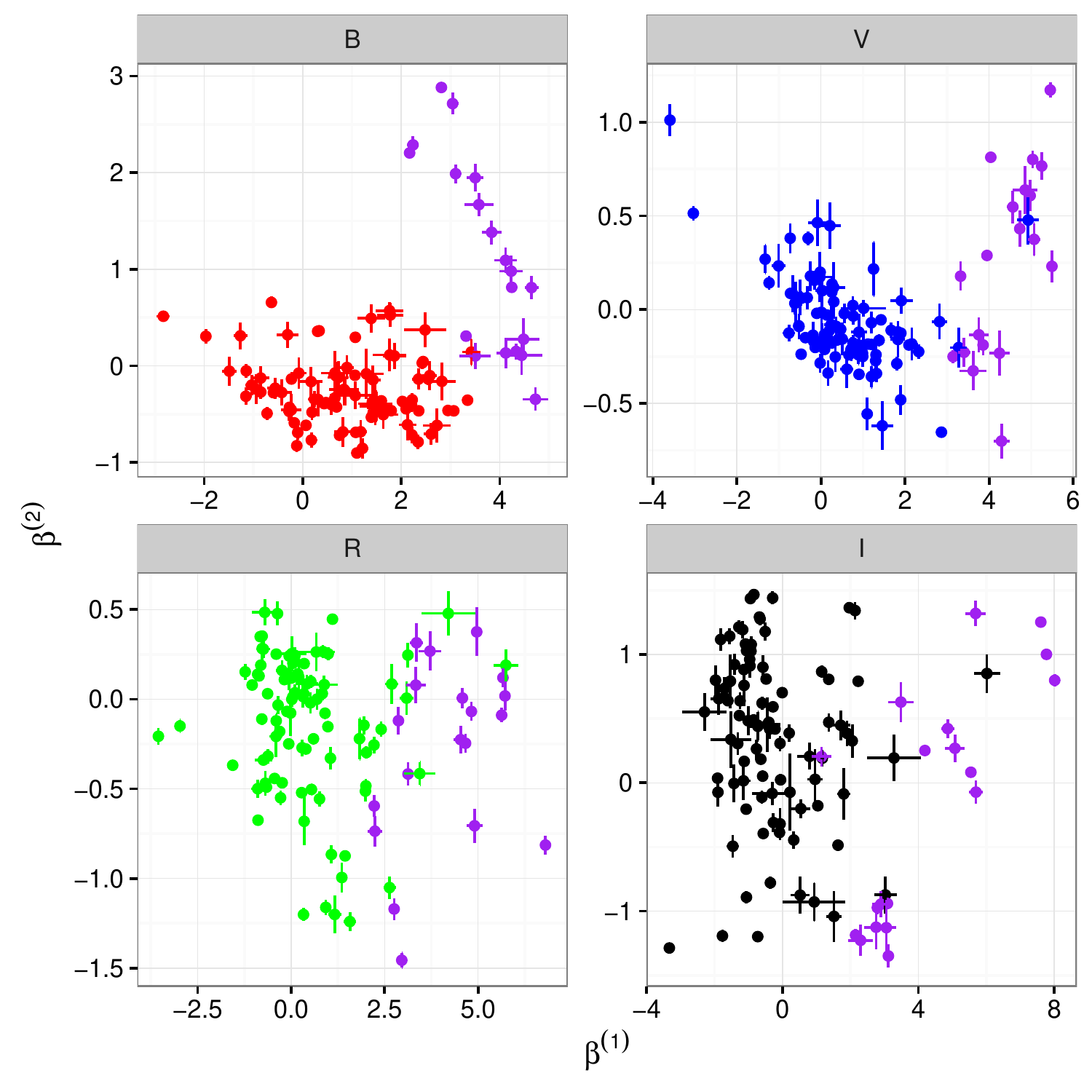}
\caption{Relation of the first two scores $\beta^{(1)}$ and $\beta^{(2)}$.
 The $B$, $V$,   $R$, $I$ band points are red, blue, green, and black,
  respectively. The purple points have $B$-band $\Delta M_{15}>1.6$.}
\label{fig:score12}
\end{figure}

The effect of higher order principal component functions on
the shape of the light curve is usually much less prominent since they
explain much smaller proportion of the total variability. 
We now focus on the first two principal components.  
The first and second principal component functions
 describe the width of different
parts of the light curve. 
The scatterplot of the first two scores is in
Figure~\ref{fig:score12}. Most of SNIa forms one cluster around the
origin in each panel. For a subgroup of SNIa with $B$-band
$\Delta M_{15}>1.6$ (purple points), their scores appear off the
central cluster and exhibits unique correlation pattern. 
 Notice this group contains SN~1991bg-like fast decliners.
 In this regard,
our light curve model has the potential of sub-grouping SNIa into
finer types.

\begin{figure*}[hbtp]
\centering
\includegraphics[width=0.48\textwidth]{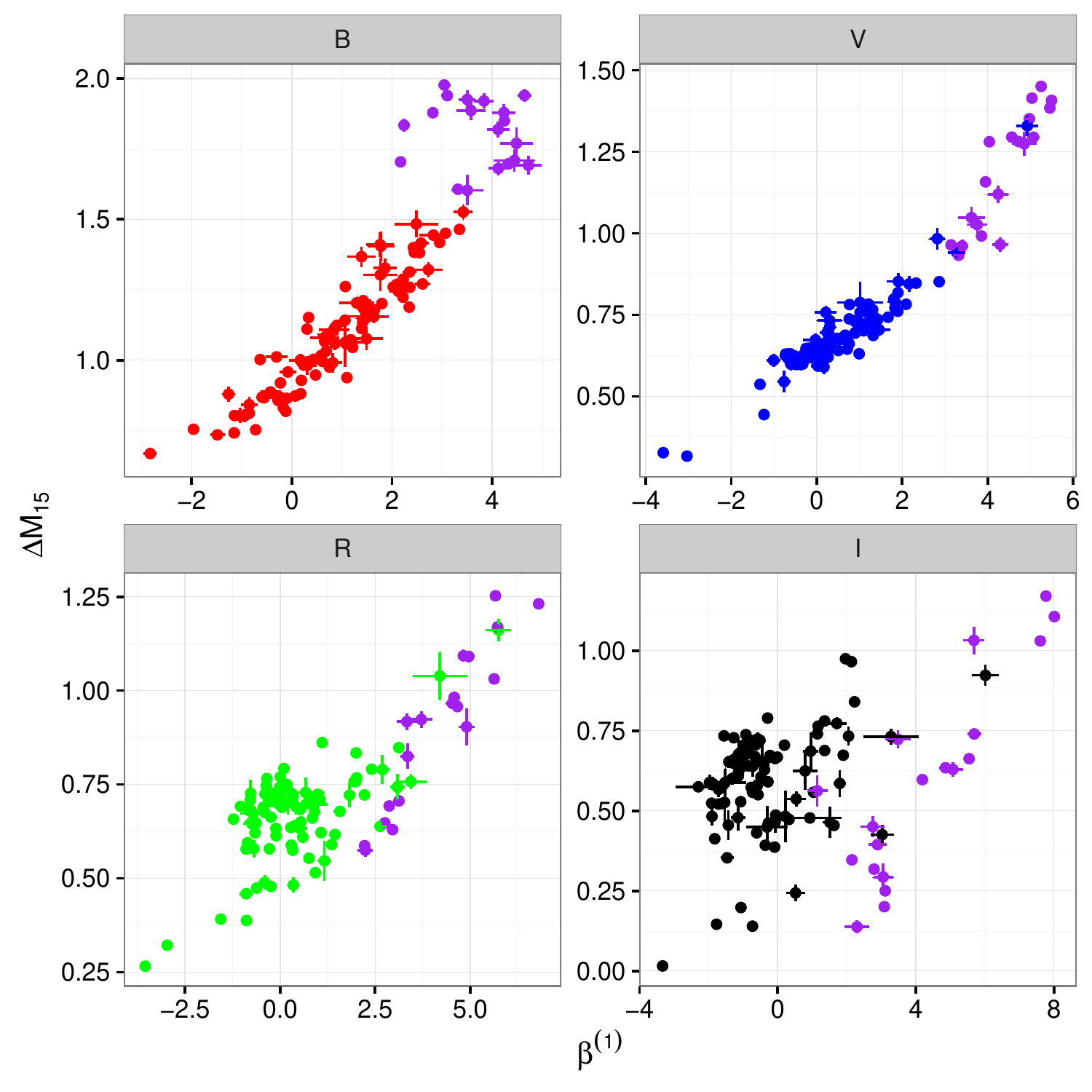}
\includegraphics[width=0.48\textwidth]{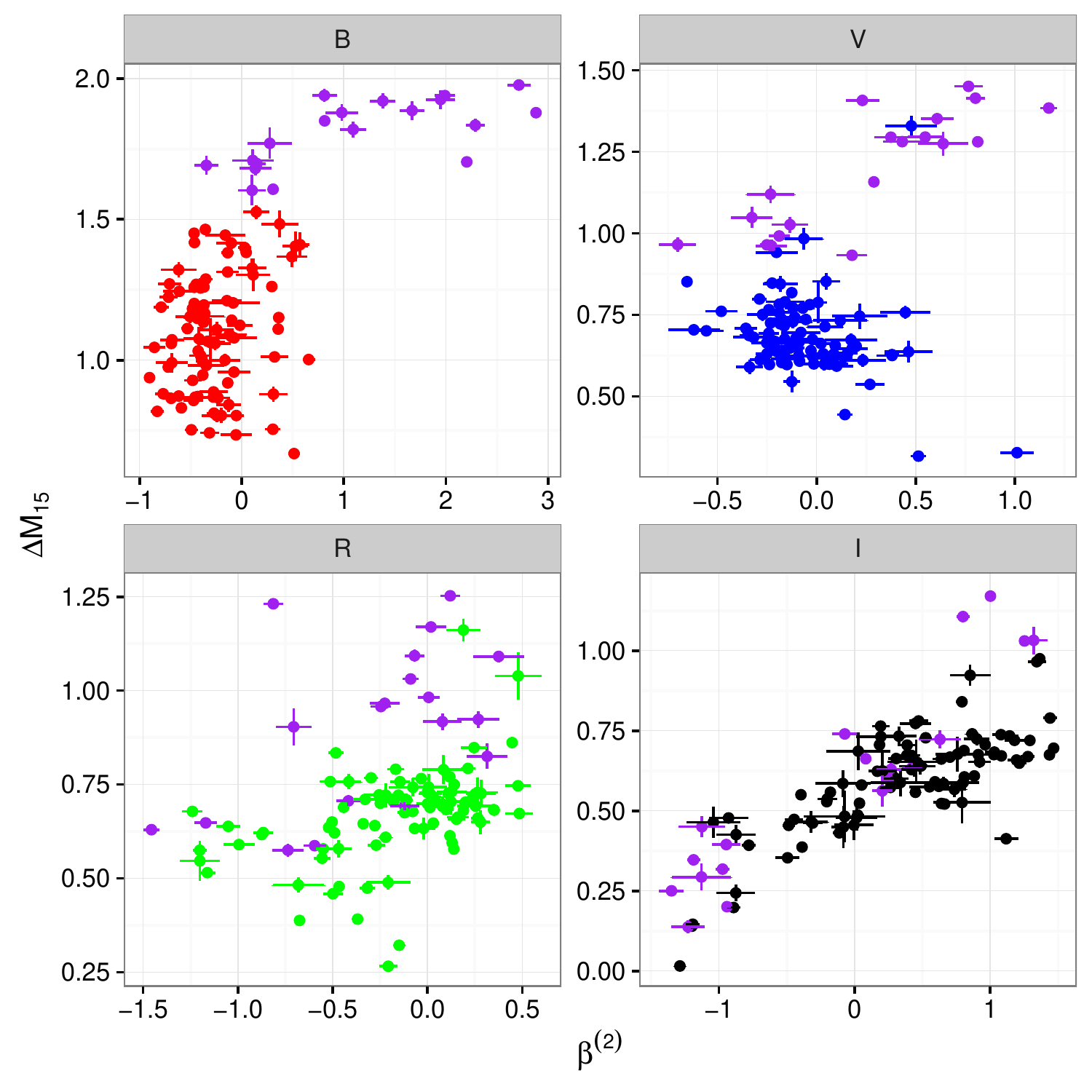}
\caption{The first two scores $\beta^{(1)}$ (left four panels), 
$\beta^{(2)}$ (right four panels) against $\Delta
  M_{15}$. Note in each panel, both the score and $\Delta
  M_{15}$ are with respect to the corresponding band. 
 The $B$, $V$,   $R$, $I$ band points are red, blue, green, and black,
  respectively. The purple points have $B$-band $\Delta M_{15}>1.6$.
\label{fig:scoredm15}} 
\end{figure*}

For comparison purpose, we also derive the light curve decline rate 
$\Delta M_{15}$ \citep{Phillips:1993ng} based on our estimated light curves. 
The $\Delta M_{15}$ parameter is calculated for the light curve of
each band. It is not a surprise that the 
scores $\beta^{(1)}$ and $\beta^{(2)}$ are highly correlated with the
$\Delta M_{15}$  parameter,  as shown in Figure~\ref{fig:scoredm15}. 
The lack of a tight correlations of light curve decline rates in different colors among themselves and with the PCA scores indicates that no single
parameter model such as $\Delta M_{15}$ is able to completely capture the family of SNIa light curve shapes, at least in linear PCA construction, although that single parameter may be found to be strongly correlated with the peak absolute magnitudes of SNeIa (see \S7 for more details).

\deleted{
What may be especially interesting is the seemingly monotonic shift of the
locus of data with filter bands. The left panel of
Figure~\ref{fig:score12} shows $\beta^{(1)}$ increases as the
wavelength increases whereas the range of $\beta^{(2)}$ remains nearly the
same. In the middle panel, the locus of data points in each filter
shows consistent correlation patterns and can be globally shifted to
approximately match each other. These characteristics suggest that in principle we
are able to derive photometric redshifts based on the scores. 
These score correlations can also
facilitate robust photometric identification of SNIa. 
Exploration of these topics is beyond the 
scope of this paper and is left for future work.  }

\begin{figure*}[hbtp]
\centering
\includegraphics[width=0.48\textwidth]{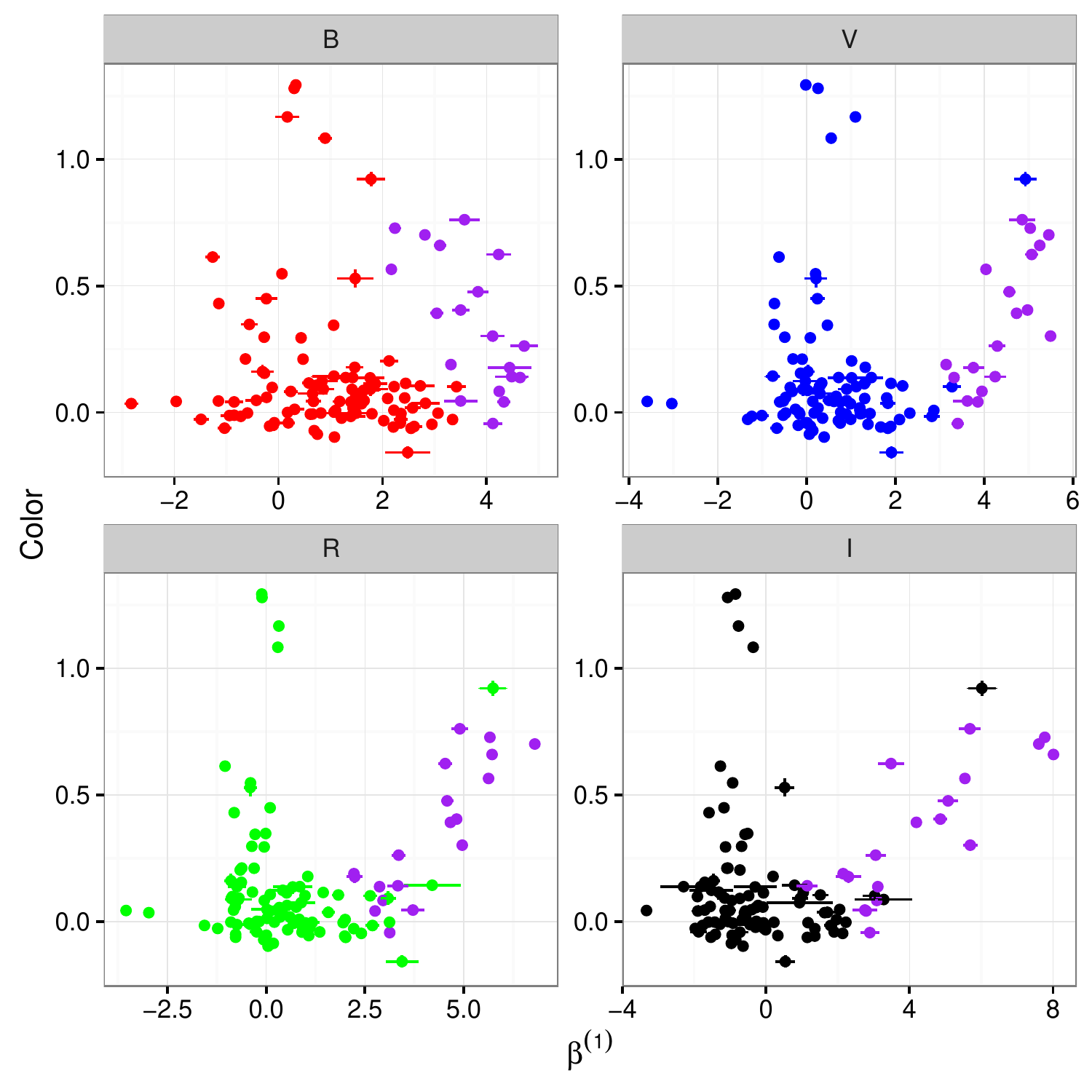}
\includegraphics[width=0.48\textwidth]{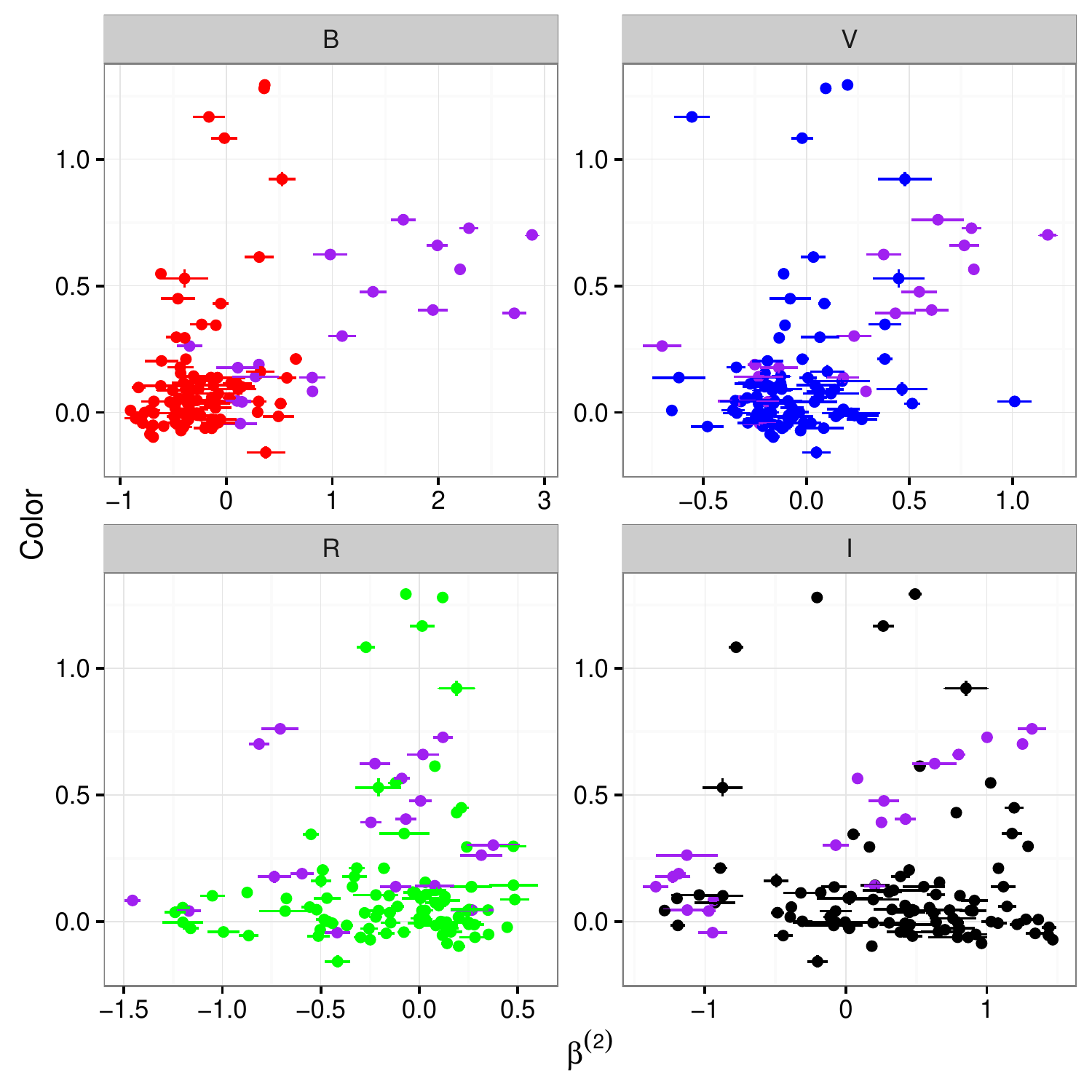}
\caption{The first two scores $\beta^{(1)}$ (left four panels), 
$\beta^{(2)}$ (right four panels) 
 against color at   $B$-band maximum, $(B-V)_{\tmax}$. 
 The $B$, $V$,   $R$, $I$ band points are red, blue, green, and black,
  respectively. The purple points have $B$-band $\Delta M_{15}>1.6$.
\label{fig:scorecolor}} 
\end{figure*}

\cite{phillips1999reddening}, \cite{nugent2002k} and \cite{wang2005dust} 
have noticed that severe reddening can shift the mean wavelength of
the filter bands and  
affect the shape of the light curves. This effect is likely to be very
small and would not significantly affect the light curve shape 
parameters we have deduced. 
\deleted{ The change of effective wavelength will be
manifested in a systematic shift of scores of the light curve model as
shown in Figure~\ref{fig:score12}. However, }
The effect of reddening is
more easily deduced by comparing the observed color with the light
curve scores. Figure~\ref{fig:scorecolor} 
shows the first score are
nonlinearly correlated with the observed color at $B$ band maximal.
For $V$, $R$ and $I$ band light curves, {\it intrinsically} redder
supernovae tend to have larger values of $\beta^{(1)}$. 
In Figure~\ref{fig:scorecolor}, the correlations between color 
and $\beta^{(1)}_V$, $\beta^{(1)}_R$, $\beta^{(1)}_I$.
are clean, and show promise for robust separation of intrinsic color
and interstellar reddening (see  \S\ref{sec:scorecolor}).  
Note that all of the quantities used in Figure~\ref{fig:scorecolor} are
measured from the shape of the light curves and do not require
information of supernova distance and reddening.

\section{Estimation of Color Excess}\label{sec:scorecolor}

\deleted{The non-linear score subsection.}

\begin{figure}[t]
\centering
\includegraphics[width=0.4\textwidth]{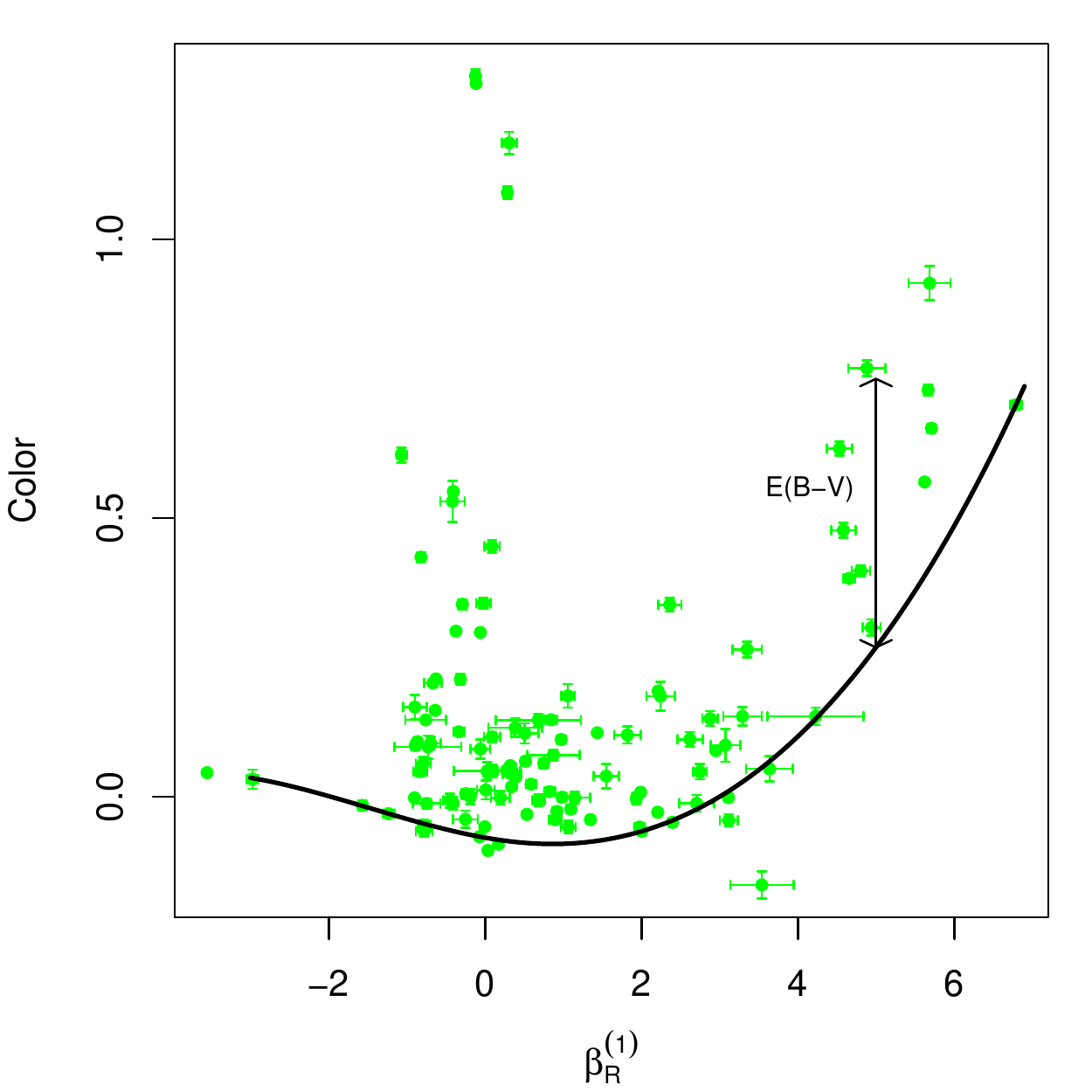}
\caption{ The lower envelope (black solid line) for the observed color and $R$
  band score $\beta^{(1)}_R$. The color excess is estimated
  by the vertical 
  distance of the observation points to the envelope line.
\label{fig:colorexcess}
} 
\end{figure}

Figure~\ref{fig:scorecolor}  reveals a relation between
the observed color $(B-V)_{\max}$ at $B$ maximum
and the first  score $\beta^{(1)}$, especially for
$V$, $R$, $I$ band light curves. This relation can be
exploited to obtain an estimate of the color excess of the supernova.
For example, here we use the relationship between 
the  $R$ band score $\beta^{(1)}_R$
and the observed color  at $B$ maximum.
In Figure~\ref{fig:colorexcess}, we show a lower
envelope (the black solid line),
 and treat it as an extinction free curve for SNIa. 
This lower envelope is estimated by
lower 10\% quantile regression with B-spline basis. The quantile
regression is iterated by removing points with large positive
residuals. This lower envelop $e_R(\beta^{(1)}_R)$ as a function of
$\beta^{(1)}_R$ serves to estimate the intrinsic color $(B-V)_0$ of the supernova.
The color excess is obtained by 
\begin{equation}\label{eqn:colorexcess}
E(B-V) = (B-V)_{\tmax} - e_R(\beta_R^{(1)})\ .
\end{equation} 
In other words, the color excess is the vertical distance from the observation
points to the lower envelope, as illustrated in Figure~\ref{fig:colorexcess}.

The same method can be used to estimate extinction using similar
relations appeared in other filter bands. The values and  
precisions from these different measurements however, can be quite
different. The lower bound to the $B$ band
$\beta^{(1)}_B$ shows very little correlation with color, and one would
get an estimate of $E(B-V)$ by approximately assuming the intrinsic
$B-V$ is nearly zero.  
In the $V$, $R$, $I$ bands, $\beta^{(1)}$ appears to produce very good
intrinsic color estimators. 

\begin{figure}[t]
\centering
\includegraphics[width=0.45\textwidth]{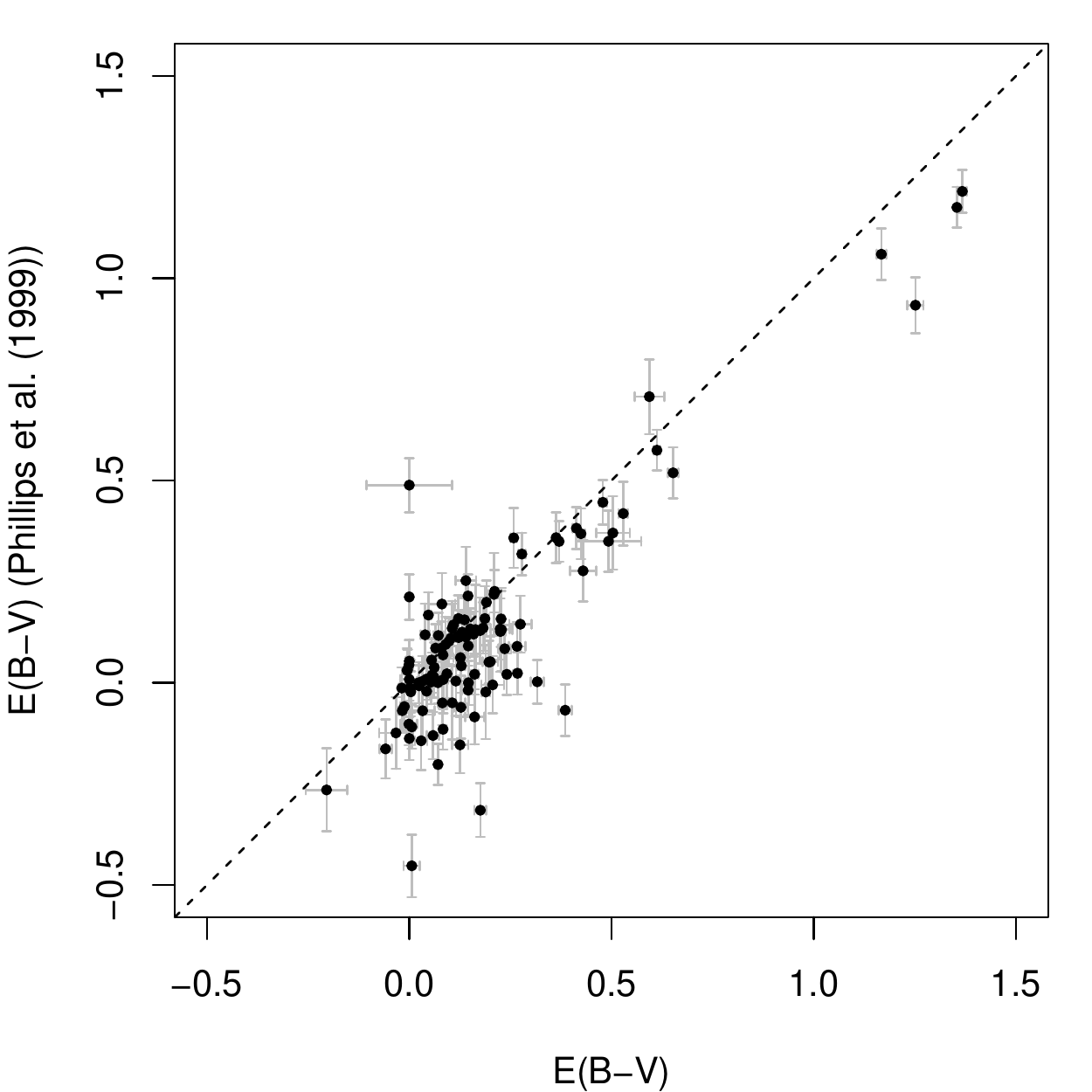}
\caption{ The y-axis is the color excess computed from
  \cite{phillips1999reddening}, and the horizontal axis is the color excess
  computed from the $R$ band score $\beta^{(1)}_R$. \added{The dashed
    line is where two measurement values agree.}
\label{fig:colorexcessCompare}} 
\end{figure}

The popular method to estimate the color excess is from the work of
\cite{phillips1999reddening}. They used the empirical linear relation 
of the intrinsic color $(B-V)_0$,
$$
(B-V)_0 = 0.725 - 0.0118\, (q_V - 60)\ ,
$$
for phase $q_V$ with respect to the $V$ band maximum. This linear
relation holds for $30\le q_V \le 90$. The color excess can be
estimated via the observed color minus the $(B-V)_0$ as above. The
observed color should be corrected by K-correction and Galactic
reddening. Figure~\ref{fig:colorexcessCompare} compares the
color excess computed via this classical method at a reference phase
$q_V = 35$ , and the color excess
estimated from the $R$ band score $\beta^{(1)}_R$. The figure shows
that 
the  color excess given by the method of \cite{1996MsT..........3L} and 
\cite{phillips1999reddening}  has negative values for a considerable portion of the supernovae. 
\cite{riess1998observational} and \cite{jha2007improved} (Equation~3) applied 
a Bayesian approach to produce non-negative color excess estimation.
The figure also shows that the uncertainty deduced from
$\beta_R^{(1)}$ is much smaller than that based on
\cite{phillips1999reddening}. This is not surprising, because the
method of \cite{phillips1999reddening} uses only light curve
observations beyond 30 days after $V$ band maximum, which are usually 
very sparse, while our estimation of $\beta_R^{(1)}$ uses
observations from the whole light curve.
The S/N ratios are vastly improved in our approach. Part of the reason that the Lira relation predicts more negative extinction is due to the low S/N ratio of color estimation. {\bf Other methods of deducing extinction estimates such as shown in \cite{2005ApJ...620L..87W} may also be interesting for further investigation.}

\section{Spectral Information}\label{sec:specfits}

This section examines the relationship between the scores in model (3)
and spectral features, and discuss the possibility of using the scores
for identifying spectral classes. There exists some analysis regarding
light curve width and  spectral features such as \sirp\ in the literature. 
With the aid of model~(\ref{equ2:basics}), we are
able to present more details on how light curve shape (not just its width)
changes with spectral features. We will also demonstrate the light
curve scores can be linked to spectroscopically different SNIa.
\deleted{, although with limited precision.} 
This linkage is important. With refined spectral subclasses,
the K-correction can be 
applied with higher precision. Identifying sub-classes of SNIa 
is of ultimate importance in assessing systematic evolutionary effect
when applying SNIa as standard candles.  

\subsection{The Scores and Spectral Features}\label{sec:scorespc}

\begin{figure}[htbp]
\centering
\includegraphics[width = 0.45\textwidth]{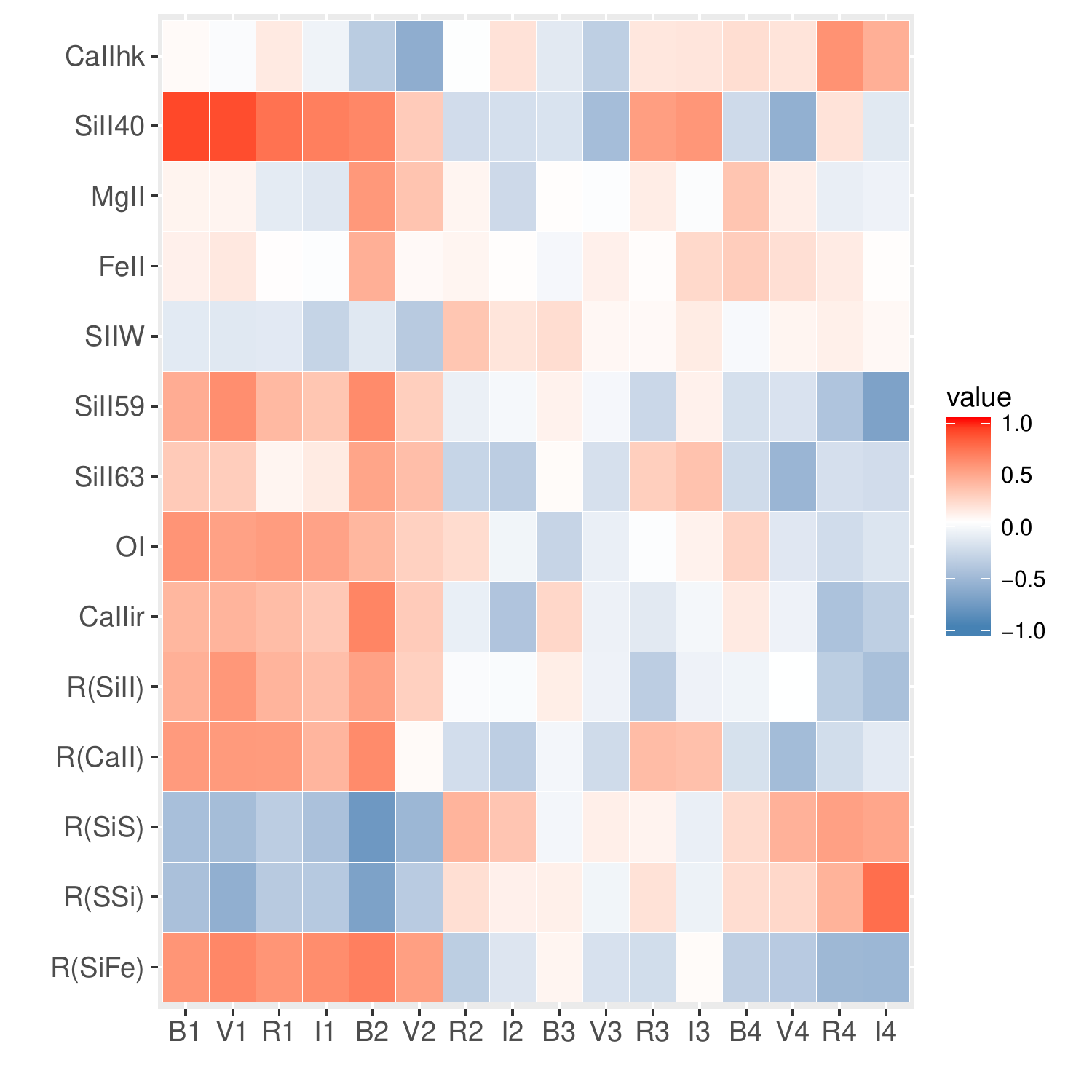}
\caption{ The heatmap showing correlation between the scores by our
  FPCA model   and the spectral features. The horizontal axis corresponds to the
  first four scores across all of the $B$, $V$, $R$, $I$ bands. The vertical
  axis corresponds to various spectral features.
} \label{fig:heatmap}
\end{figure}

\begin{figure*}[htbp]
\centering
\includegraphics[width = 0.9\textwidth]{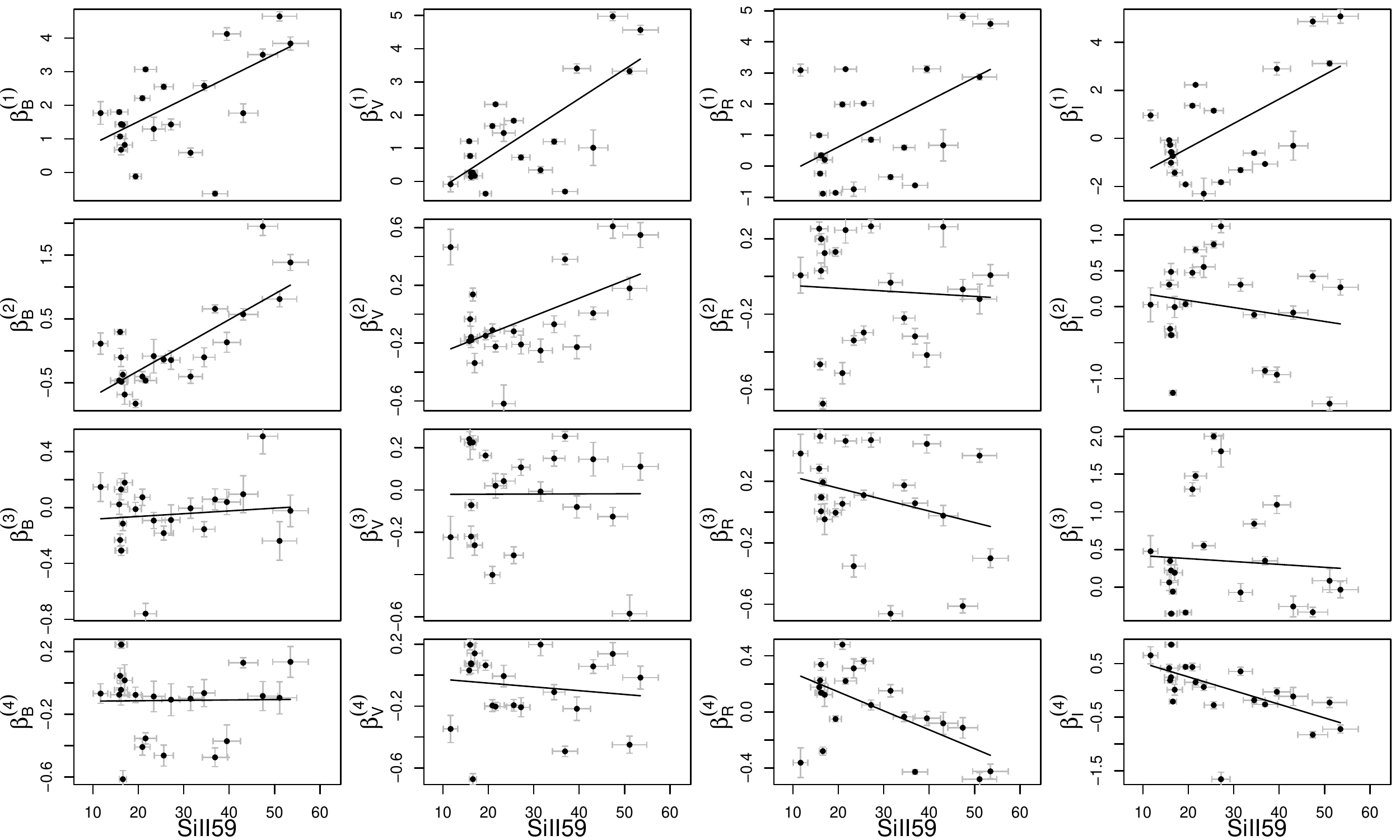}
\caption{  The scores  $\beta^{(k)}_\lambda$  against the
  pseudo-equivalent width (pEW) of \sito. The points are fitted by a
  robust linear regression (the solid line). } \label{fig:siii59}
\end{figure*}

\begin{figure}[htbp]
\centering
\includegraphics[width = 0.5\textwidth]{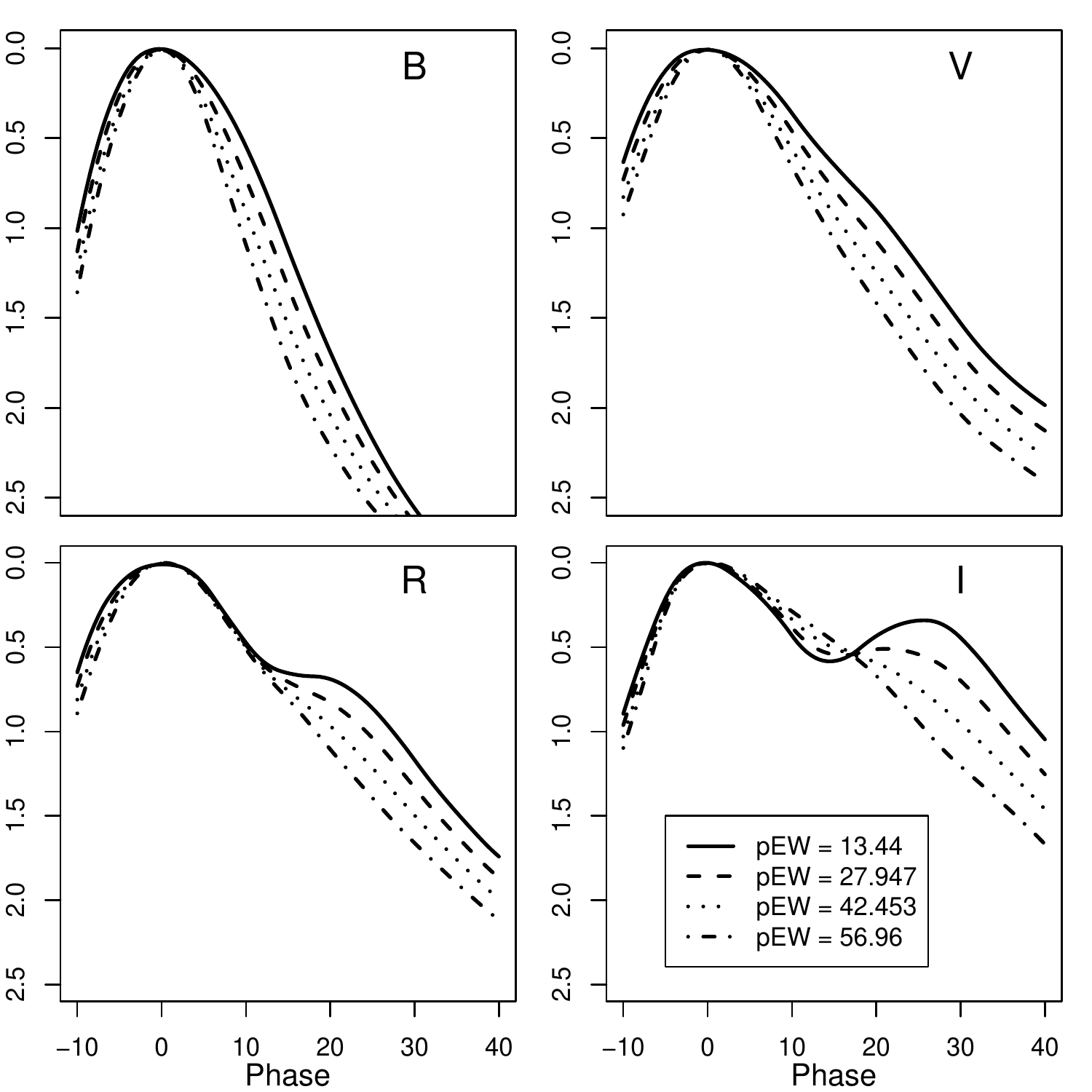} 
\caption{ The ``average'' light curve shape at four different levels of the
  pseudo-equivalent width (pEW) of \sito.
} \label{fig:siii59toshape}
\end{figure}

The dataset of SNIa with comprehensive spectral data is 
sparse, and the measurement of the strength of spectral features
usually suffers from severe systematic errors due to difficulties in
defining the level of continuum and observational noise; the latter is
usually not even available for most published SNIa
spectra. \cite{wagers2010quantifying} developed a mathematical
framework based 
on wavelet decomposition of the spectra to reconstruct the 
signal from published data. It was shown in \cite{wagers2010quantifying} that
large noise can easily bias estimate of spectral line strength, and
Monte-Carlo simulation can be used to simulate the effect and correct
the bias. However, there is no overlap of the SNIa sample in
\cite{wagers2010quantifying} and the current sample. A more recent
derivation  of spectral line strength is given in \cite{zhao2016oxygen}, but
its data sample size is small.  

High quality spectral feature measurements are scarce. 
They provide detailed information on the spectral features, which can
significantly affect the measured colors and lead to abnormal
extinction behavior. The spectral features provide
further constraints on the intrinsic properties of
supernovae and associated extinction. Measurement of spectral features
may prove to be critical for the WFIRST program
\citep{spergel2013wide} which aims at
unprecedented precision. 

We show in this paper the correlations of the
most significant spectral features with the light curve features
produced from our method. 
 Here we will only use the spectral features extracted by
\cite{silverman2012berkeley2}, including pseudo-equivalent
width (pEW), spectral feature depths, and fluxes at the center and end
points of nine spectral feature complexes. The nine spectral feature
complexes are \cahk, \sirp, \mgii, \feii, \siW,
\sito, \siye, \oit, \caii~near-IR~triplet. 
We will correlate the first four \textit{original} scores 
$\beta^{(1)}, \beta^{(2)}, \beta^{(3)}, \beta^{(4)}$ 
with the pseudo-equivalent
width (pEW) of these spectral feature.

The pEW of the spectral features within 5 days of the $B$ maximum are
obtained, and their Spearman correlation coefficients with the scores
are computed. The correlation is visualized as heatmap in
Figure~\ref{fig:heatmap}. 
These nine spectral features corresponds to the first nine rows in the figure.
Its columns from left to right correspond to the scores $\beta^{(1)}_B, \beta^{(1)}_V,
\beta^{(1)}_R, \beta^{(1)}_I, \beta^{(2)}_B, \cdots, \beta^{(4)}_I$
respectively. Saturated red (blue resp.) implies a strong positive
(negative resp.) correlation; and white color implies a very weak correlation.

The spectral feature \sirp\ and
\sito\ are important spectral luminosity indicator \citep{nugent1995evidence}.
With the light curve width parameter from the SALT~II model,
\cite{silverman2012berkeley3} noticed that these two features are correlated with light
curve width. This is also confirmed in
our dataset. Both of them have strong (positive) 
correlation with the first two scores across four optical bands.  
The exception is that $\beta^{(2)}_R$
has weak (negative) anti-correlation with these two spectral features. 
We have also shown in
Figure~\ref{fig:scoredm15} the correlation of $\Delta M_{15}$ with the first two
scores. The correlations among the spectral feature, our model
scores, and  the $\Delta M_{15}$ imply that with sufficient amount of
well calibrated data it would be possible to construct light curve
templates for different spectral sub-classes of SNIa. 

\added{Our work provides more details on how the spectral
  features correlate with light curve shapes. Figure~\ref{fig:siii59} is the
scatter plot of the  scores against the pEW of \sito.  Let's consider the
correlation for the first two dominant scores.
The first row of Figure~\ref{fig:siii59} indicates that the first scores of all
bands are positively correlated with the 
pEW of \sito. 
The second row in  Figure~\ref{fig:siii59} suggests 
that the correlation between the second score $\beta^{(2)}$ 
and pEW(\sito)  changes from positive
correlation to negative correlation as the central wavelength of the
filter increases. In particular, the $B$ band score 
$\beta^{(2)}_B$  is positive correlated with pEW(\sito); and the 
$I$ band score  $\beta^{(2)}_I$  is negatively correlated with pEW(\sito).}

\added{Combining Figure~\ref{fig:siii59} with
  Figure~~\ref{fig:bandpca}, the effect the first two scores can be
  better understood. 
For both the $B$ band and $V$ band, larger value
  of $\beta^{(1)}$ and $\beta^{(2)}$ will both shrink the light curve
  width. 
For the $I$ band, its first score $\beta^{(1)}_I$ 
increases with pEW(\sito), which makes the light curve narrower.
At the same time, its second score $\beta^{(2)}_I$ decreases with
pEW(\sito). This has the counter effect of making the light curve
wider, especially around the peak. Smaller $\beta^{(2)}_I$ also
makes the secondary peak earlier in phase. }

\deleted{Our work provides more details on how these two spectral
  features 
correlate with light curve shape. Figure~\ref{fig:siii59} is the
scatter plot of the  scores with the pEW of \sito.  
In the first row of Figure~\ref{fig:siii59}, the first scores of all
bands are positively correlated with the 
pEW of \sito. However, as the central wavelength of the
band increases, the first score $\beta^{(1)}$ becomes less sensitive
to this spectral feacture. The $I$ band score $\beta^{(1)}_I$ starts to
drop when pEW(\sirp) increase to 17; it almost remains at a
constant level for  pEW(\sirp) in the interval $[5,15]$.}

\deleted{Recall the first score mainly affects the decline rate
after 15 days from the peak. The second score affects the light curve
width around the peak, and the decline rate contrast before and after
+15 days in phase. As both $\beta^{(1)}_B$ and $\beta^{(2)}_B$
are negatively correlated with pEW(\sirp), stronger
\sirp\   will shrink the $B$ band light curve width across the 
entire phase range $[-10,50]$. 
On the other hand, in the last column of Figure~\ref{fig:siii40}, 
 the correlation pattern for $\beta^{(1)}_I$ and $\beta^{(2)}_I$
 implies that strong \sirp\ tend to make the $I$ band light
 curve wider around the peak and decrease faster after $+15$ days in phase.}

This effect on light curve shape is illustrated in
Figure~\ref{fig:siii59toshape}, which shows 
the ``average'' light curve shapes for
 each band at different levels  of pEW(\sito). 
\added{These average light curve shapes are computed as follows.}
The scores as a function of pEW(\sito) is
fitted by \replaced{a LOESS curve (which is a robust local
  regression)}{a robust linear regression},
 shown as the solid line in Figure~\ref{fig:siii59}.
Then we compute the value of $\beta^{(k)}_\lambda$ at pEW$
\approx 13.44, 27.95, 42.45, 56.96$ for $k=1,2,3,4$ and
$\lambda\in\{B,V,R,I\}$.  After that, the average light curve 
shapes are computed as $\phi_{0\lambda}(q) + \sum_{k=1}^4
\beta^{(k)}_\lambda\phi_{k\lambda}(q)$ for each filter $\lambda$. 
Of special interest is the lower right panel
of Figure~\ref{fig:siii59toshape}. As expected from the previous analysis,
when the strength of \sito\ increases, the $I$ band light curve becomes wider
around the peak, but narrower after +15 days in phase. \added{
The secondary peak gradually becomes weaker and appears earlier in phase. 
Similarly, the shrinkage for $R$ band is
  only evident 10  days after the peak.} On the other hand, the $B$
band and $V$ band  light curves become uniformly narrower across the 
entire phase range. 

A parallel result could be drawn for the spectral feature 
\sirp. Its graphical result is in the appendix. 
The correlation between the scores and \sirp\ 
is in Figure~\ref{fig:siii40}. This correlation pattern resembles that
in Figure~\ref{fig:siii59}. The ``average'' light curve shapes
corresponding to different levels of \sirp\ are plotted in 
Figure~\ref{fig:siii40toshape}.

Next, we consider five spectral ratios as defined
by \cite{silverman2012berkeley3}. The first is the Si~II ratio, which
is the pEW of \sito\ divided by the pEW of \siye,
$$
\mathcal{R}(\mathrm{Si\ II}) = 
\frac{
\mathrm{pEW(\sitom)}
}{
\mathrm{pEW(\siyem) }
}\, .
$$
The second is the ratio $\mathcal{R}(\mathrm{Ca\ II}) $
of the flux at the red and blue end of  Ca~II~H\&K,
$$
\mathcal{R}(\mathrm{Ca\ II}) = 
\frac{
\mathrm{F_r(Ca~{\small II}~H\&K )}
}{
\mathrm{F_b(Ca~{\small II}~H\&K ) }
}\, .
$$
These two spectral feature ratios are among the 
first spectral luminosity indicators
\citep{nugent1995evidence}. Three additional spectral ratios are
defined as
\begin{equation*}
\begin{split}
\mathcal{R}(\mathrm{SiS}) &= 
\frac{
\mathrm{F_r(S~{\small II}~W )}
}{
\mathrm{F_r(Si~{\small II}~\lambda6355)}
}\, , \\
\mathcal{R}(\mathrm{SSi}) &= 
\frac{
\mathrm{pEW(S~{\small II}~W )}
}{
\mathrm{pEW(Si~{\small II}~\lambda5972)}
}\, , \\
\mathcal{R}(\mathrm{SiFe}) &= 
\frac{
\mathrm{pEW(Si~II~\lambda5972)}
}{
\mathrm{pEW(Fe~II)}
}\, . \\
\end{split}
\end{equation*}
These five ratios correspond to the last five rows in
Figure~\ref{fig:heatmap}. They also have strong correlation (or
anti-correlation) with the scores across all four bands. 
Notice \deleted{the exception is the second score of the $R$ band
$\beta_R^{(2)}$. It has very weak correlation with all the spectral
features. Furthermore,} the correlation of $\beta_R^{(2)}$ and
$\beta^{(2)}_I$ tend to have 
opposite sign with the correlations 
involving $\beta^{(1)}_B, \beta^{(1)}_V, 
\beta^{(1)}_R, \beta^{(1)}_I, \beta^{(2)}_B, \beta^{(2)}_V$. 
The Figure~\ref{fig:ratio1-1}, 
Figure~\ref{fig:ratio1-2}, Figure~\ref{fig:ratio2-1} and 
Figure~\ref{fig:ratio2-2} in the appendix provide more illustrations
about the relation of the scores with $\mathcal{R}(\mathrm{Si\ II})$ and 
$\mathcal{R}(\mathrm{Ca\ II})$.

\subsection{The Scores and Spectral Classes}
\begin{figure}[hbtp]
\centering
\includegraphics[width=0.4\textwidth]{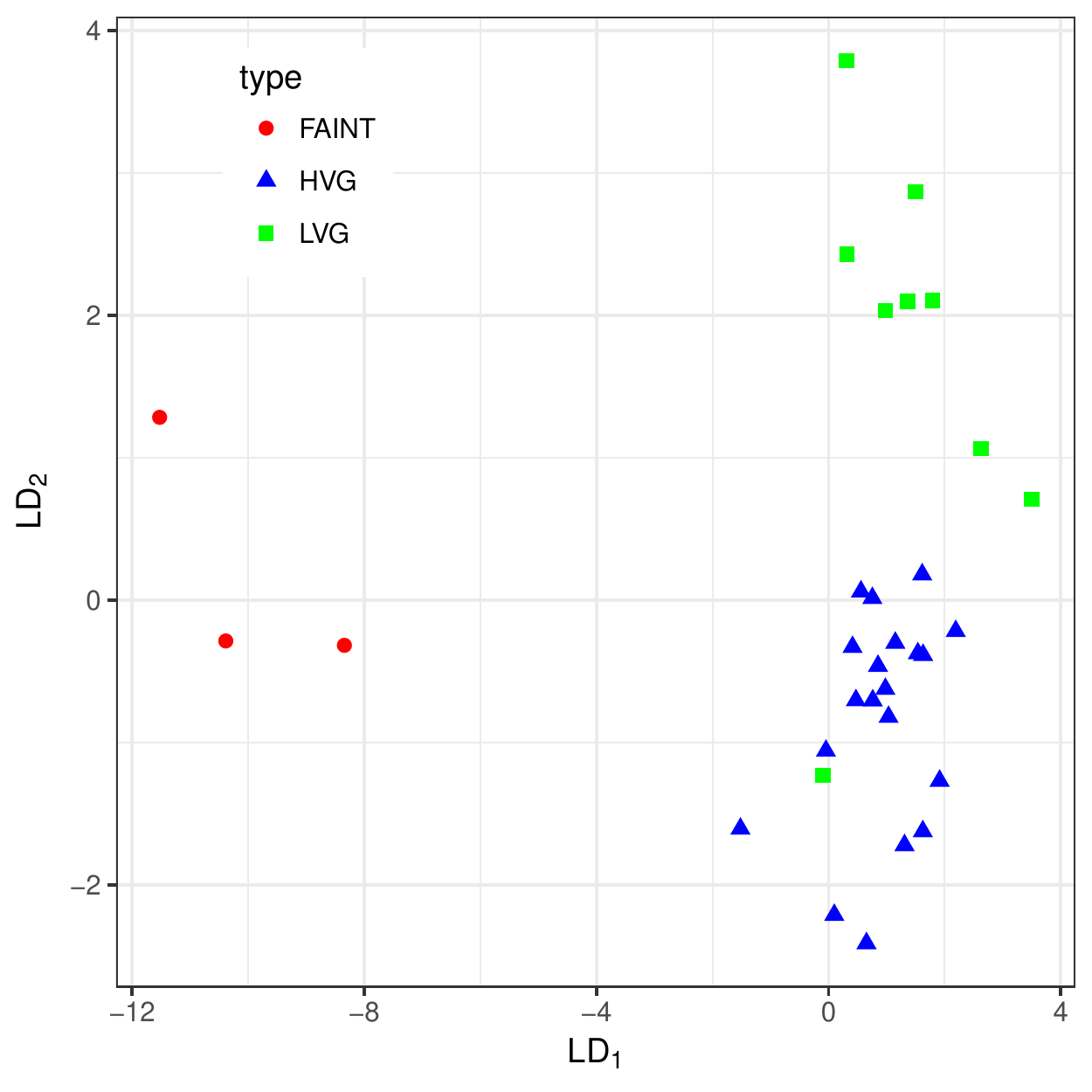}
\caption{Spectral classes separation of \cite{benetti2005diversity} 
using linear discriminant analysis of light curve principal
component scores.
The SNIa of FAINT, high
temporal velocity gradient (HVG) and low temporal velocity gradient
group (LVG) are plotted as red circle, blue
triangle, and green square, respectively.}
\label{fig:spcclass1}
\end{figure}

\begin{figure*}[hbtp]
\centering
\includegraphics[width=0.4\textwidth]{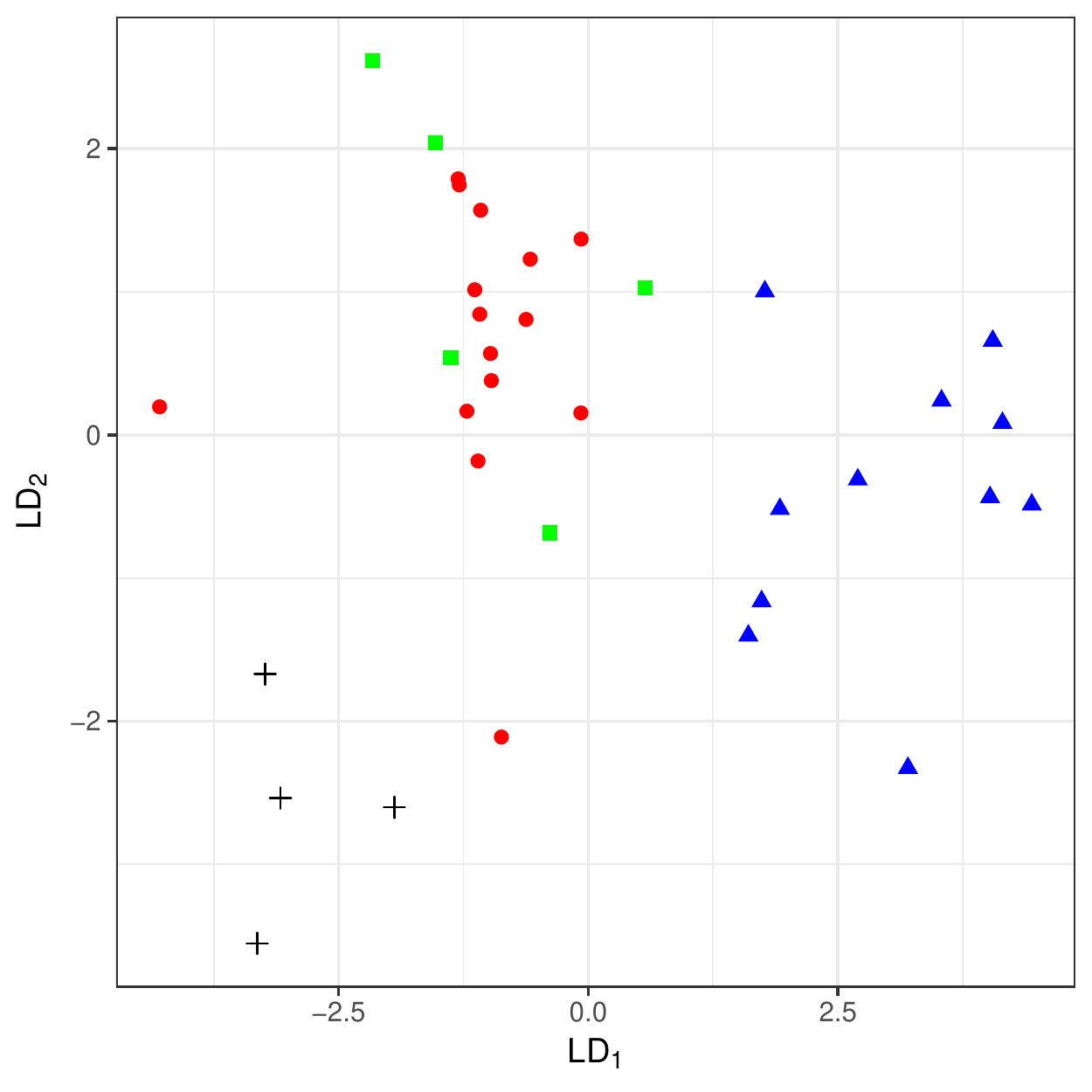}
\includegraphics[width=0.4\textwidth]{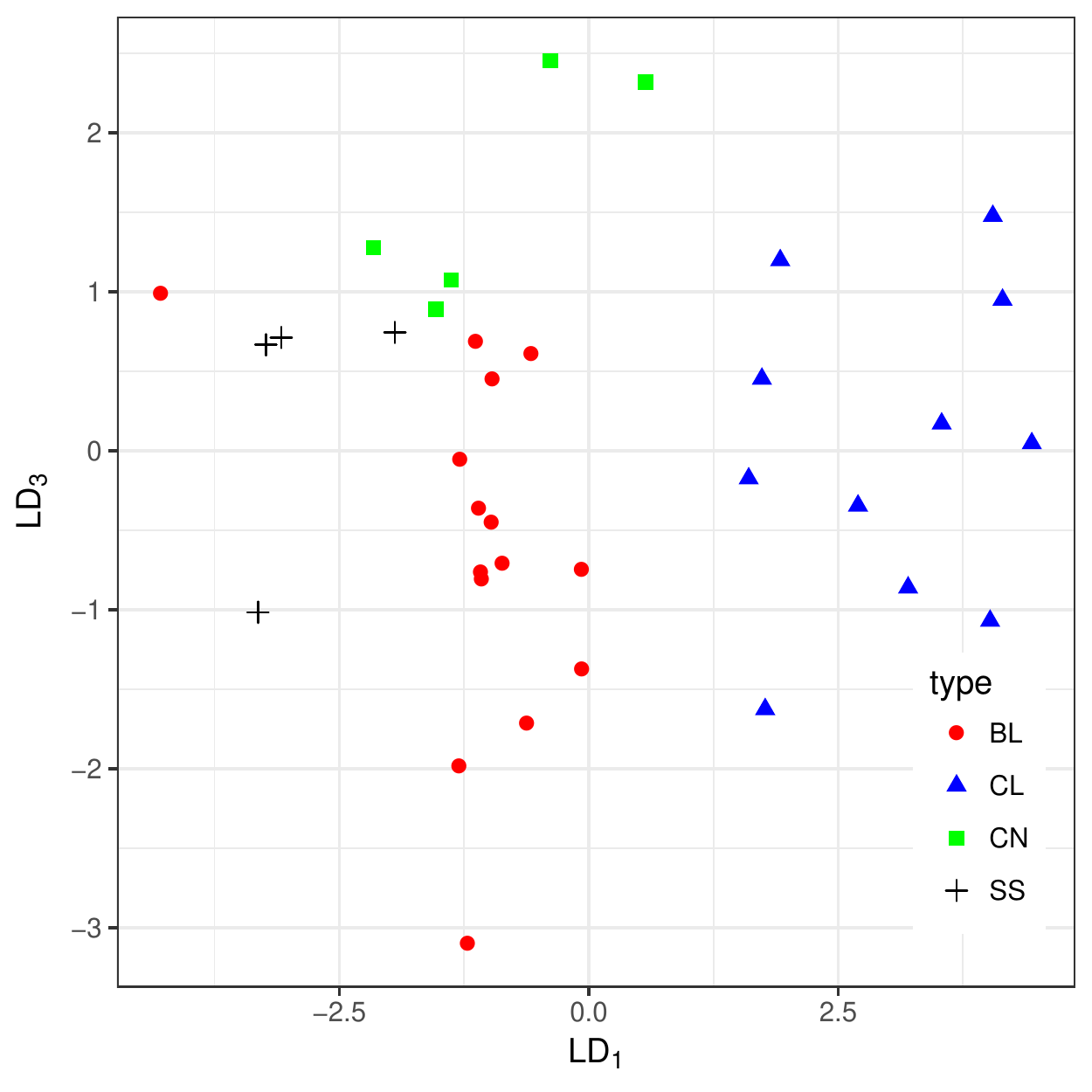}
\caption{Spectral classes separation of \cite{branch2009comparative}.
 The spectral class core normal (CN), broad line (BL), cool (CL), and shallow silicon
(SS) are shown as green square, red circle, blue triangle and black
cross, respectively. 
}
\label{fig:spcclass2}
\end{figure*}

\begin{figure}[hbtp]
\centering
\includegraphics[width=0.45\textwidth]{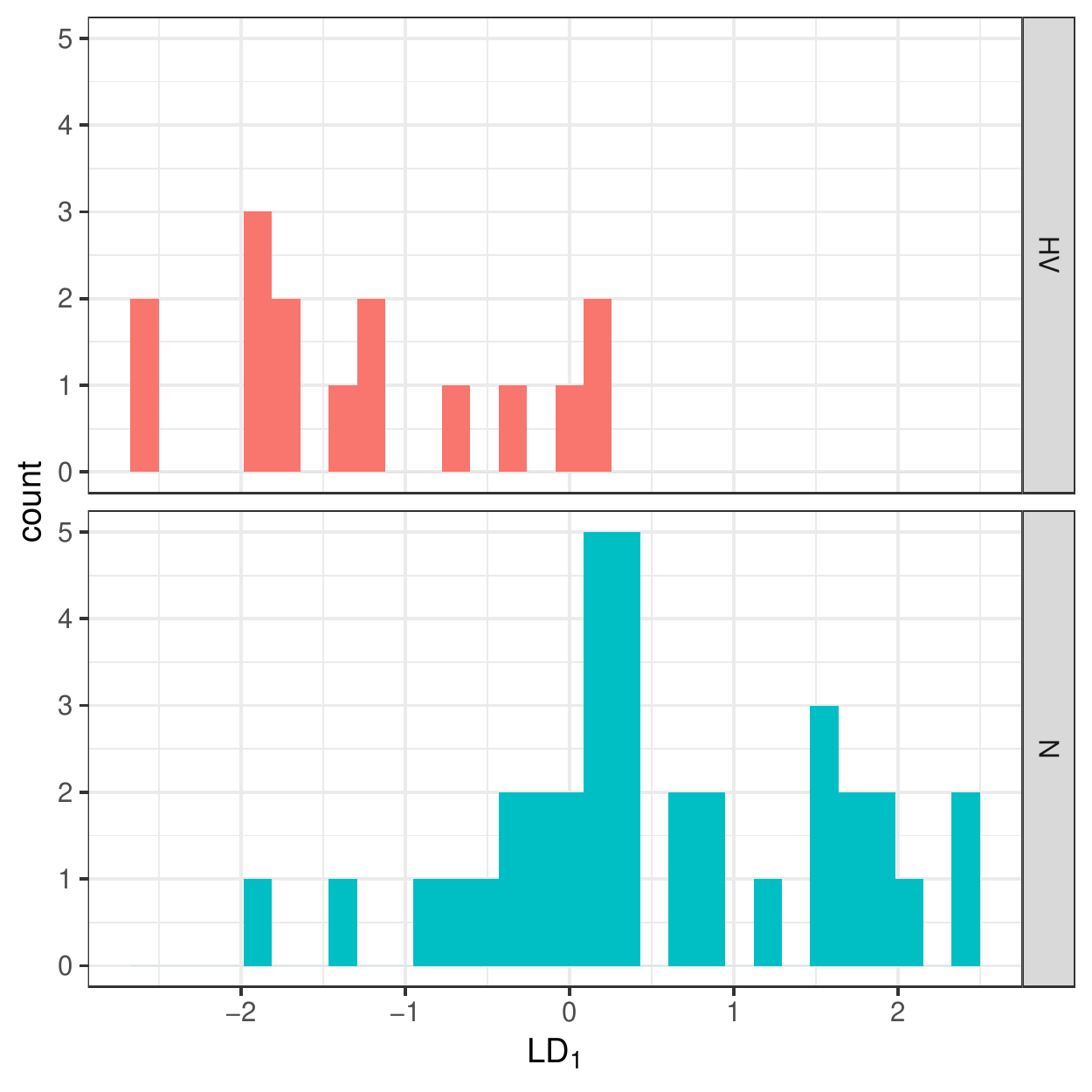}
\caption{Spectral classes separation of \cite{wang2009improved}. The
  histogram of the high velocity (HV) and normal (N) group are plotted
in the upper and lower panel, respectively.}
\label{fig:spcclass3}
\end{figure}

Section~\ref{sec:scorespc} explained that  the scores from our FPCA model provide
abundant information about spectral features. This section tries to
determine supernova spectral classes based on the scores. 
The task is a standard classification problem well-studied in statistics.
We can possibly treat spectral classes as the response, and our scores as predictors. 
We try to separate spectral classes with the aid of linear
discriminant analysis \citep[LDA,][]{murphy2012machine}. \added{When there
are $C$ spectral classes, LDA produces $C-1$ \textit{linear discriminants} as linear
combination of our scores $\beta^{(k)}_\lambda$, $k = 1,2,3,4$ and
$\lambda\in\{B,V,R,I\}$. The linear discriminants are denoted as
$\mathrm{LD}_c$ for $c=1,\cdots, C-1$ in our presentation. The linear discriminants
separate SNIa of different spectral classes in an optimal way.}
\deleted{To keep the analysis simple, we try to separate spectral classes with
only two selected scores. The two scores are selected with the aid of sparse
linear discriminant analysis (LDA). 
The sparse LDA tries to linearly
combine predictors to separate classes. The linear combining
coefficients are called loadings.  The number of predictors used
is encouraged to be small. We control the sparsity of the loading so
that only one variable is allowed in each of the first two loading vectors
(introducing more variables only gives marginal improvement). 
Then we examine the ability of the selected two
scores to separate spectral classes. }
In the following, we take the spectral
classes from \cite{benetti2005diversity}, 
\cite{branch2009comparative},  and \cite{wang2009improved}. 
We will use all the SNIa in our sample with spectral classes assigned
by these papers. 

Firstly, we consider the three spectral classes in
\cite{benetti2005diversity}. The three classes are FAINT, high
temporal velocity gradient (HVG) and low temporal velocity gradient
group (LVG). The average velocity gradients in the three groups are 87, 97,
and 37, respectively. 
\added{As there are $C=3$ spectral classes, the LDA produces $C-1=2$
  linear discriminants for each SNIa. }
The resulting $\mathrm{LD}_1$ and $\mathrm{LD}_2$ are
shown in  Figure~\ref{fig:spcclass1}. 
\deleted{The classification is carried out with the first and second score of
the $I$ band light 
curve,  $\beta_I^{(1)}, \beta_I^{(2)}$. The HVG and LVG classes have a small overlap regions.}
\added{ The SNIa of distinct spectral  classes are  well
  separated. However, 
 one LVG SNIa, SN2006et, gets  mixed in the HVG group.  }

\cite{branch2009comparative} provided spectral classification on the
basis of the absorption features near 5750\AA\ and 6100\AA. The absorption
features are measured by pseudo equivalent width. Their four groups
are core normal (CN), broad line (BL), cool (CL), and shallow silicon
(SS). The CN is a homogeneous class, and its absolute magnitude has a small
correlation with light curve width $\Delta M_{15}$.
The BL class tends to have strong absorption near 6100\AA.
For the CL class, the absorption features near 5750\AA\ and 6100\AA\ 
are both strong; and the SS class is another extreme with both
features being weak.  On average, the CL class tends to
have higher $\Delta M_{15}$ values and fainter absolute magnitude;
on the contrary, the SS class tend to have lower $\Delta M_{15}$
values and  brighter absolute magnitude.
Figure~\ref{fig:spcclass2} presents the four classes separation based
on three linear discriminants. \deleted{The CL (blue triangle) and SS (black cross) 
classes are separated in the lower
right and lower left corner of the plot. The CN (green square) and BL
(red circle) classes are mixed in the top.} \added{
The CL (blue triangle) and SS (black cross) 
classes are separated by the $\mathrm{LD}_1$ and 
$\mathrm{LD}_2$. The CN (green square) and BL
(red circle) classes are separated by the third linear discriminant
$\mathrm{LD}_3$. The three linear discriminants together demonstrate their ability
of clean separation. }

In \cite{wang2009improved},
the supernova samples are classified into
two groups Normal (N) and high velocity (HV) according to the observed
velocity of \siye. They found that the HV group has narrower
distribution in peak luminosity and decline rate, and this group also
prefers a lower extinction ratio. The distance prediction model was
applied to these two groups separately to reduce the dispersion in
their work. 
\added{Figure~\ref{fig:spcclass3} shows the spectral class
separation by one linear discriminant for the two spectral classes. 
These two classes have a considerable overlap in the histogram.  These
two spectral classes are not clearly distinguishable when applying our
method on light curves.}

When only light curve observation is available, 
Figure~\ref{fig:spcclass1} and Figure~\ref{fig:spcclass2} show that we
are able to classify the observation into their corresponding spectral
classes. However, even if a quantitative scheme proves to be difficult,
as illustrated by the class separation result in Figure~\ref{fig:spcclass3},
we could still try 
to extract a subgroup of SNIa which could be more homogeneous than the
entire sample. This is useful in controlling systematic errors
in distance determination.


\section{Distance Prediction}\label{sec:distance}

Our FPCA light curve model can produce the entire shape of a light
curve, which in turn can be used for intrinsic luminosity adjustment
in distance prediction. In this section, we consider an adjustment
using a functional linear form and compare it with the standard
adjustment using $\Delta M_{15}$.  
We consider both the peak magnitude and the CMAGIC magnitude 
as distance  indicators in separate subsections.
Note that in this comparative study, the purpose is to show the
potential advantage of using the entire light curve shape,  and we
still use our FPCA model to determine the value of $\Delta M_{15}$.  
\added{For a more comprehensive comparison, the $\Delta M_{15}$ fitted
  by SALT II and the SALT II shape parameter are also included as two
  alternative models.}

\subsection{Distance Models}\label{sec:distmodels}
Distance models fit a linear model for the distance modulus. 
The distance modulus $\mu$ is a function of redshift.
In particular, 
$$\mu(z) = 25 + 5\log_{10}(D_L(z) \mathrm{Mpc^{-1}})\, ,$$
where $D_L(z)$ is the luminosity distance under a fixed cosmology with 
$\Omega_m = 0.3$, $\Omega_\Lambda = 0.7$ and Hubble constant $h = 0.7$.
With type Ia supernova, the standard model for predicting distance is
\begin{align} 
(M1)\quad \mu = m_B - M & - \alpha (C - \langle C\rangle)  \nonumber \\
 & -  \delta (\Delta M_{15}  - \langle \Delta M_{15} \rangle).
\label{eqn:distsimple}
\end{align}
In the above, $m_B$ is the apparent $B$ band peak magnitude, $M$ is the
absolute $B$ band peak magnitude. All these are magnitudes in the
rest-frame filter. 
$C$ is the observed color at $B$ maximum, $(B-V)_{\tmax}$.
$\Delta M_{15}$ is the magnitude change 15 days after $B$ maximum for
the $B$ band light curve.

Although simple by concept, $\Delta M_{15}$ is not a quantity that can
be directly measured accurately.   
Its calculated value is highly influenced by light curve fitting errors.  
In practice, because of the rapid luminosity decay at 15 days past
optical maximum,   a small error in determining 
the peak epoch can lead to large inaccuracy 
of $\Delta M_{15}$. 


With our proposed FPCA light curve model, we consider an alternative
adjustment of intrinsic luminosity using the entire shape of the light
curve. The shape parameter $\Delta M_{15} $ is replaced by a functional
linear term
\begin{equation} \label{eqn:functionalQ}
Q = \int \delta(q) \{g_B(q) - m_B - \phi_{0B}(q)\}
  \mathrm{d}q\, ,
\end{equation}
where $\delta(q)$ is a fixed function to be determined by the data.
Note that in Equation~(\ref{eqn:functionalQ}),
the peak magnitude $m_B$ and the mean function $\phi_{0B}(q)$ 
are subtracted, so that
only the light curve shape may influence the value of $Q$. With this
functional linear form, the distance model becomes
\begin{align} 
(M2)\quad  \mu = m_B &- M - \alpha (C - \langle C\rangle) \nonumber \\
&- (Q - \langle Q\rangle). \label{eqn:distfunctional}
\end{align}
Since the light curve shape enters the model as a functional linear term,
we refer to (9) as a functional linear distance model. This model has
some similarity to the functional linear regression model studied in
statistics \citep{muller2005generalized}. 

For each supernova, the associated distance prediction uncertainty
includes the parts due to peculiar velocity, measurement error
in the apparent magnitude, and intrinsic Type Ia supernova property
variation.  
In the following we assume a peculiar velocity of $v_{\pec}
  = 300 \mathrm{\ km\ s}^{-1}$. It introduces magnitude uncertainty of
$\sigma_{\pec} = (5/\ln 10)(v_{\pec}/ cz) = 0.002173/z$, where $c$ is
the speed of light. The uncertainty of the apparent peak magnitude
$\sigma_m$ is computed from  the bootstrap method.
The associated distance prediction uncertainty is computed as
$\sigma_s^2 = \sigma^2_{\pec} + \sigma_m^2$.

The standard distance model~(\ref{eqn:distsimple})
 is trained by minimizing the $\chi^2$,
\begin{equation} \label{eqn:objM1}
\begin{split}
\sum_{s = 1}^S \frac{1}{\sigma_s^2}[\mu - m_B  &+ M +
\alpha (C - \langle C\rangle)  \\
& + \delta (\Delta M_{15}  - \langle \Delta M_{15} \rangle)]^2\, ,
\end{split}
\end{equation}
where $\Delta M_{15}$ is determined using our FPCA model. 
The minimization is taken with respect to $M, \alpha, \delta$.
A similar $\chi^2$ minimization applies to model~(\ref{eqn:distfunctional}).
The functional linear distance model~(\ref{eqn:distfunctional})
 is trained by minimizing
\begin{equation} \label{eqn:objM2}
\begin{split}
\sum_{s = 1}^S & \frac{1}{\sigma_s^2}
[\mu - m_B + M +  \alpha (C - \langle C\rangle)  \\
& + (Q - \langle Q \rangle)
 + \eta \int [\delta''(q)]^2 \mathrm{d} q.
\end{split}
\end{equation}
The minimization is taken with respect to $M$, $\alpha$ and $\delta(q)$
in $Q$.
The last term is the integration of the squared second order
derivative of $\delta(q)$. This is a roughness penalty to encourage
the smoothness of the solution of $\delta(q)$. 
The $\eta$ parameter controls the amount of penalty imposed
and is chosen by the cross-validation. In the minimization of 
Equation~(\ref{eqn:objM2}),
the solution of $\delta(q)$ is searched among the 
span of the  principal component functions
$\phi_k(q)$, i.e., $\delta(q) = \sum^K_{k=1} \delta^{(k)}
\phi_k(q)$. 
Using the principal component functions, we
only need to solve for the scalar $\delta^{(k)}$'s to
get an estimate of the function $\delta(q)$. 
\added{The details of the algorithm and computation of the 
degree of freedom of the resulting model can
be found in Appendix~\ref{sec:fittingFLDM}.}

\added{
For a complete comparison, we also consider two alternative distance models
to Equation~\ref{eqn:distsimple}. 
The $\Delta M_{15}$ value is replaced by the $\Delta M_{15}$ fitted by
SALT II, and this is denoted as model S1.
Besides, the $\Delta M_{15}$ is replaced by
the SALT II shape parameter $x_1$ \citep{guy2007salt2},
 and the resulting model is denoted as S2. }

\subsection{Comparing the Distance Prediction
  Models}\label{sec:comparemodel}

The leave-one-out cross-validation is used to compare the distance
prediction models: M1 (Equation~\ref{eqn:distsimple}), M2
(Equation~\ref{eqn:distfunctional}), the SALT II $\Delta M_{15}$ model
S1, and the SALT II shape parameter model S2. 
Cross-validation \citep[Section 6.5.3 of ][]{murphy2012machine}
 is a commonly used method in statistics to evaluate
the out-of-sample performance of a prediction model. It works as
follows.   Each time one sample (i.e. one SNIa) is removed 
from the dataset, and the remaining dataset is used to
train the distance prediction model. 
The resulted model is then applied to the removed sample to get a
predicted distance modulus $\hat{\mu}$. The cross-validated error
 of this prediction is denoted as $\Delta\mu = \mu - \hat{\mu}$.
We repeat this procedure for all SNIa samples in the dataset and
summarize the cross-validated errors by the weighted mean squared
errors (WMS), 
$$
\mathrm{WMS}_{cv} = (\sum_s \Delta\mu_s^2/\sigma_s^2)
/ (\sum_s 1/\sigma_s^2)\, ,
$$
whose square-root is $\text{WRMS}_{cv} = \text{WMS}_{cv}^{1/2}$.

Before applying the model to our dataset, we make a further selection
of the dataset. We select the SNIa observations with CMB
redshift $z>0.01$ and the observed color $(B-V)_{\tmax}<CC$
for several values of $CC$. 
The cut $0.01$ on redshift restricts the uncertainty due 
to peculiar velocity. In addiction, the cut $CC$ on the 
observed color helps us to select
a homogenous group of supernovae. 

The result is shown in Table~\ref{tbl:accuracy}.
In the table, the $\mathrm{WRMS}_{cv}$ is computed using cross-validated
errors. The $\chi^2$ is computed using in-sample
errors. \added{The degree of freedom (DOF) for this $\chi^2$ is computed
using the formulas at the end of Appendix~\ref{sec:fittingFLDM}.}
For example, the result for $CC=0.05$ is given on the first row. 
With the cut of $z>0.01$ and $(B-V)_{\tmax}<(0.05)$, there are 37 SNIa
in the remaining sample. The $\Delta M_{15}$ model~(\ref{eqn:distsimple})
results in a $\mathrm{WRMS}_{cv}$ of 0.089,  and  the functional linear distance
 model~(\ref{eqn:distfunctional}) has a close $\mathrm{WRMS}_{cv}$ of 0.091. 
\added{The $\mathrm{WRMS}_{cv}$ for two SALT II related models S1 and
  S2 are 0.101  and 0.108, respectively. }
The difference of the four models is not significant. 
With only shape and color information, all models 
appear to have approached the statistical
limits of the data in constructing a Hubble diagram with minimal
dispersion.  The dispersion is dominated by the peculiar velocity.
However, the functional linear distance
model~(\ref{eqn:distfunctional}) 
consistently produces  smaller $\mathrm{WRMS}_{cv}$ across different sample groups. As
the value of $CC$ increases, the $\mathrm{WRMS}_{cv}$ for
model~(\ref{eqn:distfunctional}) is stable at the level of $\sim
0.120$. The $\mathrm{WRMS}_{cv}$ for the other models increases to more
than $0.14$.

\begin{table*}
\centering
\caption{Comparison of the distance models\label{tbl:accuracy}}
\begin{tabular}{|cr|rr|rr|rrr|rrr|}
\hline
& & 
\multicolumn{2}{c|}{S1} &
\multicolumn{2}{c|}{S2} &
\multicolumn{3}{c|}{M1} &
\multicolumn{3}{c|}{M2} \\
$CC$ & $N$ & 
$\mathrm{WRMS}_{cv}$&  $\chi^2$ &
$\mathrm{WRMS}_{cv}$&  $\chi^2$ &
$\mathrm{WRMS}_{cv}$&  $\chi^2$ & DOF &
$\mathrm{WRMS}_{cv}$&  $\chi^2$  & DOF \\
\hline
\hline
 $0.05$ &  37 &  0.101 &  127.15 &  0.108 &  141.45 &  0.091 &  114.34 &  34.00 &  0.089 &  93.37 &  31.97\\ 
 \hline 
$0.1$ &  48 &  0.120 &  255.83 &  0.127 &  277.66 &  0.133 &  300.04 &  45.00 &  0.117 &  198.32 &  42.81\\ 
 \hline 
$0.2$ &  62 &  0.153 &  487.01 &  0.162 &  528.86 &  0.137 &  415.53 &  59.00 &  0.119 &  276.56 &  56.65\\ 
 \hline 
$0.3$ &  65 &  0.152 &  509.85 &  0.162 &  559.14 &  0.137 &  433.16 &  62.00 &  0.118 &  288.45 &  59.61\\ 
 \hline 
$0.4$ &  67 &  0.151 &  514.45 &  0.161 &  563.45 &  0.138 &  447.91 &  64.00 &  0.119 &  300.58 &  61.59\\ 
 \hline 
\end{tabular}
\tablecomments{Comparison of  the $\mathrm{WRMS}_{cv}$, $\chi^2$ and degree of freedom
  (DOF) for 
four models: S1, S2, M1 (Equation~\ref{eqn:distsimple})
and M2 (Equation~\ref{eqn:distfunctional}).  S1 is Equation~\ref{eqn:distsimple} with
SALT II $\Delta M_{15}$, and S2 is Equation~\ref{eqn:distsimple} with SALT II shape parameter.
The comparison is based on different 
 observed color cutoffs $(B-V)_{\tmax} < CC$, and 
a common redshift cut $z > 0.01$. The column $N$
  is the sample size after the cutoff. The DOF of S1 and S2 is the
  same as the DOF of M1.} 
\end{table*}

\begin{figure}[hbtp]
\centering
\includegraphics[width=0.4\textwidth]{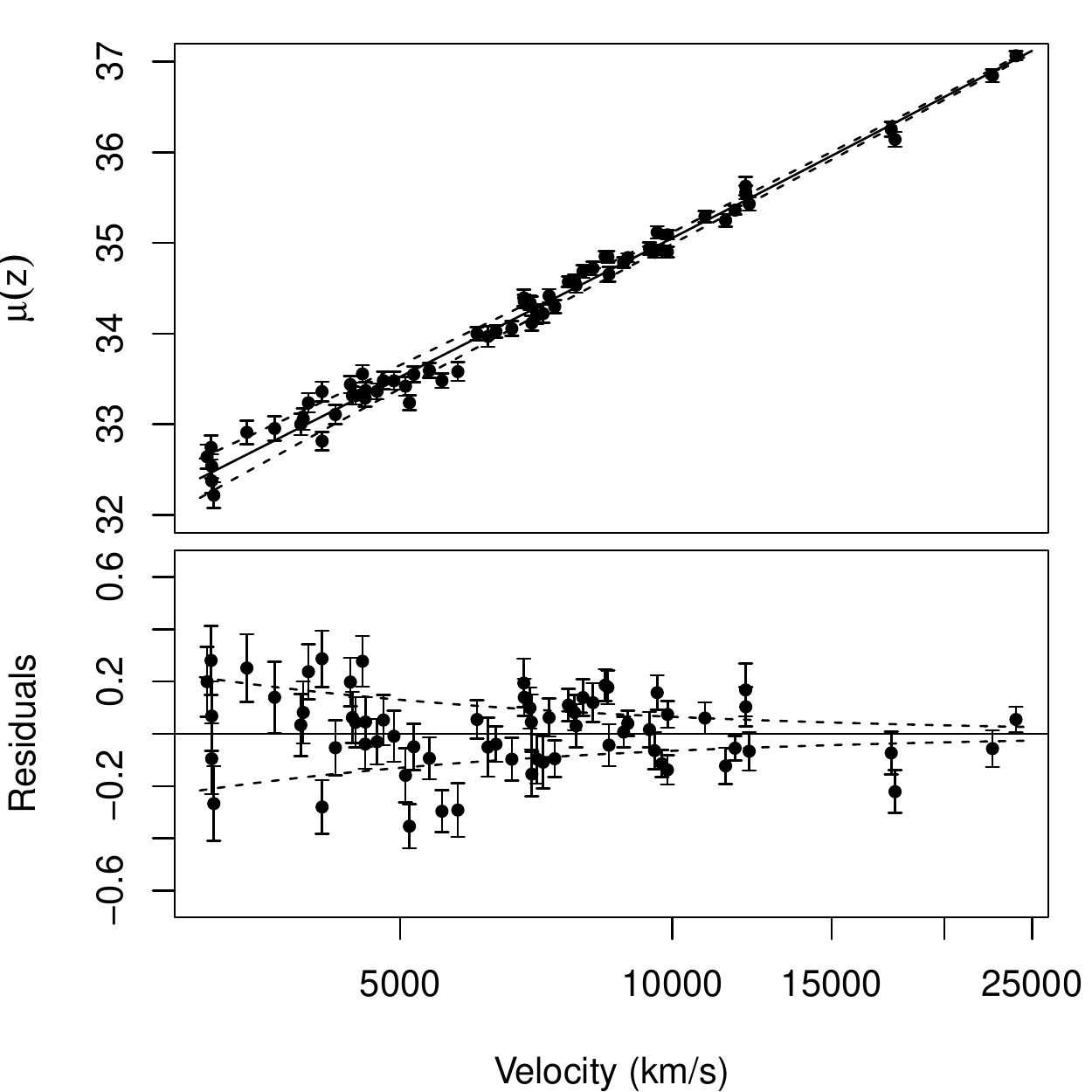}
\caption{ Predicted distance and residuals of the functional linear distance
  model. The dashed lines represent the uncertainty due to the
  peculiar velocity.
\label{fig:distancepred}
} 
\end{figure}

\begin{figure}
\centering
\includegraphics[width = 0.4\textwidth]{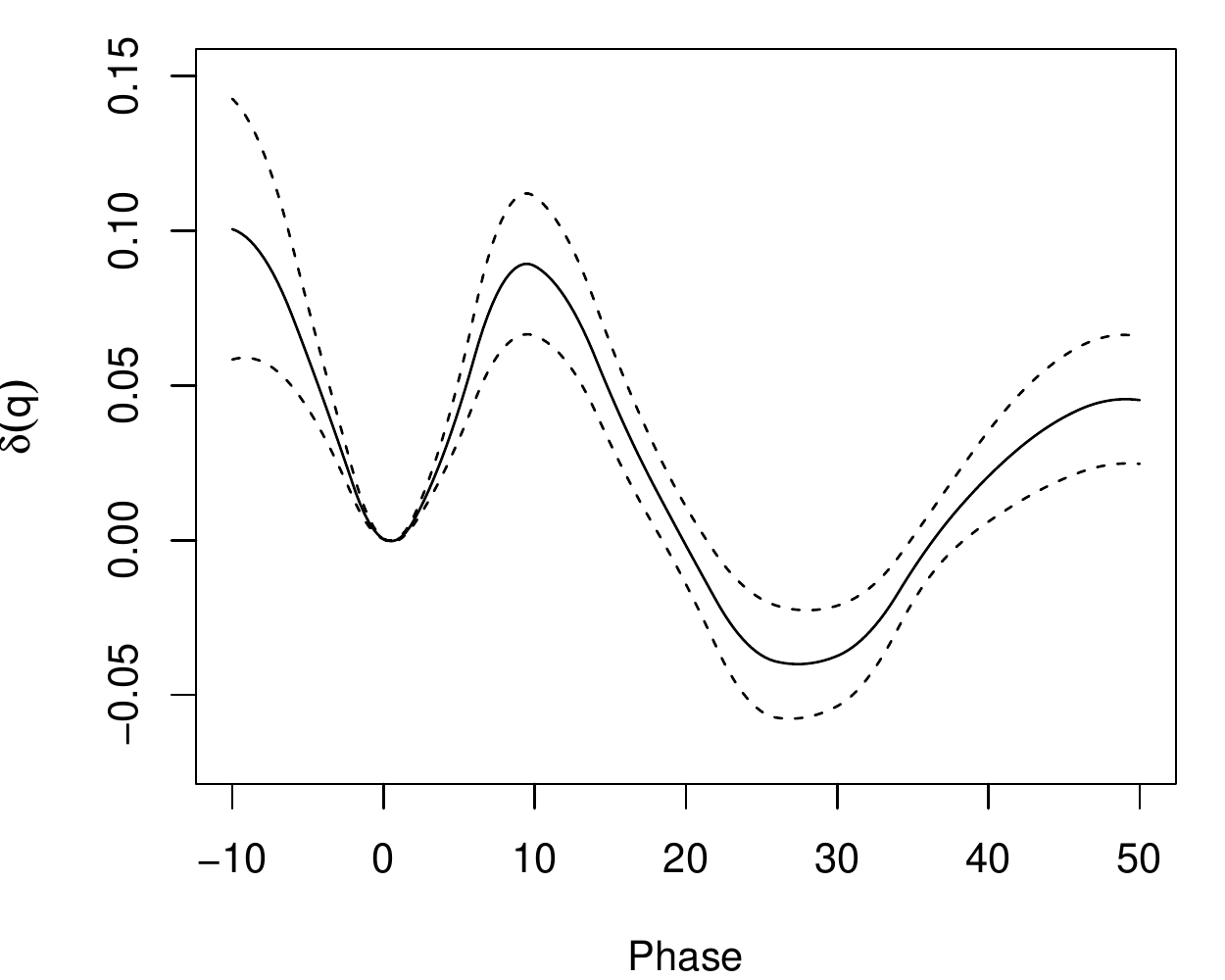}
\caption{The estimated $\delta(q)$ of the functional linear 
form~(\ref{eqn:functionalQ}). The solid line is the
estimated $\delta(q)$, the dashed lines represent one standard
deviation uncertainty.
\label{fig:deltaq}}
\end{figure}

\begin{figure}
\centering
\includegraphics[width = 0.4\textwidth]{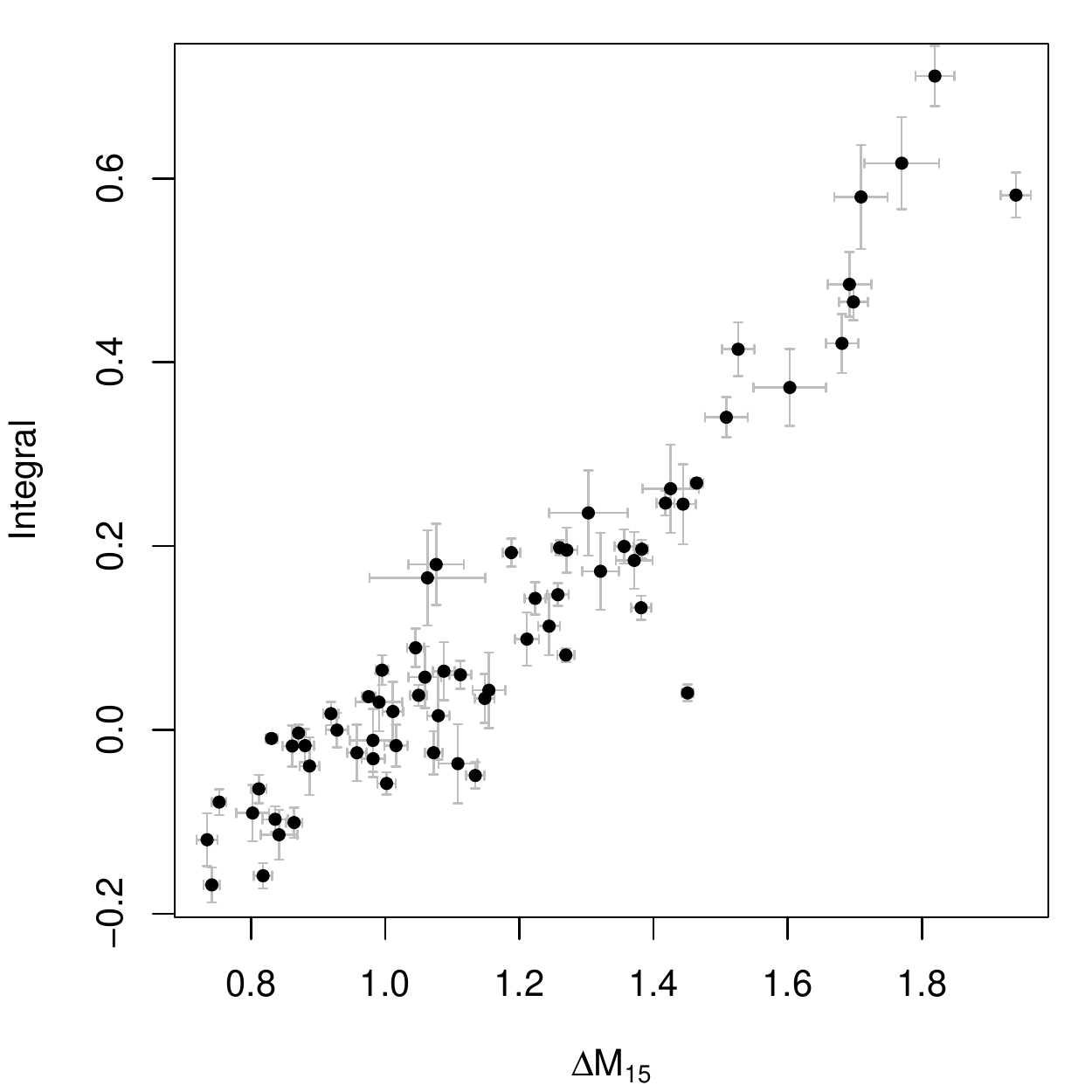}
\caption{Comparing the $\Delta M_{15}$ on the horizontal axis with
 the value of the functional linear form
$\int \delta(q) (g_B(q) - m_B -\phi_{0B}(q)) \mathrm{d}q$ on the vertical axis. 
\label{fig:integration}}
\end{figure}

At last, we present more detailed results
for model~(\ref{eqn:distfunctional}) at the color cut $CC = 0.4$,
where the sample size is the largest in our consideration.
The upper panel of Figure~\ref{fig:distancepred}
shows the predicted distance modulus versus redshift
velocity. The  lower panel of Figure~\ref{fig:distancepred}  
plots the cross-validated residuals and associated error bars.  
The dashed curves represent the uncertainty due to the assumed peculiar
velocity. Note the scatter of the residuals is dominated by 
peculiar velocity at redshifts around 300 $\mathrm{km}\cdot
\mathrm{s}^{-1}$.  
In addition, the estimated functional
coefficient $\delta(q)$ is presented in Figure~\ref{fig:deltaq}.
This functional coefficient $\delta(q)$ is positive over the phase
range $(-10,20)$. This suggests that the functional linear form
 still tries to measure the width of the light
curve in its own way, and the measurement is adjusted by the phase range
$(20,50)$. 
Figure~\ref{fig:integration} compares the quantity 
$\Delta M_{15}$ with the calculated value of the 
the functional linear form
  $\int \delta(q) (g_B(q) - m_B -\phi_{0B}(q)) \mathrm{d}q$. 

\subsection{The CMAGIC for Distance Prediction}

\begin{figure}[hbtp]
\centering
\includegraphics[width=0.4\textwidth]{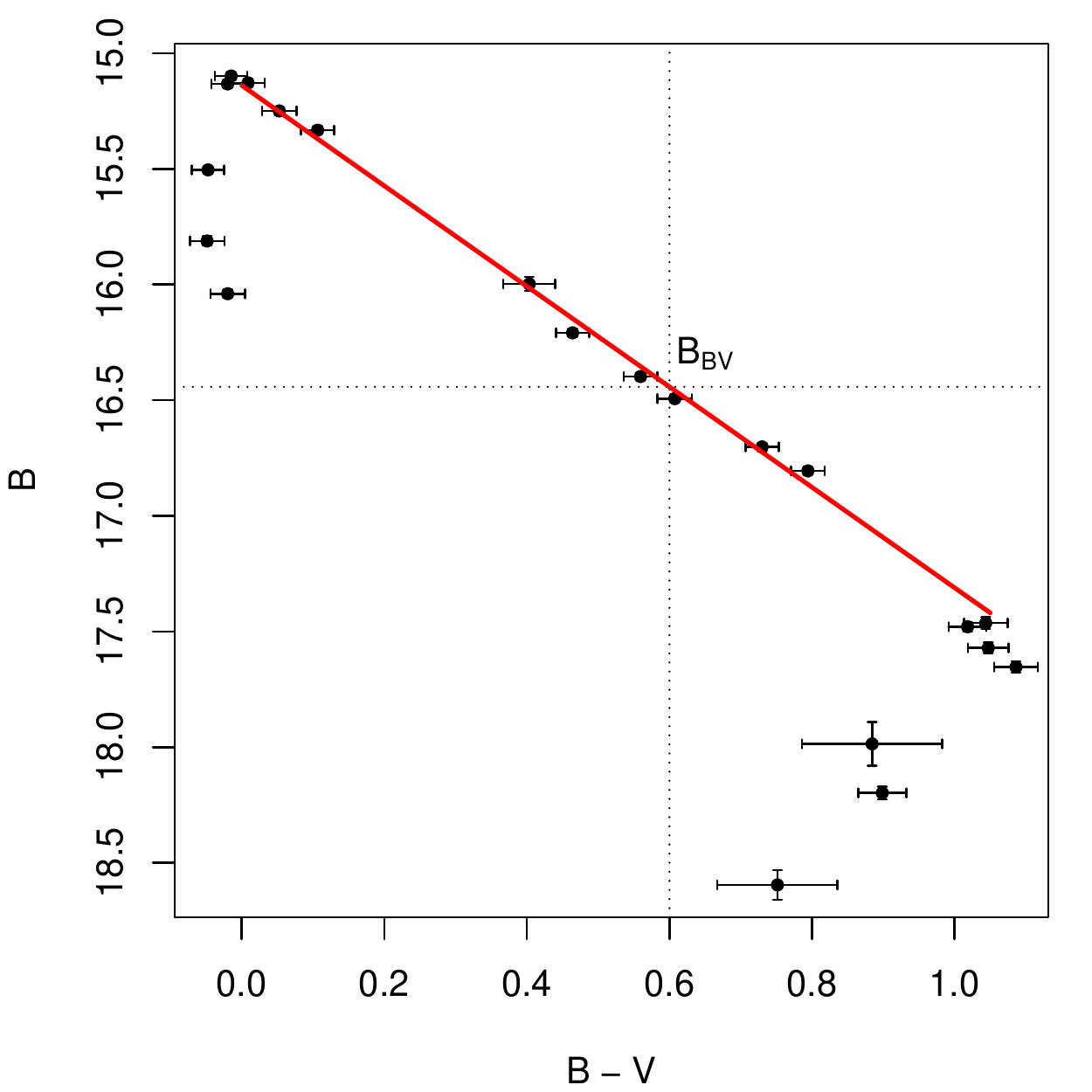}
\caption{ An example of the linear color magnitude evolution after B
  maximum. The red line is the fitted linear relation. The $B$ band 
magnitude   at $B-V = 0.6$ is extracted to get the CMAGIC magnitude
$B_{BV}$. 
} \label{fig:bbveg}
\end{figure}

We now evaluate the effectiveness of using the entire light curve
shape to adjust intrinsic luminosity for distance predication
when the CMAGIC magnitude is used as the distance indicator. 
The CMAGIC magnitude, proposed by
\cite{wang2003multicolor}, exploits the linear relation of the
color evolution for about 30 days after the $B$ maximum. 
During this phase the
$B$ and $B - V$ magnitude follow a linear trend, as
illustrated by the red line in Figure~\ref{fig:bbveg}. 
\begin{equation}
B  = B_{BV} + \gamma (B - V - 0.6)
\end{equation}
where $\gamma$ is the slope of the linear relation.
The exact starting and ending epochs of this linear evolution  vary
among supernovae with their intrinsic brightness. Some supernovae with
a small $\Delta M_{15}$ show a ``bump'' feature in the color magnitude
evolution immediately after $B$ maximum. 
For our supernova samples, the observation points in the $B$ band phase
range $[+3, + 25]$ 
are used to fit the linear relation. \cite{wang2003multicolor}
noticed that
the slope $\gamma$ has a small scattering $0.18$ around the mean of $2.07$.

The CMAGIC magnitude, denoted by $B_{BV}$, is defined as
the $B$ band magnitude when $B - V = 0.6$. 
\cite{wang2003multicolor} replaced the peak magnitude $m_B$
 in (\ref{eqn:distsimple}) by $B_{BV}$ and considered the following model 
\begin{align}
(M3)\quad \mu =& B_{BV} - M - \delta (\Delta M_{15} - \langle \Delta M_{15}
  \rangle) - (b_2 - \gamma) \times \nonumber \\
& [\frac{m_B - B_{BV}}{\gamma} + 0.6 + 1.2(\frac{1}{\gamma}-
  \langle \frac{1}{\gamma}\rangle) ]\, ,
\label{eqn:distCMAGIC}
\end{align}
where $M$, $\delta$ and $b_2$ are parameters to be estimated.
In this paper, we will use $\Delta M_{15}$ determined by our FPCA model. 
Besides, as an alternative model, the $\Delta M_{15}$ is replaced by
the functional linear form $Q$ defined in Equation~(\ref{eqn:functionalQ}), 
\begin{align}
(M4)\quad \mu =& B_{BV} - M -
(Q -\langle Q\rangle) -  (b_2 - \gamma)\times\nonumber \\
& \quad 
[\frac{m_B - B_{BV}}{\gamma} + 0.6 + 1.2(\frac{1}{\gamma}-
  \langle \frac{1}{\gamma}\rangle) ]\, .
\label{eqn:distCMAGICFunctional}
\end{align}
\added{We add two more models for completeness of comparison. 
The $\Delta M_{15}$ estimated by 
SALT~II is employed in Equation~\ref{eqn:distCMAGIC}, 
and this model is denoted as S3. Besides, the $\Delta M_{15}$ in 
Equation~\ref{eqn:distCMAGIC} is also replaced by the SALT~II shape parameter
$x_1$, and this model is denoted as S4.}


In order to fit the linear color  evolution,
with the dataset described in Section~\ref{sec:data}, 
at least five observation points is 
required in the $B$ band phase range $[+3, + 25]$. 
We select those samples with $z>0.01$ and make various
levels of cut on the observed color $(B - V)_{\tmax}$ at $B$ maximum. 
The color cut is necessary, due to the fact that the linear color
evolution and CMAGIC can be best constrained among low $(B-V)_{\tmax}$
samples. A five sigma cut is also applied to $\gamma$.
 For the $66$ samples with color cut of 0.3, 
their slopes $\gamma$ have a mean of 2.15 and standard deviation 0.20.

\begin{figure}[hbtp]
\centering
\includegraphics[width=0.48\textwidth]{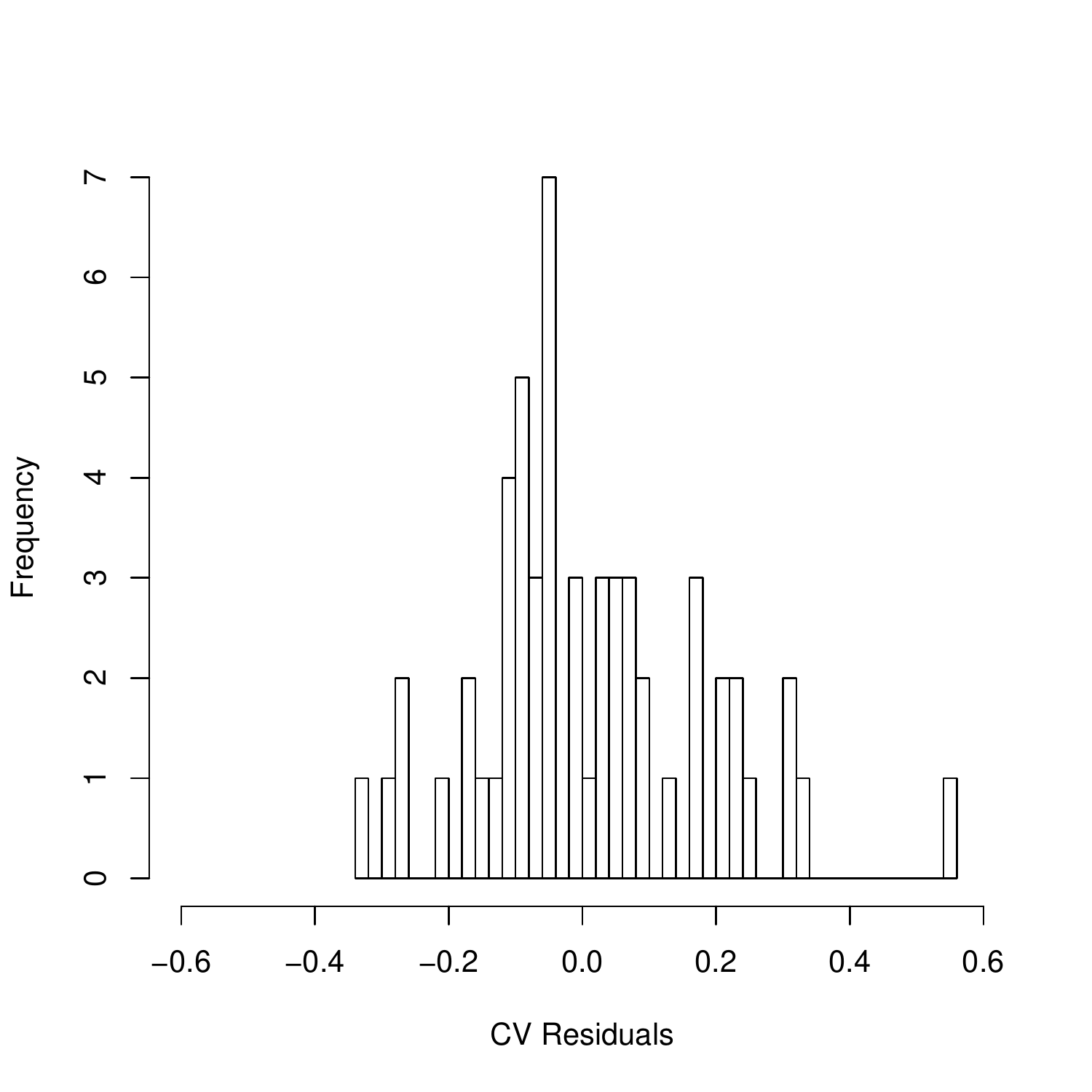}
\caption{The histogram of cross-validation residuals for M4 at color
  cut $CC=0.4$. 
} \label{fig:cmagicresiduals}
\end{figure}

Over the selected SNIa samples, four models S3, S4, M3
(Equation~\ref{eqn:distCMAGIC}), and 
M4 in (Equation~\ref{eqn:distCMAGICFunctional}) 
are tested using the  leave-one-out cross-validation 
 procedure as described in Section~\ref{sec:comparemodel}.
Table~\ref{tbl:accuracy2} presents their $\mathrm{WRMS}_{cv}$, $\chi^2$ and degree of
freedom (DOF) at 
different levels of cutoff at the observed color at $B$ maximum. 
\added{The performances of the four models are almost comparable. However,
the shape parameters produced by our model still give smaller residuals.}
At the cut $0.05$ of the color at $B$ maximum, the model S3, S4
M3 and M4 has a $\mathrm{WRMS}_{cv}$ of $0.145$, $0.149$,
 $0.119$ and $0.122$, respectively.  
At the cut $0.4$ of the color at $B$ maximum, the models
have $\mathrm{WRMS}_{cv}$ of $0.155$, $0.160$, $0.137$ and $0.135$,
respectively. The histogram of the residuals for M4 is ploted in 
Figure~\ref{fig:cmagicresiduals}. While there are a few samples with
large residuals, most samples have residuals with absolute value less
than $0.2$. 
SN 2007ca show significantly larger residual for unknown reasons. It is however the most highly extinguished supernova included in our sample.
Notice the sample size $N$ in
Table~\ref{tbl:accuracy2} is smaller than in
Table~\ref{tbl:accuracy}, because we require at least $5$
observational points  in the $B$ band phase range $[+3, + 25]$
to constrain the color evolution.

\begin{table*}
\centering
\caption{Comparison of the distance models\label{tbl:accuracy2}}
\begin{tabular}{|cr|rr|rr|rrr|rrr|}
\hline
& & 
\multicolumn{2}{c|}{S3} &
\multicolumn{2}{c|}{S4} &
\multicolumn{3}{c|}{M3} &
\multicolumn{3}{c|}{M4} \\
$CC$ & $N$ & 
$\mathrm{WRMS}_{cv}$&  $\chi^2$ &
$\mathrm{WRMS}_{cv}$&  $\chi^2$ &
$\mathrm{WRMS}_{cv}$&  $\chi^2$ & DOF &
$\mathrm{WRMS}_{cv}$&  $\chi^2$  & DOF \\
\hline
\hline
 $0.05$ &  32 &  0.145 &  184.13 &  0.149 &  195.27 &  0.119 &  151.12 &  29.00 &  0.122 &  111.34 &  26.32\\ 
 \hline 
$0.1$ &  40 &  0.156 &  270.95 &  0.160 &  281.62 &  0.137 &  233.32 &  37.00 &  0.130 &  165.53 &  34.45\\ 
 \hline 
$0.2$ &  52 &  0.149 &  351.53 &  0.154 &  371.20 &  0.130 &  292.66 &  49.00 &  0.130 &  243.81 &  46.41\\ 
 \hline 
$0.3$ &  55 &  0.149 &  371.92 &  0.155 &  397.16 &  0.129 &  306.25 &  52.00 &  0.128 &  255.46 &  49.28\\ 
 \hline 
$0.4$ &  56 &  0.155 &  413.60 &  0.160 &  435.51 &  0.137 &  347.93 &  53.00 &  0.135 &  289.69 &  50.28\\ 
 \hline 
\end{tabular}
\tablecomments{Comparison of  the $\mathrm{WRMS}_{cv}$, $\chi^2$ and degree of freedom
  (DOF) for 
four models: S3, S4, M3 (Equation~\ref{eqn:distCMAGIC})
and M4 (Equation~\ref{eqn:distCMAGICFunctional}).  S3 is Equation~\ref{eqn:distCMAGIC} with
SALT II $\Delta M_{15}$, and S4 is Equation~\ref{eqn:distCMAGIC} with SALT II shape parameter.
The comparison is based on different 
 observed color cutoffs $(B-V)_{\tmax} < CC$, and 
a common redshift cut $z > 0.01$. The column $N$
  is the sample size after the cutoff. The DOF of S3 and S4 is the
  same as the DOF of M3.} 
\end{table*}


\section{Results from the fv-FPCA model}~\label{sec:joint}

\begin{figure}
\centering
\includegraphics[width=0.23\textwidth]{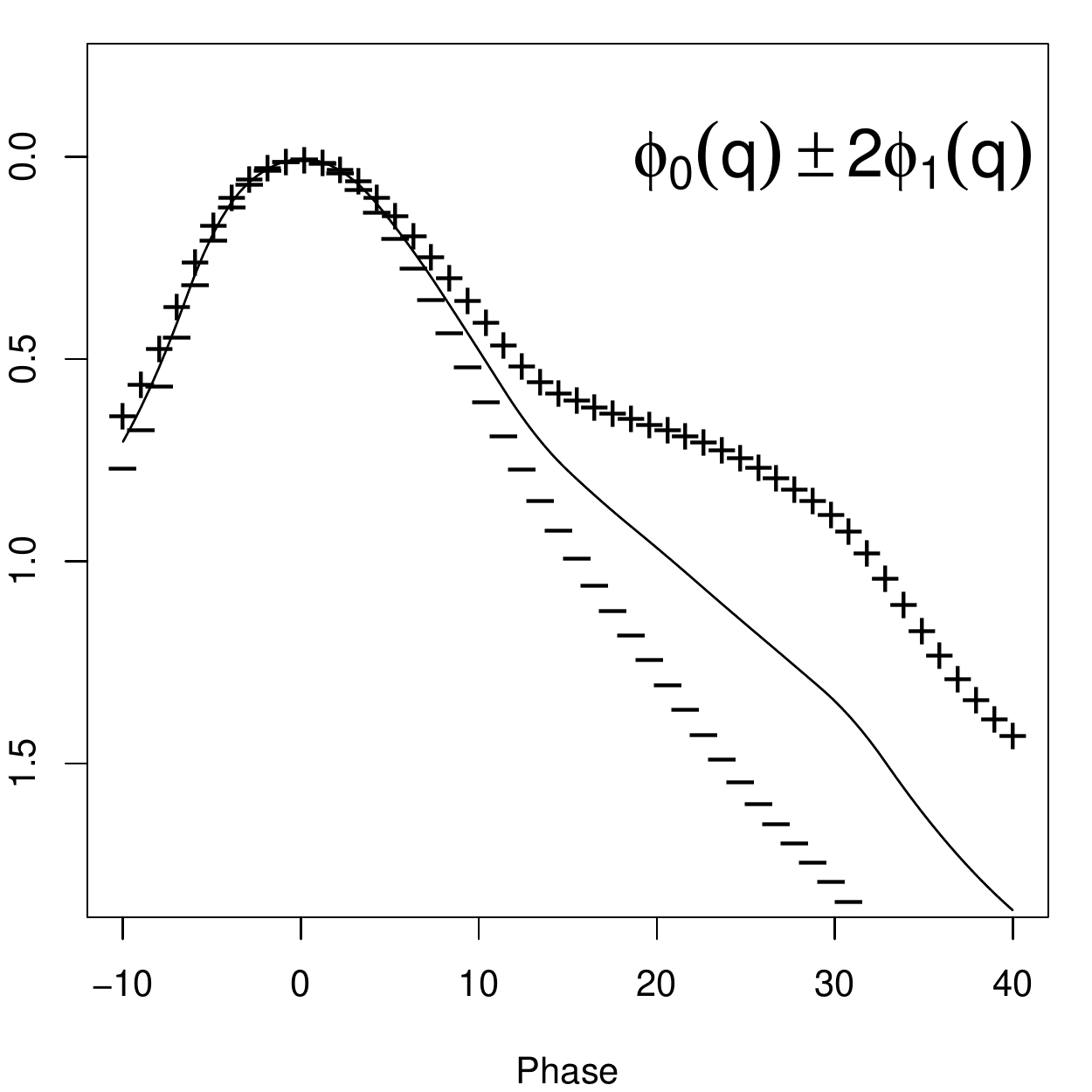}
\includegraphics[width=0.23\textwidth]{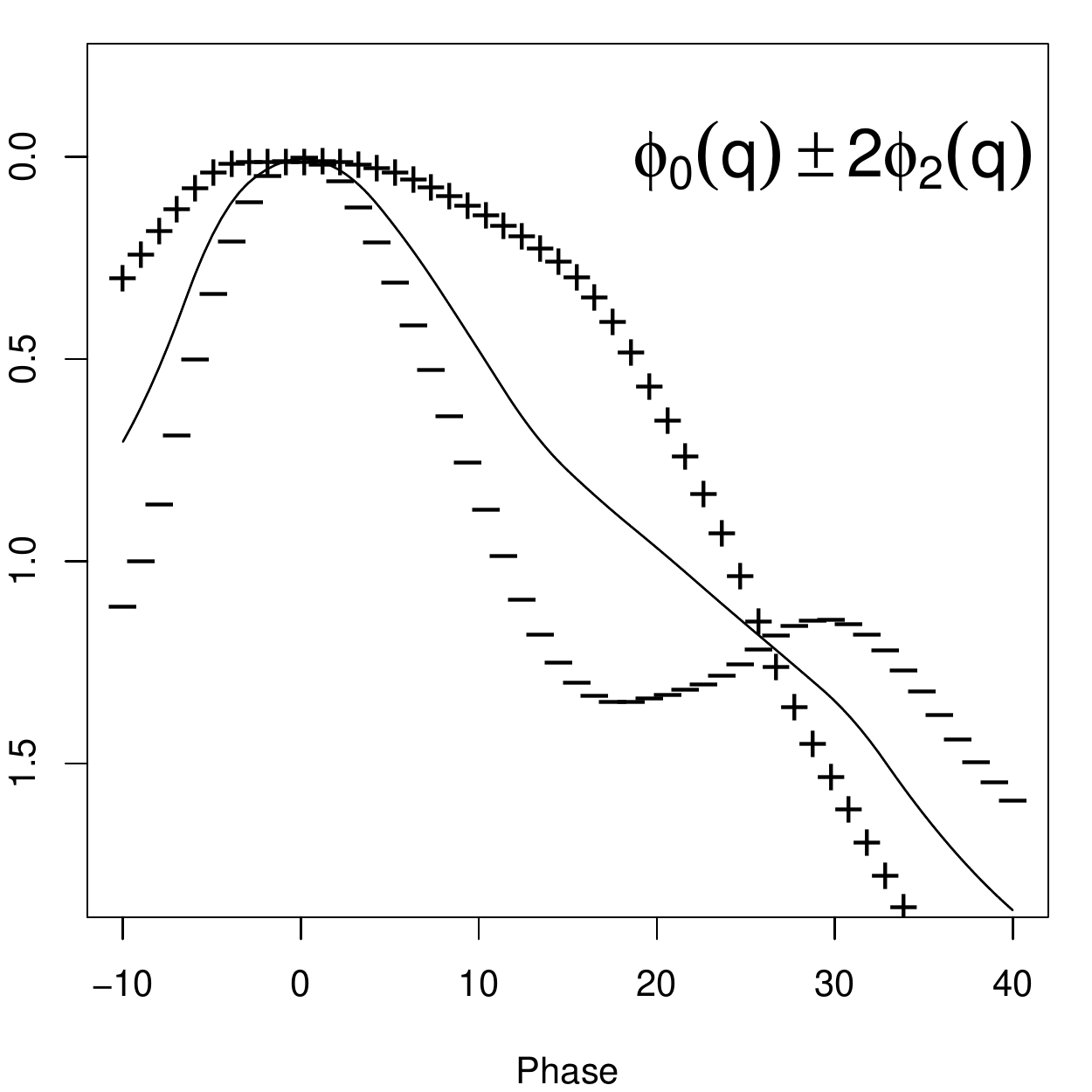}
\includegraphics[width=0.23\textwidth]{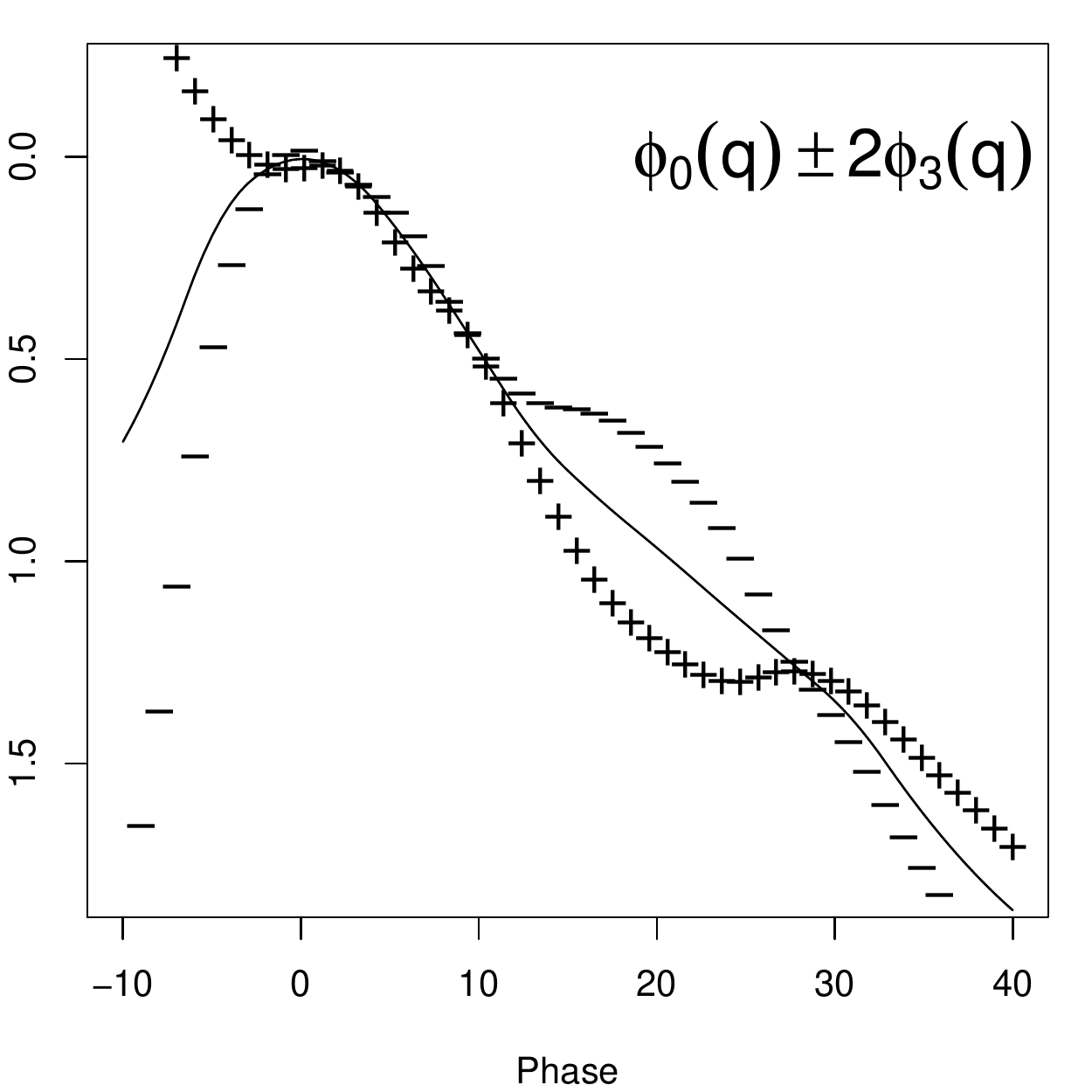}
\includegraphics[width=0.23\textwidth]{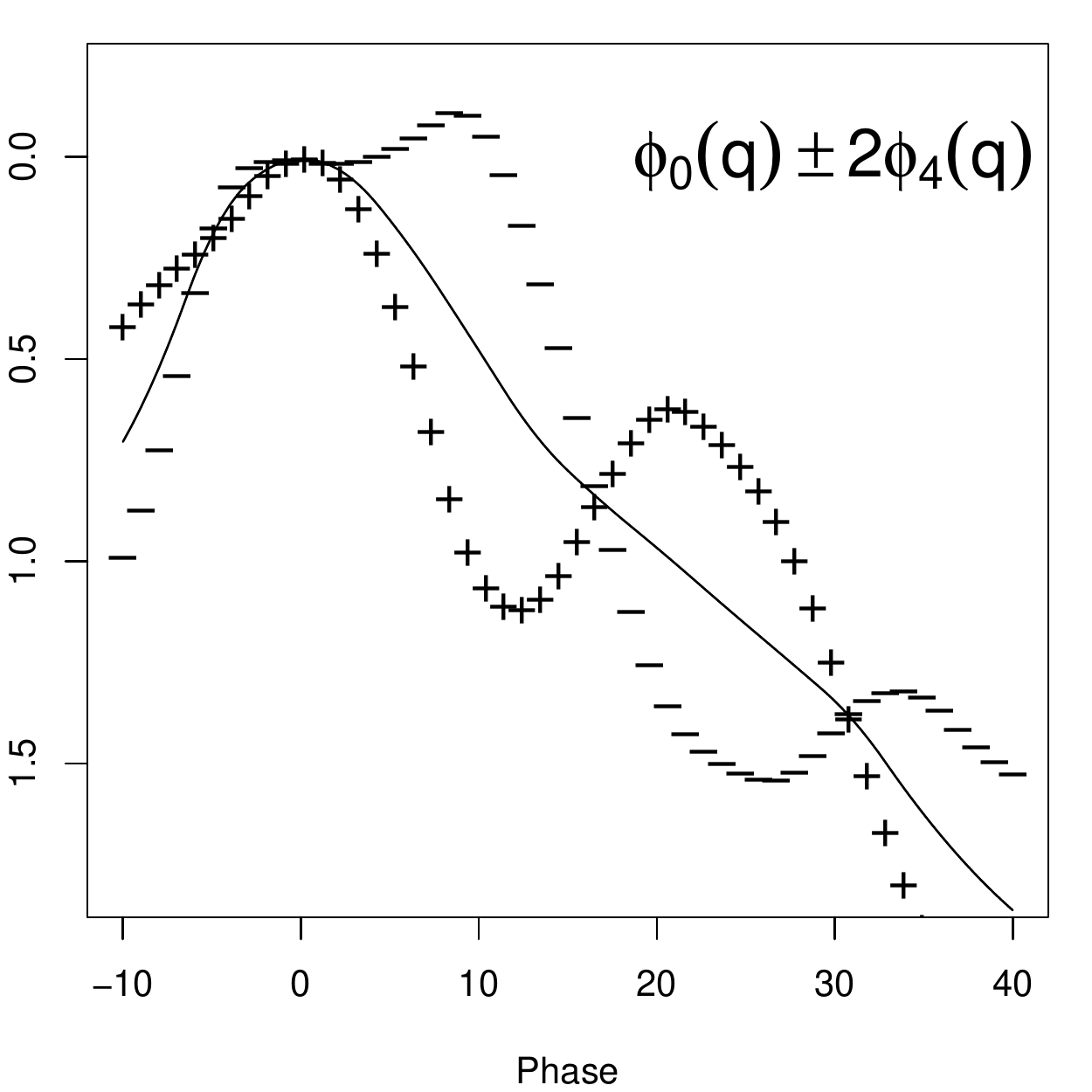}
\caption{Results from the fv-FPCA model. Plots of $\phi_0(q)\pm 2\phi_j(q)$  for the first four
principal components functions. This shows how these functions
 change the shape of light curve. The solid
  line in each panel is the mean function. The ``$+$'' points
  represent $\phi_0(q) 
  + 2\phi_j(q)$, and the ``$-$'' points represent $\phi_0(q) -
  2\phi_j(q)$. The vertical scale is given in magnitudes. 
  }  \label{fig:tpcfuns}
\end{figure}

\begin{figure*}[hbtp]
\centering
\includegraphics[width=0.32\textwidth]{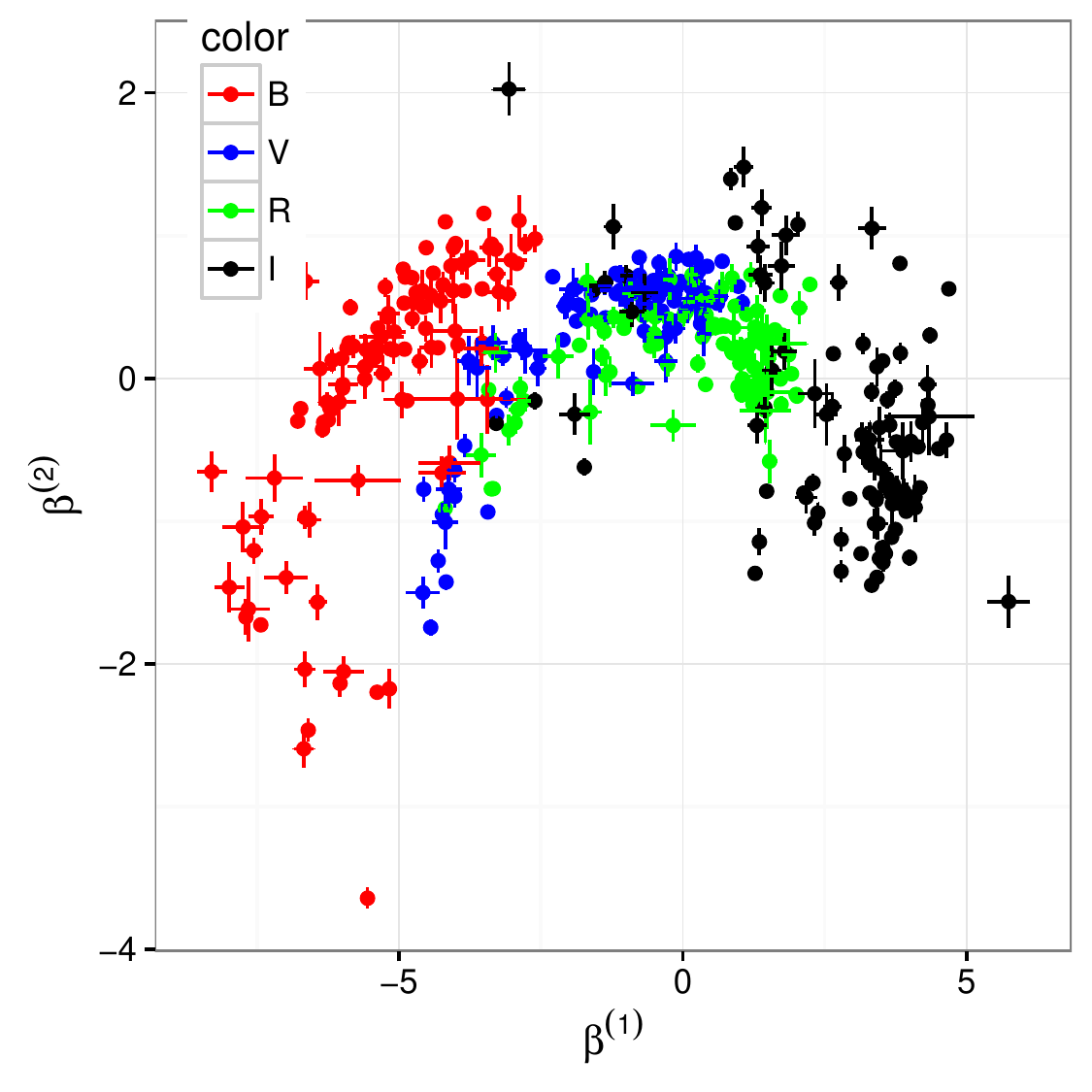}
\includegraphics[width=0.32\textwidth]{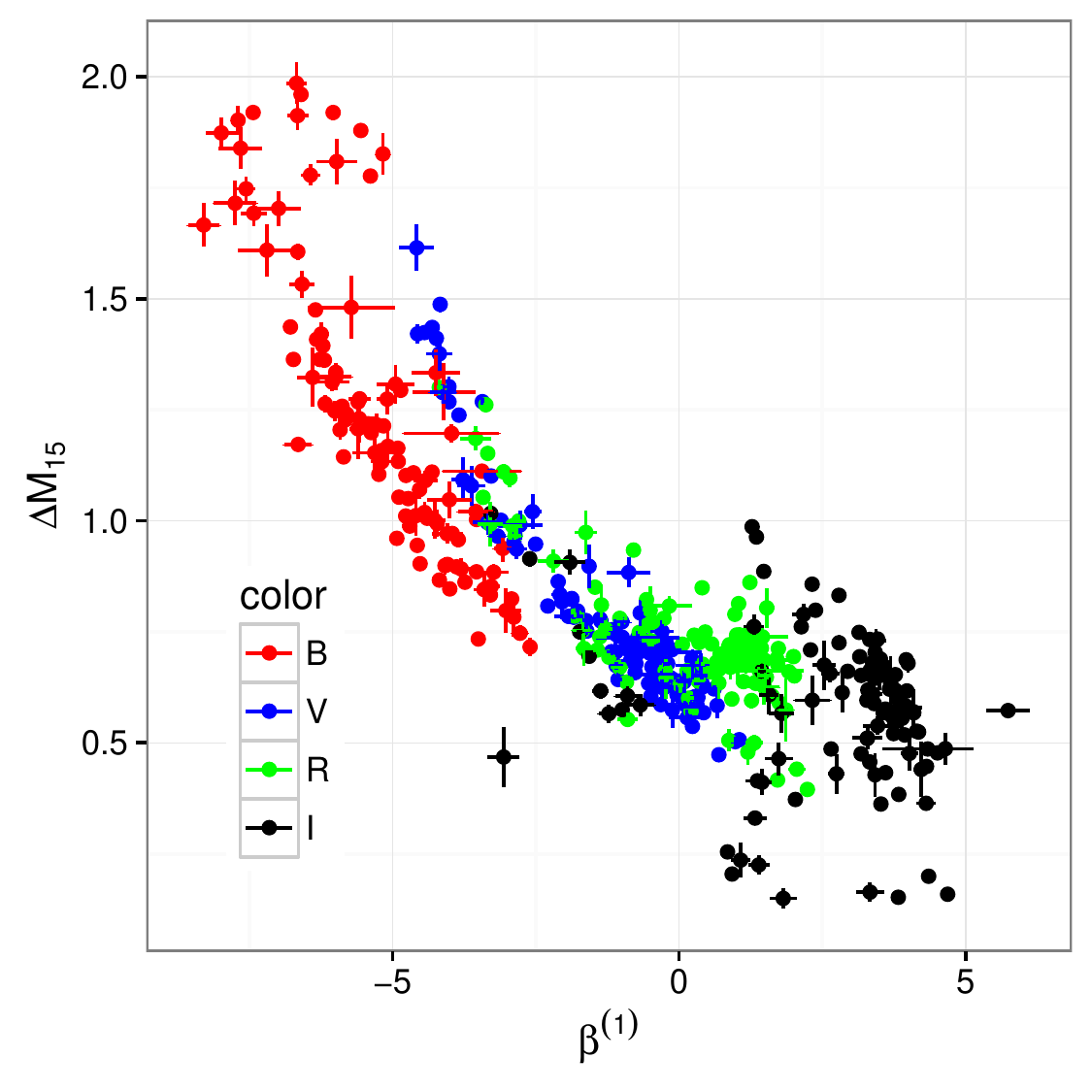}
\includegraphics[width=0.32\textwidth]{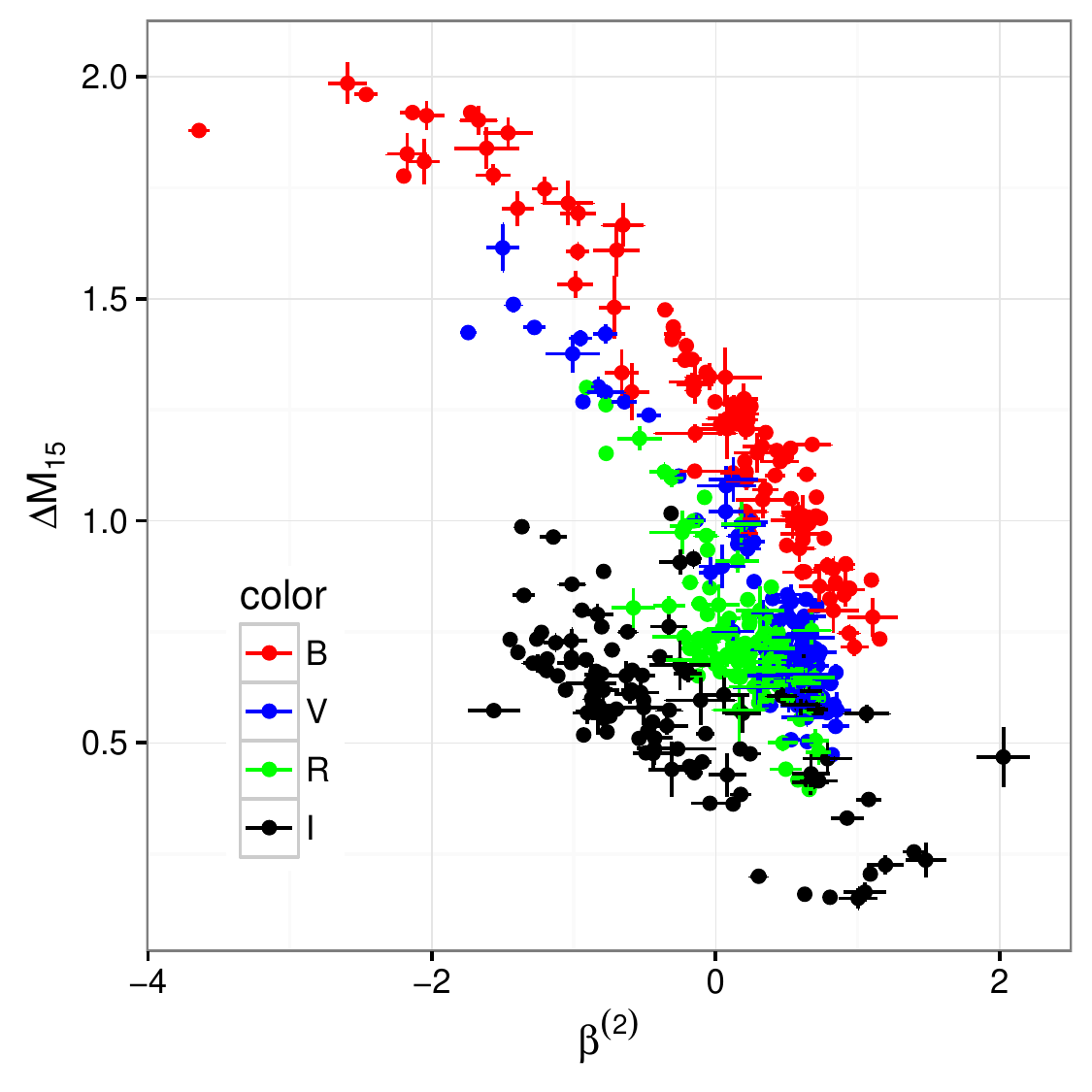}
\caption{Results from the fv-FPCA model. Relation of the first two scores $\beta^{(1)}, \beta^{(2)}$, and $\Delta
  M_{15}$ for each color band. The left panel is the plot of $\beta^{(1)}$ and $\beta^{(2)}$; the
  middle panel is the plot of $\beta^{(1)}$ and $\Delta M_{15}$; and the
  right panel is the plot of $\beta^{(2)}$ and $\Delta M_{15}$. The $B$, $V$,
  $R$, $I$ band points are red, blue, green, and black,
  respectively. }
\label{fig:tpcscores}
\end{figure*}

\begin{figure}[hbtp]
\centering
\includegraphics[width=0.45\textwidth]{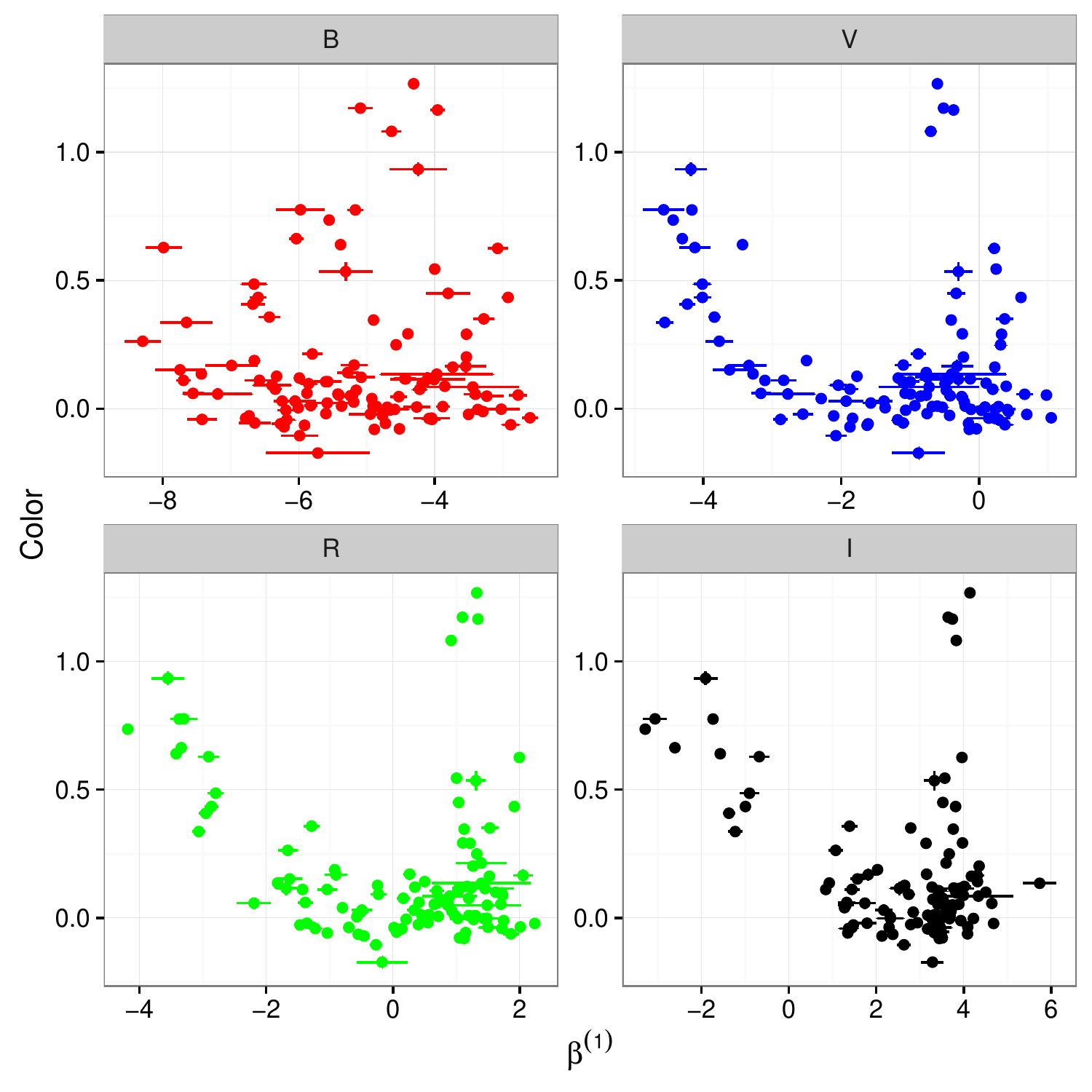}
\caption{Results from the fv-FPCA model. Plot of the first score $\beta^{(1)}$ against color at
  $B$-band maximum, $(B-V)_{\tmax}$. 
\label{fig:tpcscore1color}} 
\end{figure}

\begin{figure}[hbtp]
\centering
\includegraphics[width=0.45\textwidth]{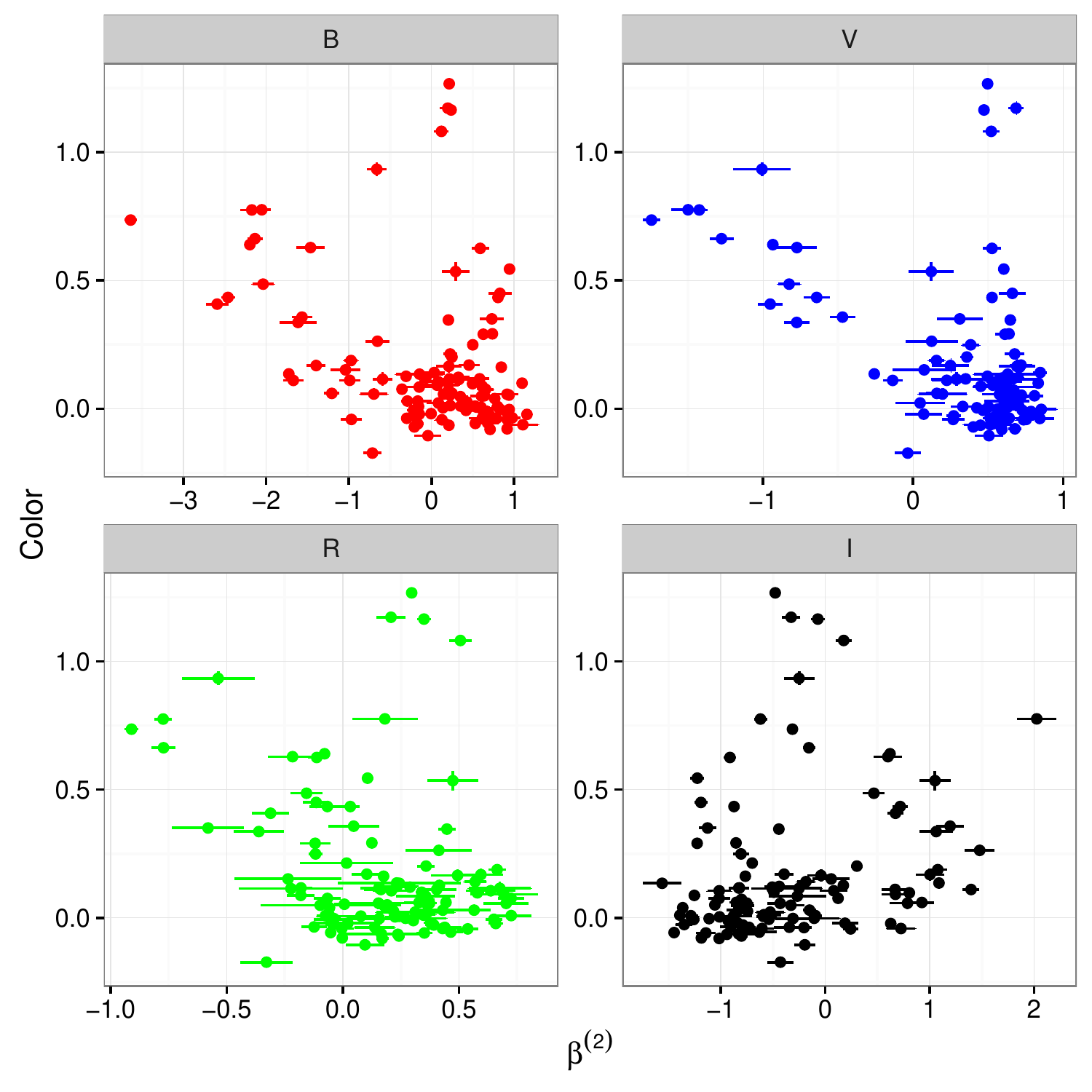}
\caption{Results from the fv-FPCA model. Plot of the second score $\beta^{(2)}$ against color
at $B$-band maximum,   $(B-V)_{\tmax}$. \label{fig:tpcscore2color}}  
\end{figure}

In the previous sections, fs-FPCA models (Equation~\ref{equ2:basics}) are trained
separately using observations from different well-defined optical filters. This
approach has a limitation that one has to correct observations to a
rest-frame filter before light curve fitting. However, our FPCA method
can be employed in another way to overcome this
limitation. Specifically, we can pool data from several bands together
to train a filter-vague FPCA (fv-FPCA) model (Equation~\ref{eqn:jointbandmodel}), 
where a single set of mean function and
principal component functions are used to describe the light curves
of all bands. When fitting a new light curve, the 
fv-FPCA model can be directly applied to data in their observed filter,
and no correction to rest-frame is necessary initially.  
We illustrate this approach to the same dataset used in previous
sections, where the K-corrections have already been applied to the data.  
Not surprisingly, most of the results remain the same as presented in previous
sections. This demonstrates that both approaches are effective in
capturing light curve information. 
We only present below results that are different from previous sections.

The four panels of Figure~\ref{fig:tpcfuns} show the estimated
mean function and the effects of a single principal component function. The
solid line represents the mean function $\phi_0(q)$. The ``$+$'' points
represent $\phi_0(q) + 2\phi_j(q)$, and the ``$-$'' points represent 
$\phi_0(s) -  2\phi_j(s)$ for $j = 1,2,3,4$. 
The first principal component function $\phi_1(q)$ (shown in the
supper left panel) reflects
decline rate about 15 days after  the peak. 
The second  principal component function
$\phi_2(q)$ (shown in the upper right panel) is sensitive to the
light curve width around the peak. Meanwhile, it also reflects a contrast of decline rate
before and after $\sim$ 20 days in phase. 
The third principal component function (shown in the lower left panel) 
adjusts the bump around 20 days after the
peak. The fourth principal component function (shown in the lower right panel) exhibits
more complex fluctuations. For the proportion of variability, 
the first principal component function explains 91.69\% of the total variability, 
the first two principal component function explains 96.31\%, and 
the first four together explain 99.24\%.

As the scores are measured by the same set
of principal functions, their values are comparable and can be plotted
in one panel.  The first two scores are correlated in a
nonlinear pattern, as shown in the left panel of
Figure~\ref{fig:tpcscores}. 
What may be especially interesting is the seemingly monotonic shift of the
locus of data with filter bands. The left panel of
Figure~\ref{fig:tpcscores} shows $\beta^{(1)}$ increases as the
wavelength increases, whereas the range of $\beta^{(2)}$ remains nearly the
same. The locus of data points in each filter
shows consistent correlation patterns and can be globally shifted to
approximately match each other. These characteristics suggest that in principle we
are able to derive photometric redshifts based on the scores. 
These score correlations can also
facilitate robust photometric identification of SNIa. 

The relation between $\beta^{(1)}$, $\beta^{(2)}$ and $\Delta M_{15}$
are presented in the middle and right panel of
Figure~\ref{fig:tpcscores}. Besides,
Figure~\ref{fig:tpcscore1color} and Figure~\ref{fig:tpcscore2color}  
show that the first and second scores are
nonlinearly correlated with the observed color at $B$ band maximal.
From here, the color excess can be deduced in a similar way as in Section~\ref{sec:scorecolor}.
In addition, all other analysis conducted previously can be carried out, including the spectral information
correlation, spectral classification and distance prediction. The
messages remain the same as in the previous sections.

\begin{figure}
\centering
\includegraphics[width=0.5\textwidth]{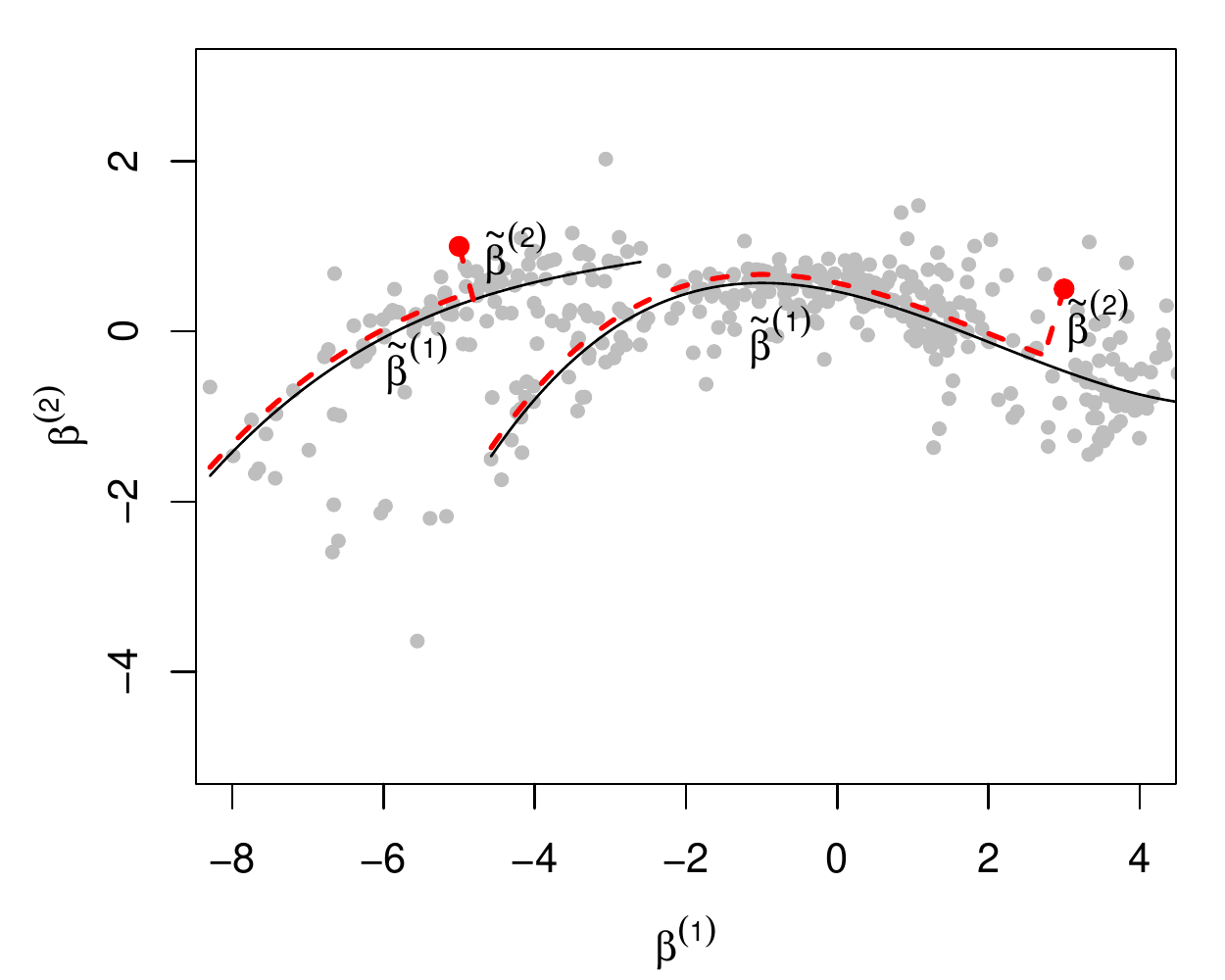}
\caption{Results from the fv-FPCA model. The nonlinear scores $\tbeta^{(1)}, \tbeta^{(2)}$
are calculated via fitting curves to the original scores
 $\beta^{(1)}, \beta^{(2)}$.}
\label{fig:newscore}
\end{figure}

\begin{figure*}[htbp]
\centering
\includegraphics[width = 0.31\textwidth]{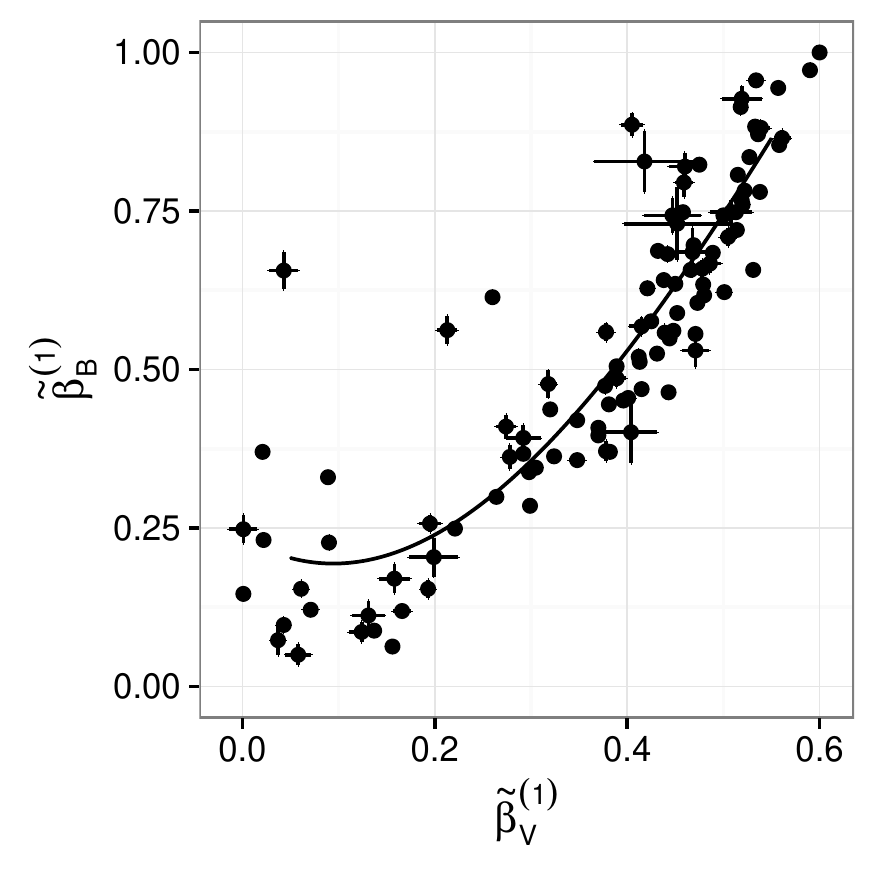}
\includegraphics[width = 0.31\textwidth]{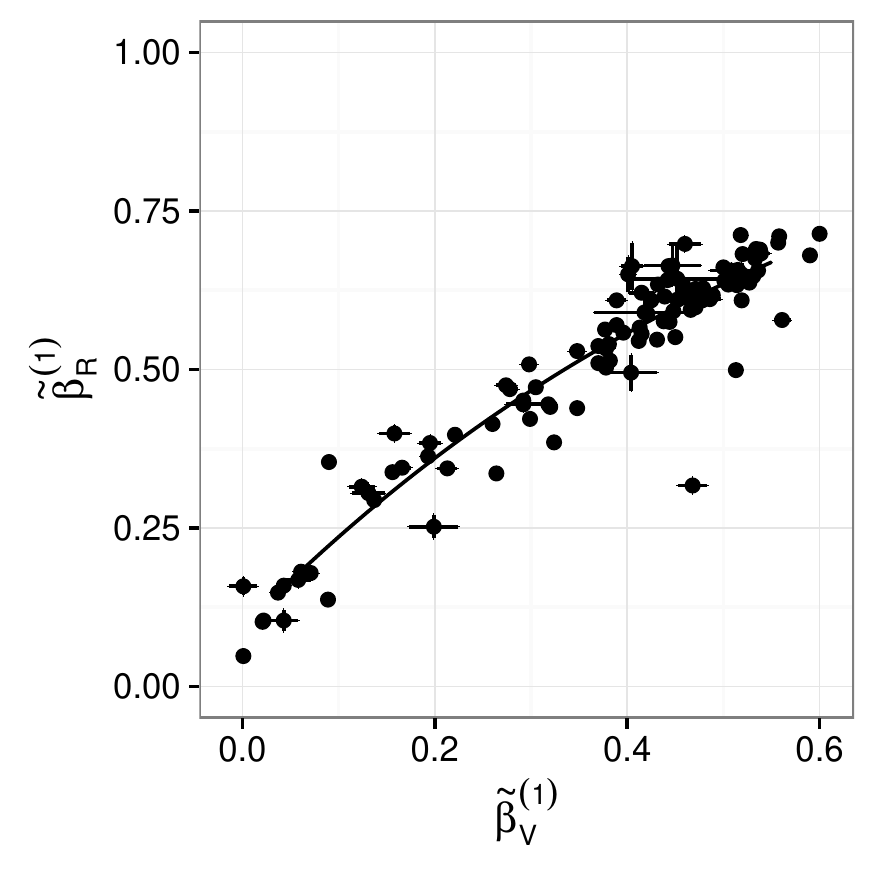}
\includegraphics[width = 0.31\textwidth]{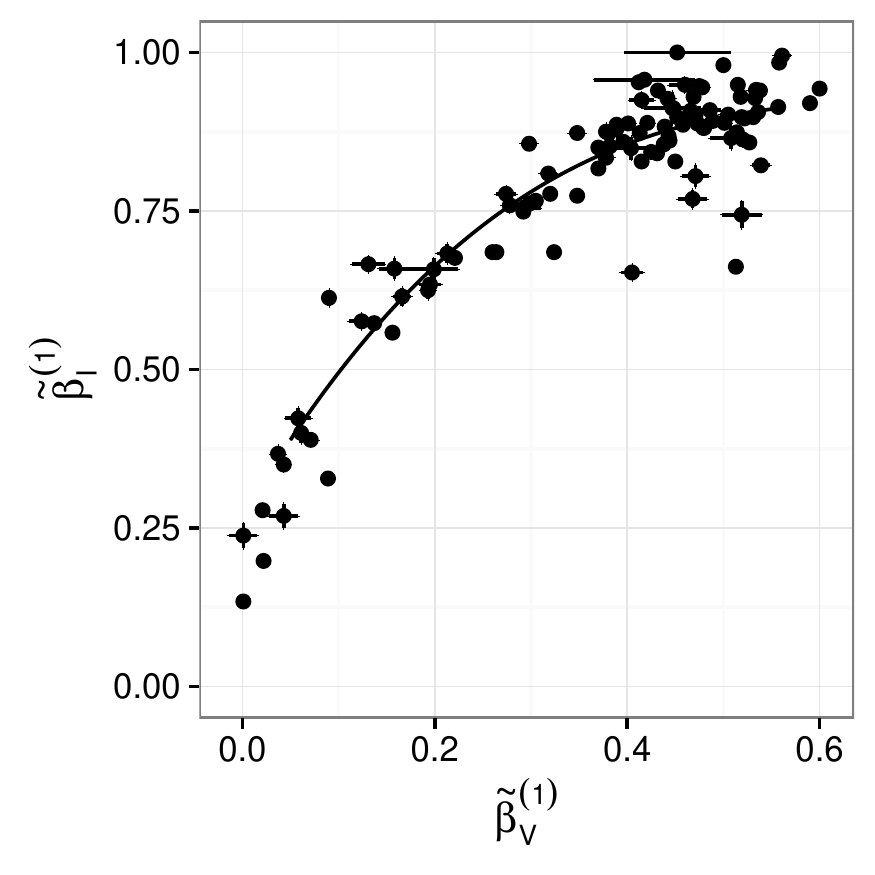}
\caption{ Results from the fv-FPCA model. The relation between the first nonlinear $V$ band score
  $\tilde{\beta}^{(1)}_V$ (on horizontal axes) with
$\tilde{\beta}^{(1)}_B$, $\tilde{\beta}^{(1)}_R$, 
$\tilde{\beta}^{(1)}_I$ (on vertical axes). 
} \label{fig:newbeta1}
\end{figure*}

\begin{figure}[htbp]
\centering
\includegraphics[width = 0.45\textwidth]{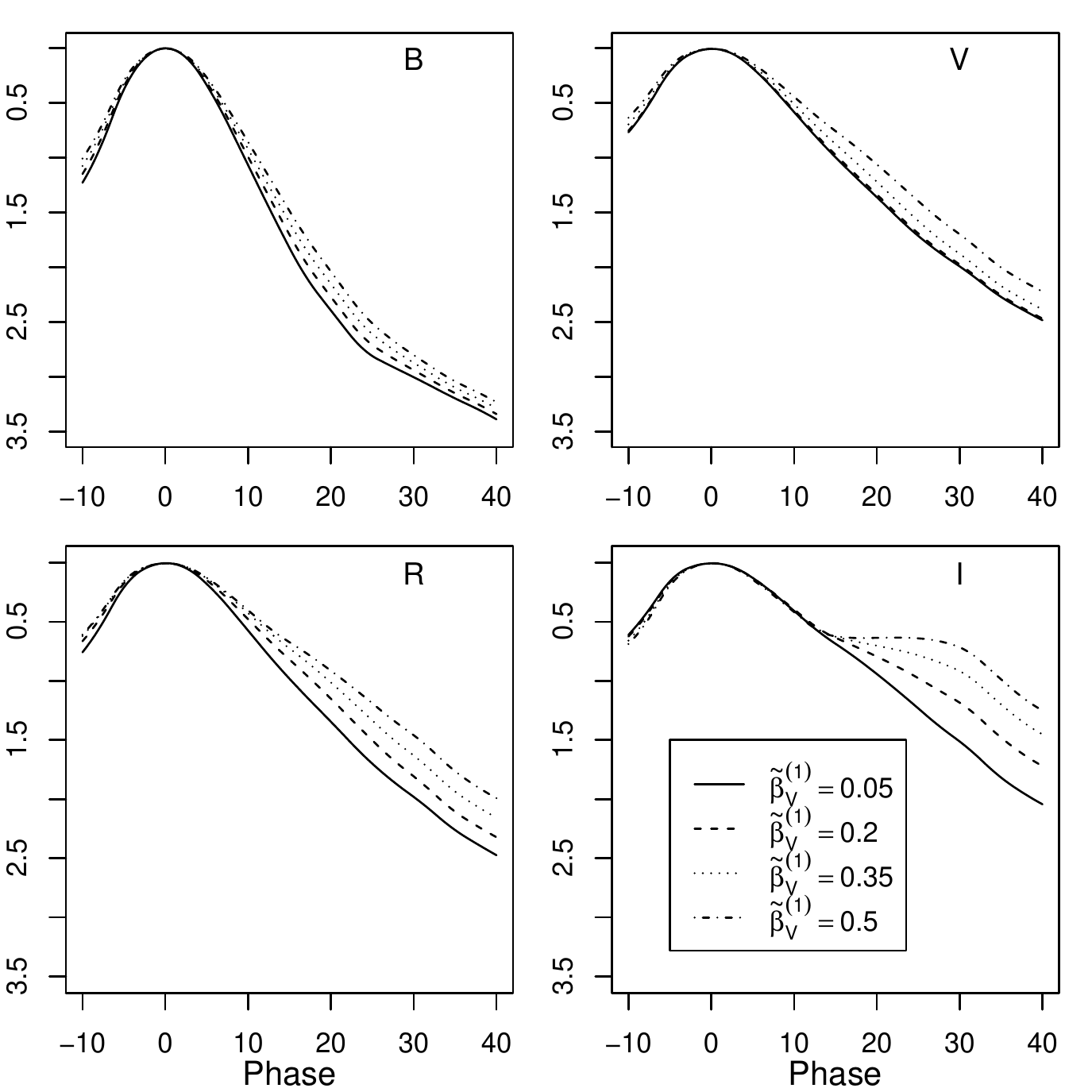}
\caption{Results from the fv-FPCA model. Average light curve shapes parametrized by the nonlinear
  score $\tilde{\beta}_V^{(1)}$. 
} \label{fig:newbetashape}
\end{figure}

A final point to notice is the potential of more effective dimension reduction
with nonlinear dimension reduction techniques. 
This is due to the nonlinear relation between the first two dominant scores
 $\beta^{(1)}, \beta^{(2)}$ in Figure~\ref{fig:tpcscores}.
For simplicity, consider the two dimensional space of
$\beta^{(1)}$ and $\beta^{(2)}$. These first two dimensions alone
already account for 96.31\% cumulative variance of the dataset. 
We adapt the concept of principal ridge \citep{ozertem2011locally}. 
A curve is fitted for $\beta^{(1)}_B$ and $\beta^{(2)}_B$ of $B$ band, 
as the left black solid line in Figure~\ref{fig:newscore}. 
This nonlinear curve is treated as the first nonlinear dimension
for $B$ band scores. 
The second nonlinear dimension is the one locally perpendicular to
the curve. Given the original scores $\beta^{(1)}, \beta^{(2)}$ from
our model, the new \textit{nonlinear scores} $\tbeta^{(1)}, \tbeta^{(2)}$ are calculated
as follows.  Consider the left red point in Figure~\ref{fig:newscore}, it
is projected onto the $B$ band curve. The projection is identified by the nearest
point on the curve. The nonlinear score $\tbeta^{(1)}$ is the geodesic
distance from 
the leftmost point of the curve to the projection point.  The
geodesic distance along the curve is indicated by the red dashed
curve. The new score $\tbeta^{(2)}$ is the usual Euclidean distance of
the original point to the projection point, as indicated by the
red vertical dashed line. In a similar manner, a curve is fitted for
$V$, $R$, $I$ band together (the right solid curve in Figure~\ref{fig:newscore}), 
and the linear scores for $V$, $R$, $I$
band are projected onto this curve to obtain the new nonlinear score. 

In the original linear system, 
the first dimension explains 91.69\% of the
total variability. The first two dimension together
explains 96.31\% of the total variability.
Now in the nonlinear system, the explained proportion
of the first nonlinear score 
should be larger than 91.69\%, but smaller than 96.31\%. More light
curve information is absorbed into the first dimension. The first
score alone can provide adequate fit to SNIa light curves.

The more thorough dimension reduction can be made by describing the
correlation among $\tbeta^{(1)}_B, \tbeta^{(1)}_V, \tbeta^{(1)}_R, \tbeta^{(1)}_I$.
The shape of light curves across these optical bands should be well
correlated due to their common spectral evolution.  
The relation of $\tbeta^{(1)}_V$  with $\tbeta^{(1)}_B,
\tbeta^{(1)}_R, \tbeta^{(1)}_I$ is plotted in Figure~\ref{fig:newbeta1}.
For most of the cases, the $V$ band nonlinear score $\tbeta^{(1)}_V$  is a
reliable predictor of 
$\tbeta^{(1)}_B, \tbeta^{(1)}_R$ and  $\tbeta^{(1)}_I$. Therefore we
can use a single parameter, $\tbeta^{(1)}_V$, to describe all the
SNIa  light curve shapes. For each $\tbeta^{(1)}_V$, the predicted values
of $\tbeta^{(1)}_B, \tbeta^{(1)}_R$ and  $\tbeta^{(1)}_I$ are obtained
by the fitted curves in Figure~\ref{fig:newbeta1}. This single
parameterization of light curve shape is especially useful for fitting
high redshift supernovae with sparse and noisy observations, 
because only one parameter is required to be
constrained by the data.
Using this scheme with a single parameter $\tbeta^{(1)}_V$, 
Figure~\ref{fig:newbetashape} depicts all four band light curve shapes
for $\tbeta^{(1)}_V = 0.05, 0.2, 0.35, 0.5$. These light curves 
are the expected shapes accounting for first order correction of the
mean curve with $\tbeta^{(1)}_V$.

\section{Discussions and Conclusions}\label{sec:discussions}

We have presented in this paper an empirical model for 
SNIa light curves. Using this model, the entire light curve of a SNIa
can be represented by a few scores.  
These scores characterize light curve shape, intrinsic color, and color
excess for SNIa.
Some light curve scores are even correlated with
spectral features measured independently of SNIa light curves.
In previous studies, the absorption features
of  SNIa spectra have been empirically
compared with the color and the light curve width parameter. 
For example, \cite{silverman2012berkeley3} 
showed the strength of \sirp\ is anti-correlated with
SALT II width parameter and uncorrelated with color. 
This anti-correlation only implies stronger \sirp\ correlated with
narrower light curve shape. However it is interesting to
explore further how the light curve shape in multi-bands changes with
the strength of this line.
The score parameters from our model reveal more such information. 
Regarding this, we have presented a  more detailed morphology analysis
of light curve with respect to the feature strength.

\deleted{Moreover, by examining the correlation among scores,
  especially the first two dominant scores, we find the SNIa light
  curve resides in a tight nonlinear subspace. A more dramatic
  dimension reduction is possible by nonlinear dimension reduction
  techniques. This tight nonlinear subspace specifies plausible
  parameter domains for SNIa light curves. A new light curve can be
  classified as SNIa if its scores are inside this subspace. On the
  other hand,  a photometrically different SN can also be identified
  if its score vector is far
away from this subspace. Our algorithm may produce a quantitative
photometric classification scheme for supernovae.These classification
and outlier detection task have been carried out previously by hsiao
2014 and Ishida 2013.In this paper, an
initial analysis of nonlinear dimension reduction shows some
promises. More thorough  work is left for future study.}

Beyond these empirical investigations, 
the proposed model embraces more potential in cosmology
model fitting. 
Although the primary light curve shape parameter such as $\Delta
M_{15}$, the stretch parameter, or the parameter  
in SALT II is effective, 
it is still worthwhile to explore other constructions using the shape of
the entire light curve.  
Estimation of $\Delta M_{15}$ is sensitive to local observations
around the peak and around the +15 days in phase. If we lack enough
observations to constraint light curve shapes around these days in
phase, the estimated $\Delta M_{15}$ would have large uncertainty. 
Besides, the $\Delta M_{15}$  parameter only captures the declining part of the
light curves, and fails to capture 
the light curve shape at the rising side. The stretch parameter
may not be applicable for SN wavelength bands longer than $I$ band,
and may not always fit well for both the rising and falling part of a
SNIa light curve.  
A product of our FPCA model is to replace the $\Delta M_{15}$ 
term in the existing distance prediction models by a
functional linear term, which provides a more flexible
and data-driven way to adjust the light curve
shape for distance prediction.  
\replaced{By comparing with the previous distance models using
$\Delta M_{15}$ adjustment, we have demonstrated that using the
functional linear form of the entire light curve consistently 
gives smaller residual scattering and robust distance prediction.}{
Comparison with the previous distance models using $\Delta M_{15}$ adjustment
and SALT II shape parameter suggests that the functional linear form
of the entire light curve has the potential to give smaller residual
scattering and robust distance prediction.  
}

Among the effort to reduce distance prediction scatter, 
one common conjecture is that SNIa is not a homogeneous group. There
exist subclasses of their own characteristics. Each subclass has
its own dust correction and K-correction. Picking out a more 
homogeneous subclass help to further reduce the dispersion.
Some works endeavor to identify subclasses based on spectral data 
\citep{benetti2005diversity, branch2009comparative, wang2009improved}.
The dispersion reduction is more significant
by pairing supernovae with identical spectral features and applying
pairwise dust correction \citep{fakhouri2015improving}.
Our study finds that, when only light curve data is available,  
the scores extracted from the light curve can still help to
determine spectral classes, although with limited precision. 
Therefore, it is possible to reduce the dispersion by
fitting the distance and dust correction model
within a subclass of SNIa observations.
Another potential application of this result is to improve
the precision of K-correction, as the spectral template from the
corresponding spectral class can be applied 
for this subclass of observations.

\added{The filter-vague FPCA model can be applied to light curve data with
unknown redshift. If needed, K-correction can be applied to the fitted
light curve parameters (i.e., scores), instead of to each point of the
light curve data. This approach can be useful for large surveys where
redshift can only be approximately estimated through photometric
redshift determined by colors of the host galaxy.  The filter-vague FPCA
is useful for deriving photometric redshift through supernova light
curve data. This approach is more useful when the redshift is entirely
unknown, e.g., at the initial stage of a survey, for which correction
to rest-frame is impossible. It also provides a solution for
photometric classifications of transients from wide field supernova
surveys using light curve shapes, which can be particularly
interesting for LSST and WFIRST. } {\bf Precise redshift will be needed if the filter-vague FPCA is used for cosmological distance determination}. 
 In future experiments with LSST or WFIRST, the sample size will be significantly larger than today, more parameters can  be deduced for controls of systematic effects of SNIa,  the PCA approach allows for accurate quantification of information loss in the light curve fitting procedures.

\acknowledgements {\bf Part of this work has received funding from the Key Research Program of Frontier Sciences (QYZDY-SSW-SLH010).} The work of He and Huang is partially supported by
Texas A\&M-NSFC Collaborative Research Grant Program. 

\section*{Supplementary Materials}

The supplementary materials are available to download at
\url{https://github.com/shiyuanhe/supern}.
The supplementary materials contain the following: 
\begin{enumerate}
\item \added{The SNIa FPCA templates in text format. The template files include both the 
filter-vague model and filter-specific model. }
  \item The SNIa data table. The data table contains all SNIa samples
    in this paper with their scores, spectral line strength and
    spectral class. \added{The table is illustrated in Table~\ref{tbl:suptable}.}
  \item The FPCA software. The software provides a web-based user
    interface for light curve fitting, 
    intrinsic color estimation, spectral line strength estimation and
    spectral classes determination. It also provide the probability
    that the submitted sample belongs to SNIa. 
\end{enumerate}


\appendix

\section{Algorithm} \label{appendix:algorithm}
This section presents the details of the model training algorithm used
for model training in Section~\ref{sec:training}. 
The presentation focuses on the fs-FPCA. 
If data from several bands are pooled together to train
the fv-FPCA model~(\ref{eqn:jointbandmodel}),
the algorithm needs to be modified slightly in an obvious fashion. 
All light curves are registered with the
transformation,   $q_{s\lambda j}=( t_{s\lambda j}-b_{s\lambda
})/(1+z_{s\lambda })$. This 
transformation aligns all the light curves such that their peaks are
at phase zero. However, the peak epoch $b_{s\lambda }$ is unknown.  An
initial estimate of the 
peak magnitude $m_{s\lambda }$ and peak epoch $b_{s\lambda }$
are obtained by a local quadratic regression. With this initial
 estimate, a two-step procedure is carried out for model training, i.e., learning the 
 mean function $\phi_{0\lambda}(q)$ and learning the principal
 component functions   $\phi_{k\lambda}(q)$'s ($k\ge 1$).

\subsection{Learning  the Mean Function}
If we define
$\mathbf{B}_{s\lambda }=\big(\mathbf{b}(q_{i1}),\cdots,\mathbf{b}(q_{in_i})\big)^T$,
the model (\ref{equ2:splineform}) in matrix form is 
\begin{equation} \label{eqn:matrixNotation}
  \mathbf{y}_{s\lambda }=m_{s\lambda}\mathbf{1}_{n_{s\lambda }}+
\mathbf{B}_{s\lambda}\boldsymbol{\theta}_{0}+\mathbf{B}_{s\lambda }
  \boldsymbol{\Theta}_\phi\boldsymbol{\beta}_{s\lambda}
+\mathbf{W}_{s\lambda }\boldsymbol{\epsilon}_{s\lambda }\, ,
\end{equation}
where $\mathbf{1}_{n_{s\lambda }}$ is a vector of ones with length
$n_{s\lambda }$,  
$\boldsymbol{\epsilon}_i$ is a random vector of length $n_{s\lambda }$
 following a standard normal distribution,
and $\mathbf{W}_{s\lambda }=\text{diag}\{\sigma_{s\lambda
  1},\cdots,\sigma_{s\lambda   n_{s\lambda}}\}$.  
This representation includes all the light curve observations of the
$s$-th supernova  with filter $\lambda$.

The last two terms in~(\ref{eqn:matrixNotation}) has an expectation of
zero. Although their covariance matrix is unknown, the general least
square estimate is still consistent with least square estimation. 
With the estimated peak magnitude $\widehat{m}_{s\lambda}$ and
peak epoch $\widehat{b}_{s\lambda}$, 
we estimate $\vtheta_{0 \lambda}$ by solving
\begin{equation}\label{eqn:learnMean}
\min_{\vtheta_{0\lambda}}  \sum_{s} \frac{1}{n_{s\lambda}}
  \Vert(\mathbf{W}_{s\lambda})^{-1}(\mathbf{y}_{s\lambda}-
\widehat{m}_{s\lambda}\mathbf{1}_{n_{s\lambda}}-\mathbf{B}_{s\lambda}
\boldsymbol{\theta}_{0\lambda})\Vert^2_2
+\eta\, \text{tr}\big(\boldsymbol{\theta}_{0\lambda}^T\boldsymbol{\Omega}
\boldsymbol{\theta}_{0\lambda}\big)
\end{equation}
\added{for each filter $\lambda$,} where $ \boldsymbol{\Omega}=\int \mathbf{b}''(t)^T \mathbf{b}''(t) \textrm{d}t $.
The last term is the roughness penalty to encourage a smooth
solution, and $\eta$ is the tuning parameter.
Essentially, the smoothness is achieved by controlling the integral of 
the squared second order derivative of the solution, i.e.,
$\int (\phi_{0 \lambda}''(q))^2 \mathrm{d} q =
\text{tr}\big(\boldsymbol{\theta}_{0 \lambda}^T\boldsymbol{\Omega}
\boldsymbol{\theta}_{0 \lambda}\big)$.
Suppose $\widehat{\boldsymbol{\theta}}_{0 \lambda} $ is the solution of the
optimization problem~(\ref{eqn:learnMean}), then
the resulting mean function is $\hat{\phi}_{0 \lambda}(q) = \mathbf{b}(q)^T
\widehat{\boldsymbol{\theta}}_{0 \lambda}$.

\subsection{Learning the Principal Component Functions}

Now align the peak and subtract the mean function from the observed
light curves,
$\widetilde{\mathbf{y}}_{s\lambda} =
\mathbf{y}_{s\lambda}-\widehat{m}_{s\lambda}\mathbf{1}_{n_{s\lambda}}-
\mathbf{B}_{s\lambda}\widehat{\boldsymbol{\theta}}_{0 \lambda}\, . $
This is the remaining difference to be fitted by the principal
component functions $\phi_{k \lambda}(q)$'s. 

Let $\mathbf{s}_{s\lambda}=\boldsymbol{\Theta}_\phi
\boldsymbol{\beta}_{s\lambda}$ 
and put them  in \replaced{a $P\times 4$ matrix 
$\mathbf{S}_s = (\mathbf{s}_{sB}, \mathbf{s}_{sV}, \mathbf{s}_{sR},
\mathbf{s}_{sI})$ 
for the $s$-th supernova. Combine all of them  in a $P\times (4S)$
matrix 
$\mathbf{S}=\big(\mathbf{S}_1,\mathbf{S}_2, \cdots, 
\mathbf{S}_S \big)$.}{a $P\times S$ matrix $\mathbf{S}_{\lambda} =
\left(\mathbf{s}_{1 \lambda}, \mathbf{s}_{2\lambda}, \cdots,
  \mathbf{s}_{S \lambda}\right)$. }  We estimate $\mathbf{S}_{
\lambda}$ by solving 
the following  optimization problem \added{for each filter $\lambda$}
\begin{equation}\label{main2}
\min_{\mathbf{S}_{\lambda}}  \sum_{s}
  \frac{1}{n_{s\lambda}}
\Vert(\mathbf{W}_{s\lambda})^{-1}(\widetilde{\mathbf{y}}_{s\lambda} 
-\mathbf{B}_{s\lambda}\mathbf{s}_{s\lambda})\Vert^2_2+
\eta_1 \Vert \mathbf{S}_{\lambda} \Vert_{S_1}
+\eta_2\text{tr}\big(\mathbf{S}_{\lambda}^T
\boldsymbol{\Omega}\mathbf{S}_{\lambda}\big)\, .
\end{equation}
In the above, $\Vert \cdot\Vert_{S_1}$ is the nuclear norm penalty,
which is the summation of all singular values of a matrix. This penalty
encourages a low-rank solution of $\mathbf{S}_{\lambda}$, 
and thereby encourages a small number $K$ of  principal component functions.  
A roughness penalty
$\text{tr}\big(\mathbf{S}_{\lambda}^T\boldsymbol{\Omega}
\mathbf{S}_{\lambda}\big)$
is also imposed to ensure each column of the solution $\mathbf{S}_{\lambda}$ is smooth. 
The tuning parameters, $\eta_1$ and $\eta_2$, are selected by
cross-validation. The rank of $\mathbf{S}_{\lambda}$ is automatically
determined by the algorithm.
The optimization problem (\ref{main2}) is solved by the ADMM
algorithm \citep{boyd2011distributed}  combined with the singular value
soft-thresholding operator \citep{cai2010singular}. The ADMM algorithm
breaks the optimization into two easy-to-solve parts: one part involves the
quadratic loss and the other involves the nuclear norm. The algorithm
then iteratively updates the two parts and the Lagrangian multiplier
until convergence. 

Now, let $\widehat{\mathbf{S}}_{\lambda}$ be the solution of the
optimization problem~(\ref{main2}) for filter $\lambda$, and  
$\widehat{\mathbf{S}}_{\lambda}=
\widehat{\mathbf{U}}_{\lambda}
\widehat{\mathbf{D}}_{\lambda}\widehat{\mathbf{V}}^T_{\lambda}$
be its SVD decomposition. 
Suppose  $\widehat{\mathbf{u}}_{i\lambda}$ is the $i$-th column of
$\widehat{\mathbf{U}}_{\lambda}$.
Let  $\widehat{\boldsymbol{\Theta}}_{\phi \lambda} = 
\left(\widehat{\mathbf{u}}_{1 \lambda},  \widehat{\mathbf{u}}_{2\lambda},
\cdots,\widehat{\mathbf{u}}_{K \lambda}\right)$ be
estimated by the first $K$ columns of
$\widehat{\mathbf{U}}_{\lambda}$, then the 
estimated principal component functions are 
$\widehat{\boldsymbol{\phi}}(s)^T_{\lambda}=\mathbf{b}(s)^T
\widehat{\boldsymbol{\Theta}}_{\phi \lambda}$.

\section{Fitting Functional Linear Distance Model}
\label{sec:fittingFLDM}

\added{
In this section, we discuss the mathematical details for fitting 
the functional linear distance model~(\ref{eqn:distfunctional}).
First of all, the functional linear term~(\ref{eqn:functionalQ}) is
expanded by the principal  component functions
\begin{equation*}
\begin{split}
Q  & = \int \delta(q) (g_B(q) - m_B - \phi_{0B}(q))
  \mathrm{d}q 
 = \int \left[\sum_{k=1}^K \delta^{(k)} \phi_{kB}(q) \right]
\left[\sum_{k=1}^K \beta^{(k)}_B \phi_{kB}(q)  \right]
\mathrm{d}q =\sum_{k=1}^K \delta^{(k)}\beta^{(k)}_B\, . 
\end{split}
\end{equation*}
The last equation use the orthonormality of the principal component
functions. From here, the functional linear term eqauls the inner
product between the score vector $\vbeta$ and the vector 
$\vdelta = (\delta^{(1)} \cdots, \delta^{(K)})^T$. The roughness
penalty in Equation~(\ref{eqn:objM2}) can also be expanded as 
$$
\int [\delta''(q)]^2 \mathrm{d} q = 
\sum_{k,k'} \delta^{(k)} \delta^{(k')}
\int \phi_{kB}(q) \phi_{k'B}(q) \mathrm{d} q = 
\vdelta^T \mathbf{\Gamma} \vdelta \, ,
$$
where $ \mathbf{\Gamma} = \left(\int \phi_{kB}(q) \phi_{k'B}(q) \mathrm{d}
  q \right)$ is a $K$-by-$K$ matrix. After obtaining the functional
principal components, this matrix is  known and fixed. 

In order to solve Equation~(\ref{eqn:distfunctional}), 
for the $s$ supernova sample, define 
$\vgamma = (M, \alpha, \vdelta^T)^T$, 
$y_s = \mu(z_s) - m_{s,B}$, and
$$\mathbf{x}_s = (1, C_s - \langle C\rangle, 
\beta_{sB}^{(1)} - \langle \beta_{B}^{(1)} \rangle, \cdots,
\beta_{sB}^{(K)}  - \langle \beta_{B}^{(K)} \rangle)^T\, .$$ 
Furthermore, define a $(K+2)$-by-$(K+2)$ matrix
$$ \mathbf{\Omega} = \left(
\begin{array}{cc}
\vzero & \vzero \\
\vzero & \mathbf{\Gamma} \\ 
\end{array}
\right)\, . $$
This is an enlargement of the matrix $\mathbf{\Gamma}$ by adding two
columns and two rows of zeros. With these notations,
Equation~(\ref{eqn:distfunctional}) becomes 
\[
\sum_{s} \frac{1}{\sigma^2_s}\left[
y_s - \mathbf{x}_s^T \vgamma
\right]^2 +\eta \vgamma^T \mathbf{\Omega} \vgamma\,.
\]
Stacking the $y_s$'s into a vector $\vy = (y_1, \cdots, y_S)$, and the
$\vx_s$'s into a matrix $\vX = (\vx_1, \cdots, \vx_S)^T$. With a
diagonal matrix $\vW = \mathrm{diag}(1/\sigma_1^2, \cdots,
1/\sigma_S^2)$, the $\vgamma$ has an explicit solution, 
$ \hat{\vgamma} = (\vX^T\vW\vX + \eta \mathbf{\Omega})^{-1}
 \vX^T\vW\vy $. From here, we find
 $\mathbf{H} = \vX (\vX^T\vW\vX + 
\eta \mathbf{\Omega})^{-1} \vX^T\vW$ is the hat matrix satisfying 
$\hat{\vy} = \mathbf{H}\vy$. The degree of freedom for this fitted
model is simply the trace of the hat matrix, i.e.,
$\mathrm{tr}(\mathbf{H})$. This result can be found in 
\citet[Section 7.2]{hastie09elements}.  The degree of freedom of the
residual $\chi^2$  equals the sample size minus this quantity.
The vector  $\vdelta$ can be extracted from the corresponding entries
of the estimated $\hat{\vgamma}$.  Thereby, we have also obtained the
estimated function $\delta(q)$ in the functional linear 
term~(\ref{eqn:functionalQ}), i.e., $\hat{\delta}(q) = \sum_{k=1}^K
\hat{\delta}^{(k)}\phi_{kB}(q)$. 
}

\section{Supplementary Table}

\begin{table}[ht]
\centering
\caption{Example Supplementary Table \label{tbl:suptable}}
\begin{tabular}{rllrlllrrrrrr}
  \hline
 & SNe & Survey & redshift & Type1 & Type2 & Type3 & Bmax & Vmax & B\_scores1 & V\_scores1 & R\_scores1 & I\_scores1 \\ 
  \hline
1 & SN1998de & LOSS & 0.0157 &  &  &  & 17.3320 & 16.6420 & 4.2330 & 5.0700 & 4.5330 & 3.4840 \\ 
  2 & SN1998dh & LOSS & 0.0077 &  &  &  & 13.8900 & 13.8220 & 1.5960 & 0.9880 & 0.3940 & -0.6010 \\ 
  3 & SN1998ef & LOSS & 0.0171 &  &  &  & 14.8560 & 14.8900 & 2.3430 & 0.7600 & 0.9070 & -0.7250 \\ 
  4 & SN1999ac & LOSS & 0.0098 & HVG & CN & N & 14.1040 & 14.0570 & 1.0680 & 0.7680 & -0.2390 & -0.2690 \\ 
  5 & SN1999by & LOSS & 0.0027 &  &  &  & 13.5360 & 13.0720 & 3.0430 & 4.7290 & 4.6650 & 4.1910 \\ 
  6 & SN1999cl & LOSS & 0.0081 &  &  & N & 14.8640 & 13.7560 & 0.8980 & 0.5540 & 0.2940 & -0.3570 \\ 
  7 & SN1999cp & LOSS & 0.0103 & LVG & BL & N & 13.9470 & 13.9630 & 1.2130 & 0.3600 & 1.1060 & -0.2900 \\ 
  8 & SN1999da & LOSS & 0.0121 & FAINT & CL &  & 16.5980 & 16.0310 & 3.8350 & 4.5650 & 4.5800 & 5.0750 \\ 
  9 & SN1999dk & LOSS & 0.0141 &  &  &  & 14.8280 & 14.7550 & 1.3930 & 0.7620 & 0.1030 & -0.5770 \\ 
  10 & SN1999dq & LOSS & 0.0137 & HVG & SS &  & 14.4790 & 14.3650 & -0.1200 & -0.3700 & -0.8570 & -1.9100 \\ 
   \hline
\end{tabular}
\tablecomments{
Ten SNe Ia and twelve representative columns for the supplementary table. 
The columns Type1, Type2 and Type3 are the SNe Ia Type from \cite{benetti2005diversity}, 
\cite{branch2009comparative}, and \cite{wang2009improved}, respectively. The columns Bmax and Vmax
are the maximal magnitude for $B$ and $V$ band. The last four columns corresponds to the first FPCA score
for the $B,V,R,I$ band, respectively.
}
\end{table}
\section{Supplementary Figures}

\begin{figure*}[h]
\centering
\includegraphics[width = 0.8\textwidth]{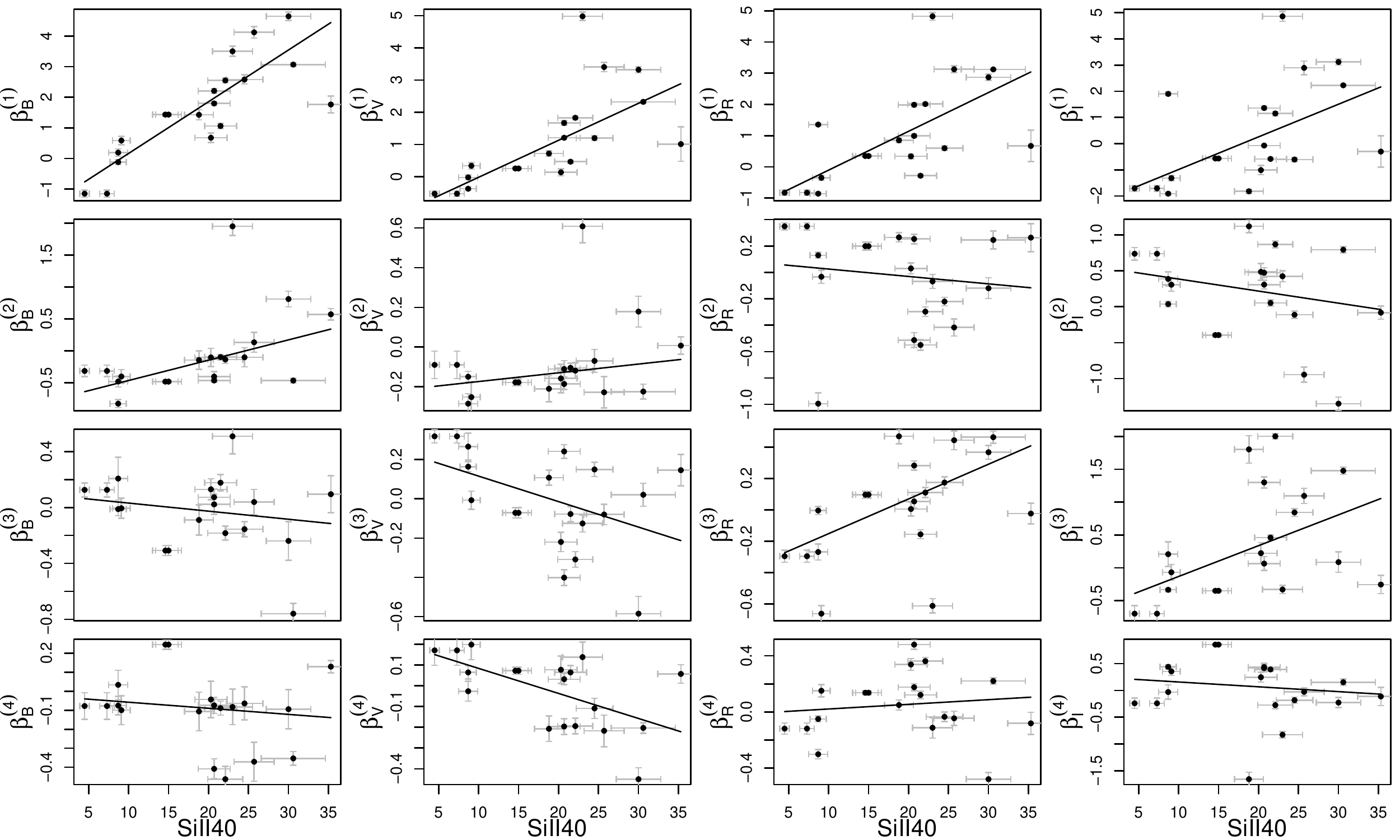}
\caption{ The scores $\beta^{(k)}_\lambda$ against the
  pseudo-equivalent width (pEW) of \sirp.
} \label{fig:siii40}
\end{figure*}

\begin{figure}[h]
\centering
\includegraphics[width = 0.45\textwidth]{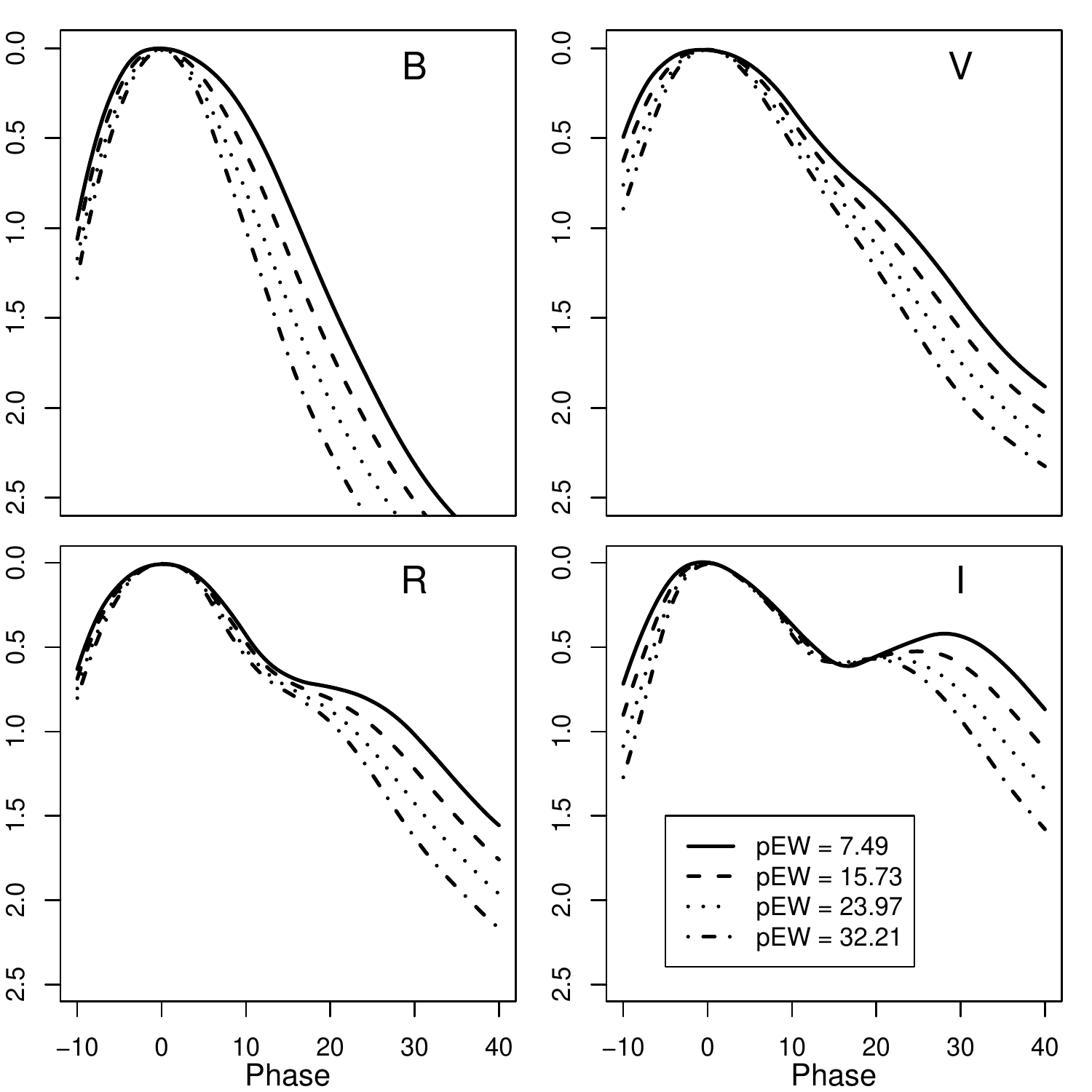}
\caption{ The ``average'' light curve shape at four different levels of the
  pseudo-equivalent width (pEW) of \sirp.
} \label{fig:siii40toshape}
\end{figure}

\begin{figure*}[h]
\centering
\includegraphics[width = 0.8\textwidth]{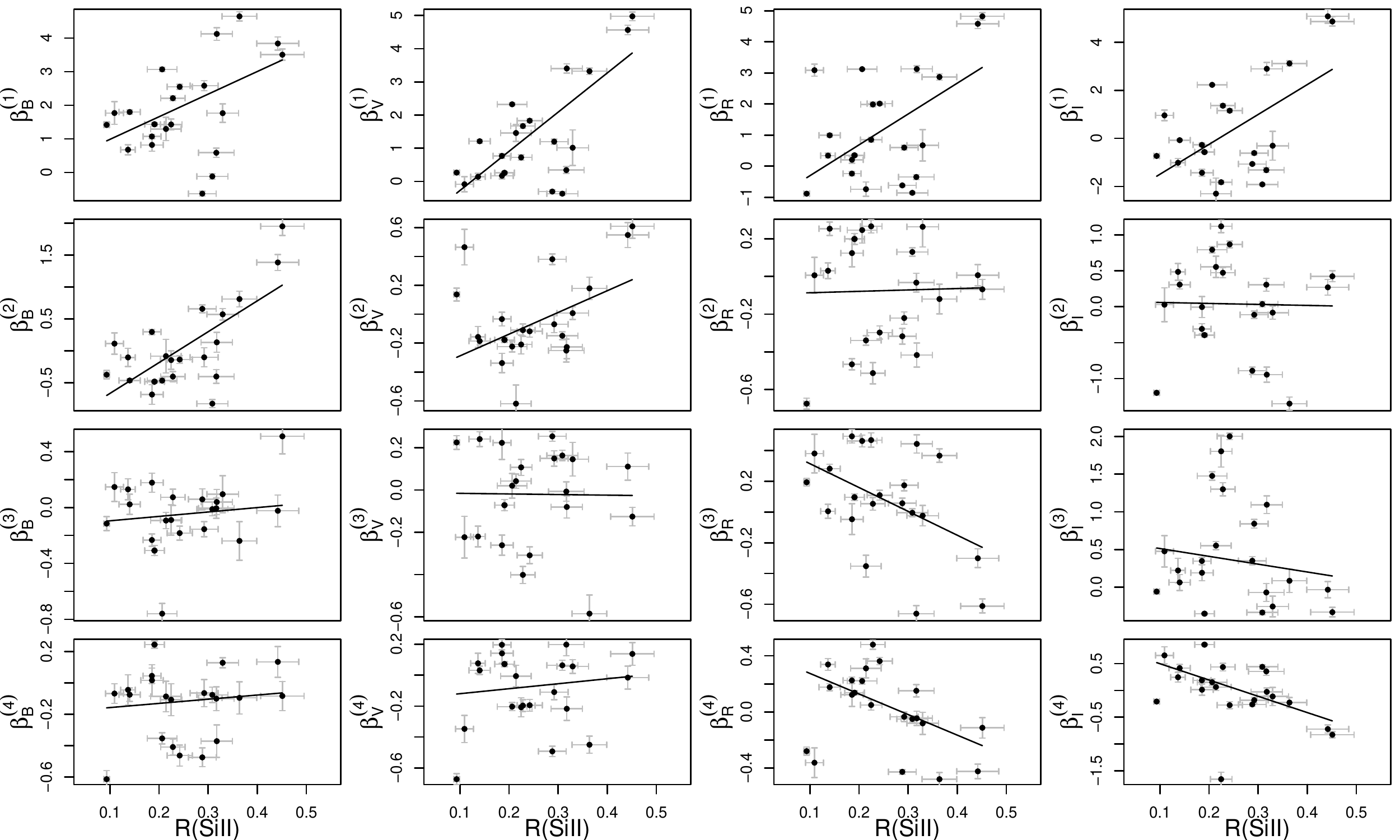}
\caption{  The scores $\beta^{(k)}_\lambda$ against the
  pseudo-equivalent width (pEW) of 
$\mathcal{R}(\mathrm{Si\ II})$.
} \label{fig:ratio1-1}
\end{figure*}

\begin{figure}[h]
\centering
\includegraphics[width = 0.45\textwidth]{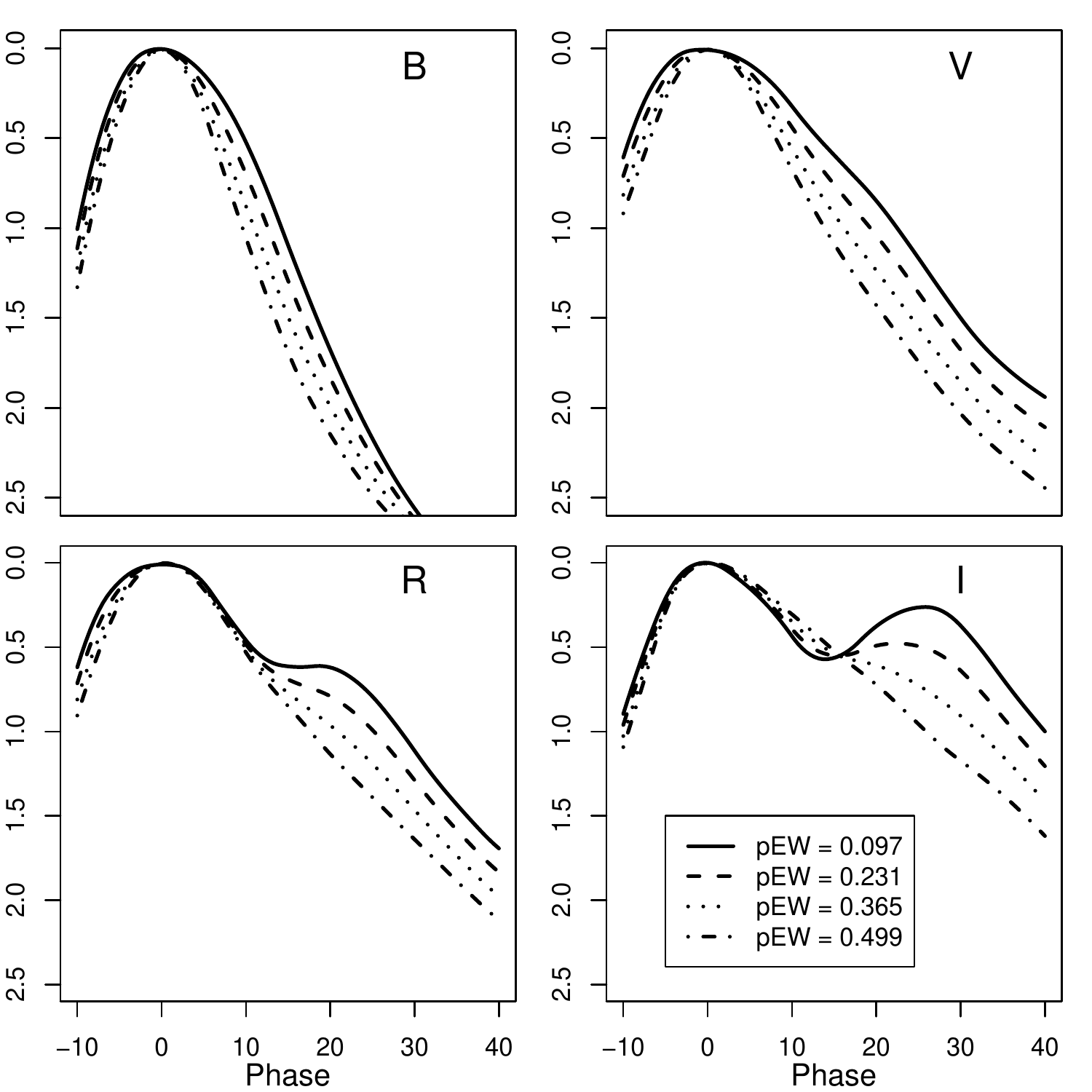}
\caption{ The ``average'' light curve shape at four different levels of the
  pseudo-equivalent width (pEW) of
$\mathcal{R}(\mathrm{Si\ II})$.
} \label{fig:ratio1-2}
\end{figure}

\begin{figure*}[h]
\centering
\includegraphics[width = 0.8\textwidth]{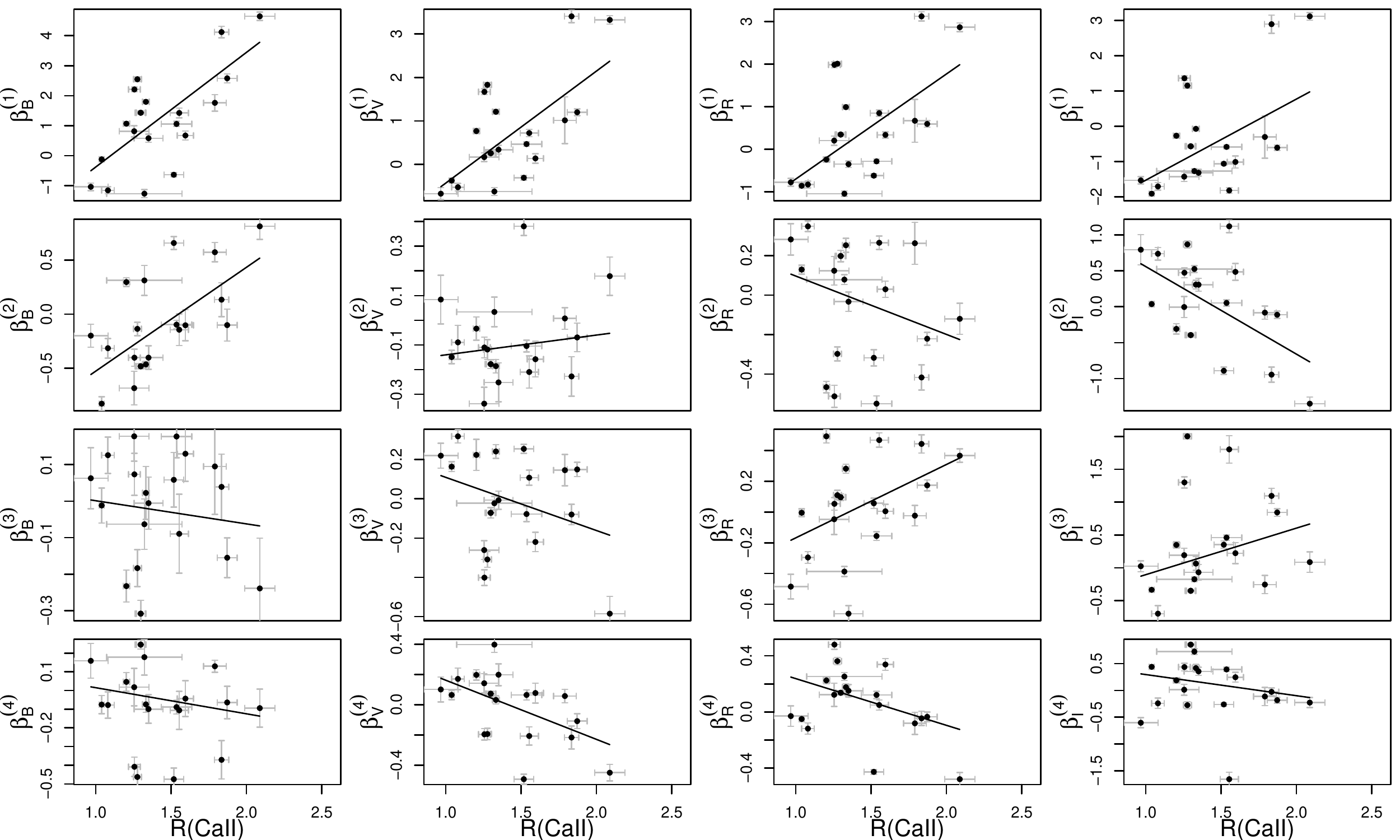}
\caption{  The scores $\beta^{(k)}_\lambda$ against the
  pseudo-equivalent width (pEW) of 
$\mathcal{R}(\mathrm{Ca\ II})$.
} \label{fig:ratio2-1}
\end{figure*}

\begin{figure}[h]
\centering
\includegraphics[width = 0.45\textwidth]{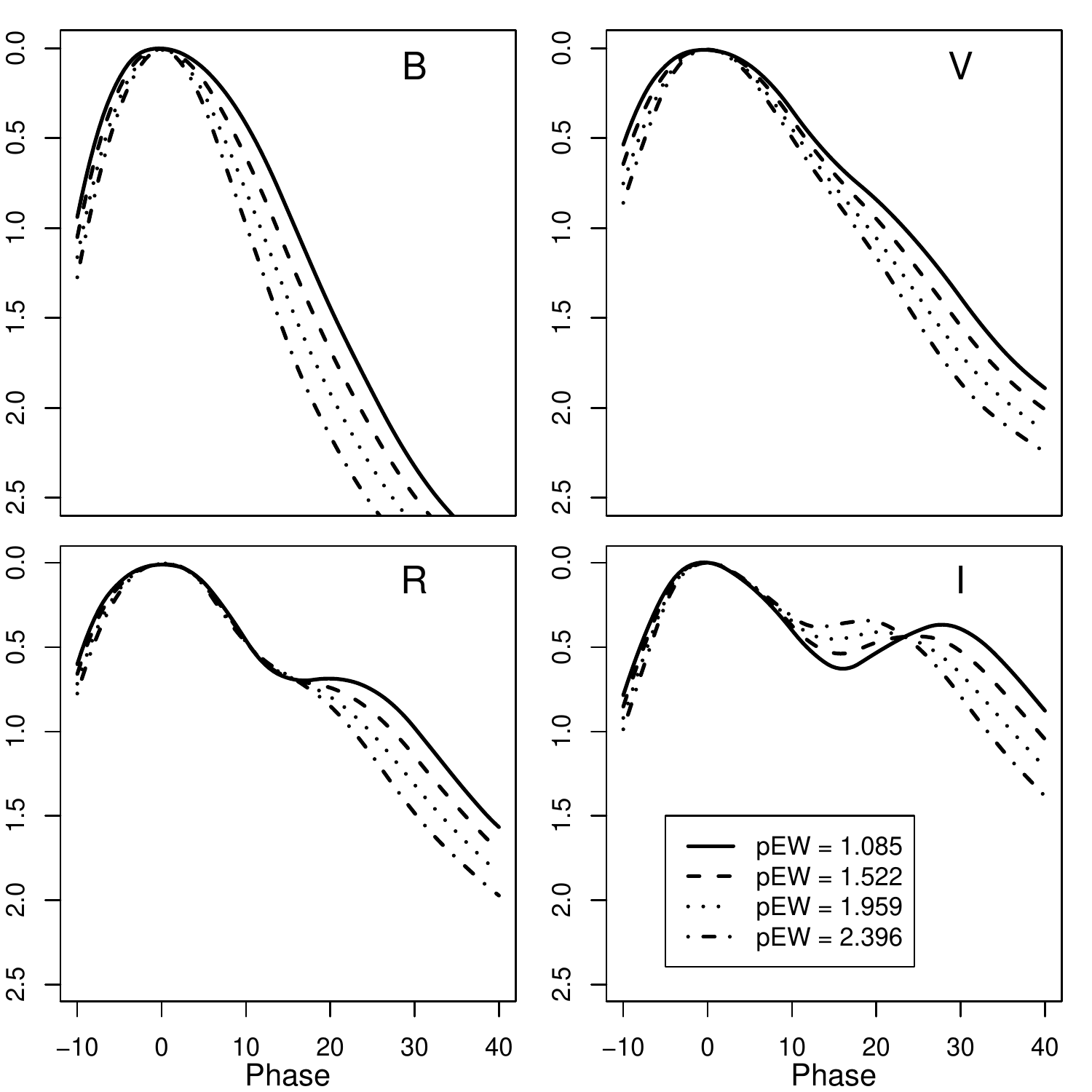}
\caption{ The ``average'' light curve shape at four different levels of the
  pseudo-equivalent width (pEW) of 
$\mathcal{R}(\mathrm{Ca\ II})$.
} \label{fig:ratio2-2}
\end{figure}

\end{document}